\documentclass[%
reprint,
nofootinbib,
amsmath,amssymb,
aps,
preprintnumbers
]{revtex4-1}

\usepackage{graphicx,color,amsmath}
\usepackage[export]{adjustbox}
\usepackage{slashed}
\usepackage{amssymb}
\usepackage{mathrsfs}
\usepackage{adjustbox}
\usepackage{hyperref}
\usepackage{booktabs}
\usepackage{amsfonts}
\usepackage{verbatim}
\usepackage{multirow}
\usepackage{comment}
 \usepackage{url}
\usepackage{graphicx}% Include figure files
\usepackage{bm}% bold math
\usepackage{slashed}
\usepackage{color}
\usepackage{amssymb}
\usepackage[dvipsnames]{xcolor}
\usepackage{tikz}
\usetikzlibrary{decorations.pathmorphing}
\usetikzlibrary{decorations.markings}
\usepackage{multirow}
\usepackage[thicklines]{cancel}
\usepackage{makecell}
\usepackage{bbold}
\usepackage{cancel}
\usepackage{xcolor,colortbl}
\usepackage{amssymb}% http://ctan.org/pkg/amssymb
\usepackage{pifont}% http://ctan.org/pkg/pifont

\newcommand{\spl}[1]{\left|{#1}\right\rangle}
\newcommand{\spr}[1]{\left|{#1}\right]}
\newcommand{\splf}[1]{\left\langle{#1}\right|}
\newcommand{\sprf}[1]{\left[{#1}\right|}
\newcommand{\scpl}[2]{{\left\langle #1\,#2\right\rangle}}
\newcommand{\scpr}[2]{{\left[ #1\,#2\right]}}

\newcommand{\Mpl}{{M_{\rm Pl}}}

\newlength{\Oldarrayrulewidth}
\newcommand{\Cline}[2]{%
  \noalign{\global\setlength{\Oldarrayrulewidth}{\arrayrulewidth}}%
  \noalign{\global\setlength{\arrayrulewidth}{#1}}\cline{#2}%
  \noalign{\global\setlength{\arrayrulewidth}{\Oldarrayrulewidth}}
}

\usepackage{tikz}
\usepackage{tkz-euclide}
\usetkzobj{all}
\usetikzlibrary{decorations.pathmorphing}	% To create gluons and vector bosons
\tikzset{
    v/.style={decorate, decoration={snake, segment length=3mm, amplitude=0.75mm}, draw},
    f/.style={draw=black, postaction={decorate},
        decoration={markings,mark=at position .6 with {\arrow[very thick]{latex}}}},
    fb/.style={draw=black, postaction={decorate},
        decoration={markings,mark=at position .4 with {\arrowreversed[very thick]{latex}}}},
    fnar/.style={draw=black},
    g/.style={decorate, draw=black,
        decoration={coil,amplitude=3pt, segment length=3.5pt}},
    s/.style={dashed,draw=black, postaction={decorate},
        decoration={markings,mark=at position .55 with {\arrow[very thick]{latex}}}},
    sb/.style={dashed,draw=black, postaction={decorate},
        decoration={markings,mark=at position .55 with {\arrowreversed[draw=black,very thick]{latex}}}},
    snar/.style={dashed,draw=black,line width =1.25pt},
}

\definecolor{mypurple}{RGB}{164,64,214}

\newcounter{qnumber}

\definecolor{beaublue}{rgb}{0.74, 0.83, 0.9}
\definecolor{darkseagreen}{rgb}{0.56, 0.74, 0.56}
\definecolor{emerald}{rgb}{0.31, 0.78, 0.47}
\definecolor{columbiablue}{rgb}{0.61, 0.87, 1.0}
\definecolor{lightgray}{rgb}{0.83, 0.83, 0.83}
\definecolor{cornellred}{rgb}{0.7, 0.11, 0.11}
\definecolor{navyblue}{rgb}{0.0, 0.0, 0.5}
\definecolor{amethyst}{rgb}{0.6, 0.4, 0.8}
\definecolor{yellow}{rgb}{1.0, 1.0, 0.0}
\definecolor{firebrick}{rgb}{0.7, 0.13, 0.13}
\definecolor{tangerineyellow}{rgb}{1.0, 0.8, 0.0}
\definecolor{deepfuchsia}{rgb}{0.76, 0.33, 0.76}
\definecolor{amber}{rgb}{1.0, 0.75, 0.0}

\hypersetup{colorlinks,bookmarksopen,bookmarksnumbered,citecolor=navyblue,
linkcolor=cornellred,pdfstartview=cornellred,urlcolor=cornellred}

\begin{document}

\begin{flushright}
IPMU\,19-0091
\end{flushright}

\title{On amplitudes, resonances and the ultraviolet completion of gravity}

\author{Rodrigo Alonso}
\email{rodrigo.alonso@ipmu.jp}
\affiliation{Kavli Institute for the Physics and Mathematics of the Universe (WPI) \\ University of Tokyo, Kashiwa, Chiba, 277-8583 Japan.}
\author{Alfredo Urbano}
\email{alfredo.urbano@sissa.it}
\affiliation{INFN, sezione di Trieste, SISSA, via Bonomea 265, 34136 Trieste, Italy; \\ 
IFPU, Institute  for  Fundamental Physics  of  the  Universe,  Via  Beirut  2,  34014 Trieste, Italy.}
%\affiliation{IFPU, Institute  for  Fundamental Physics  of  the  Universe,  Via  Beirut  2,  34014 Trieste, Italy.}
\date{\today}

\begin{abstract}
%We  
This letter constructs, making use of the on-shell %amplitude methods%
spinor-helicity formalism, a possible  ultraviolet completion of gravity following a ``bottom-up'' approach. %perspective.
 The assumptions of locality, unitarity and causality  {\it i)} require an infinite tower of 
resonances with increasing spin and quantized mass, {\it ii)} introduce  a  duality relation among crossed scattering channels, and {\it iii)} 
dress all gravitational amplitudes in the Standard Model with a form factor that closely resembles either the Veneziano or the Virasoro-Shapiro amplitude in string theory.
As a consequence
of unitarity, the theory {\it predicts} leading order  deviations from General Relativity %that become manifest %when 
in the coupling of gravity to fermions %in the infrared
  that %{\color{BlueViolet} do arise} %
  could be explained
   if space-time has torsion in addition to curvature.
%We  discuss a possible  ultraviolet completion of gravity following a ``bottom-up'' perspective.
%We first exploit the on-shell spinor-helicity formalism to recast all tree-level Standard Model four-point gravitational amplitudes 
%in spinor variables, and derive the  contribution from massive spin $J$ resonance exchange. 
%We then combine these ingredients under the assumptions of unitarity, locality and causality.
%The proposed  ultraviolet completion of gravity  {\it i)} requires an infinite tower of 
%resonances with 
%increasing spin and quantized mass, {\it ii)} introduces  a  duality relation among crossed channels, and {\it iii)} 
%dresses all the gravitational amplitudes with a form factor that closely resembles either the Veneziano or the Virasoro-Shapiro amplitude in string theory.
%To conclude,  we summarize  spin, quantum numbers and couplings of the resonances. 
%As a corollary of unitarity, the theory {\it predicts} leading order deviations from General Relativity that become manifest when gravity is coupled to fermions,
%  and that could be explained if space-time has torsion in addition to curvature.
\end{abstract}

\maketitle

%%%%%%%%%%%%%%%%%%%%%%%%%%%%%%%%%%%%%%%%%%%%%%%%%%
%%%%%%%%%%%%%%%%%%%%%%%%%%%%%%%%%%%%%%%%%%%%%%%%%%
\section{Introuduction}\label{sec:Intro}
%\paragraph*{1.\,Introduction:} 
Prior to the Large Hadron Collider turn on, our tested theory of Nature was a gauge effective field theory (EFT) with a scale $v$ at which unitarity was lost perturbatively. 
Unitarity, however, is sacred and its guardian -- as present data seem to indicate -- is the Higgs boson.
Despite all the apparent differences, the categorization above applies word for word to gravity when thought of as an EFT with scale $M_{\rm Pl} \equiv G_N^{-1/2}$ and diffeomorphisms as gauge transformations, even though as for who its `Higgs(es)' is (are), there is no experimental evidence at present.
% -- that is a gauge EFT
% with scale $M_{\rm Pl} \equiv G_N^{-1/2}$ -- even though as for who its `Higgs(es)' is (are), there is no experimental evidence at present.
Exposing the need for a ultraviolet (UV) completion of gravity by means of its similarities with the Standard Model (SM)
 is preeminently a particle physicist ``bottom-up'' approach, and it is not to say that this is the sole issue in gravity that needs addressing 
  (%to name a few more: the possibility to interpret gravity, unified with the other elementary interactions, in a full non-perturbative framework without any divergences;
  %the conceptual understanding of space-time singularities, relevant for both early cosmology and black hole physics; 
% the lack of a proper comprehension of the fundamental principles underlying black hole thermodynamics
 open problems range from a non-perturbative formulation to the understanding of singularities relevant in cosmology and astrophysics~\cite{Woodard:2009ns}). 
   %the singularity theorems and the ensuing breakdown of GR demonstrate, a fundamental understanding of the early Universe—in particular, its initial conditions near the ‘big bang’—and of the final %stages of black-hole evolution requires an encompassing theory. 
 It is, nonetheless, the point of view that rules the course of this letter.
\begin{figure}[!ht!]
	\centering
	\includegraphics[width=0.485\textwidth]{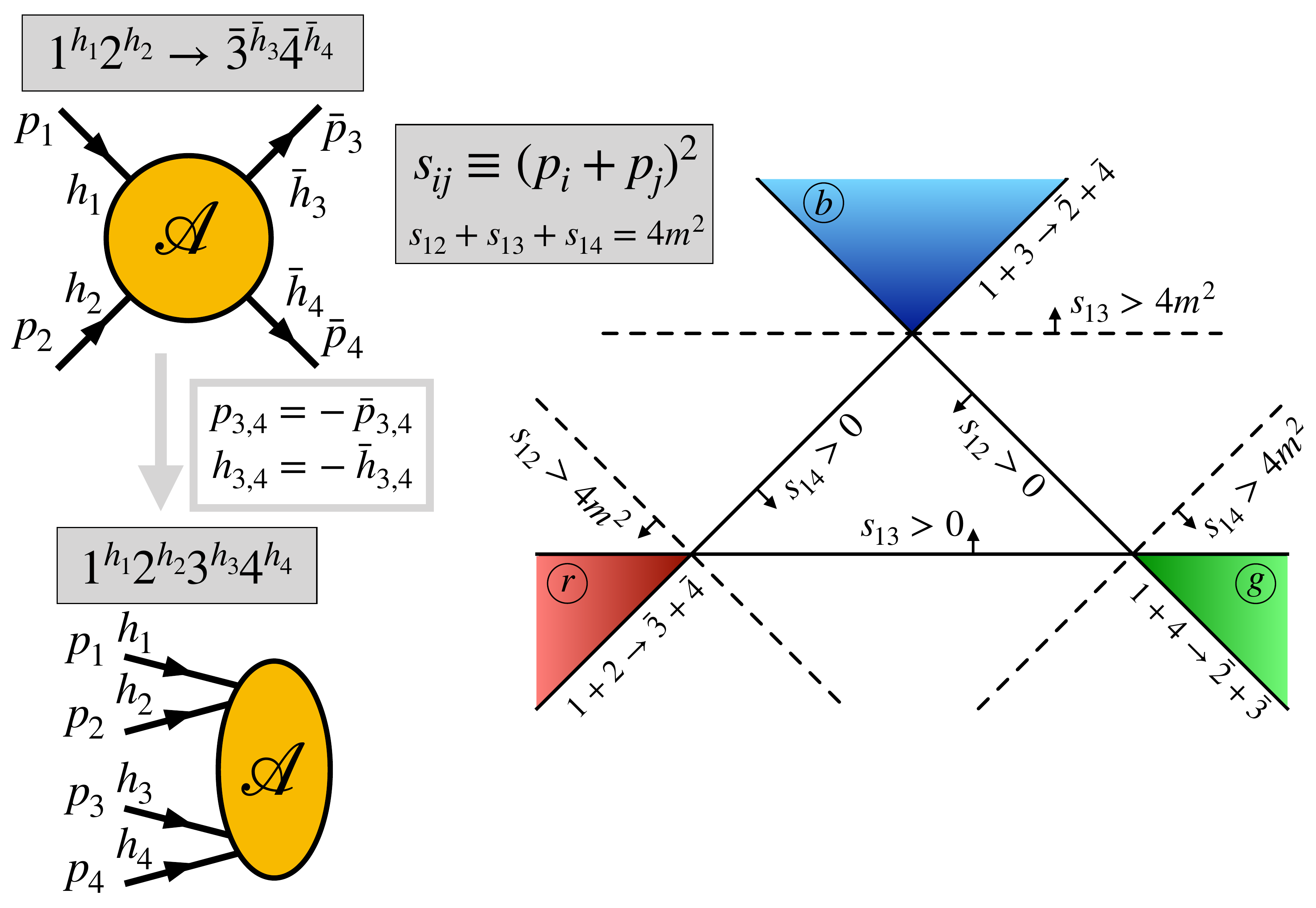}\vspace{-0.3cm}\\
	\begin{align}%\label{eq:Domains}
{\color{red}	\mathcal {A:}}& & s_{12}&\to s &s_{13}&\to t & s_{14}&\to u &h_{3,4}&\to-h_{3,4}\label{eq:Domains1}\\
{\color{blue}	\mathcal {A:}}& & s_{13}&\to s &s_{12}&\to t & s_{14}&\to u&h_{2,4}&\to-h_{2,4}\label{eq:Domains2}\\
{\color{ForestGreen}	\mathcal {A:}}& & s_{14}&\to s &s_{12}&\to t & s_{13}&\to u&h_{2,3}&\to-h_{2,3}\label{eq:Domains3}
	\end{align}
	\caption{\label{fig:schematic} Four-point amplitude with ingoing initial- and outgoing finale-state particles (top-left diagram) and with all particles taken incoming (bottom-left diagram). 
	On the right side, we show the three physical scattering regions {\color{red}{(r)}}, {\color{blue}{(b)}}, {\color{ForestGreen}{(g)}} as function of the kinematic invariants $s_{ij}$ and, in eqs.~(\ref{eq:Domains1},\,\ref{eq:Domains2},\,\ref{eq:Domains3}), the corresponding substitutions in terms of the conventional Mandelstam variables.
		For massless external particles the inner triangle collapses into a point.}
\end{figure}

One can further elaborate that the above formulation, even within particle physics and EFT, is obscured by (of course fundamental) differences in the gauge symmetry, treatment of the massless mediator, the universality of gravity etc. 
In order to sidestep these differences, the best suited subject of study are on-shell amplitudes; they circumvent gauge redundancies, field redefinitions and gauge fixing present at the Lagrangian level while making ostensible the high-energy behavior. 
Furthermore, as for practical implications, the derivation of these amplitudes with conventional Feynman rules is greatly involved (the three- and four-graviton vertex derived from the Einstein-Hilbert action 
have, respectively, about 100 and 2500 terms when fully expanded~\cite{DeWitt:1967uc})  to finally collapse in a %astonishingly 
%(or embarrassingly, for who embarks in such computation)  
%simple resulting
single-term remarkably simple amplitude~\cite{Berends:1974gk,Grisaru:1975bx,Sannan:1986tz}.

This is part of the evidence that supports postulating a theory with amplitudes as the starting and building blocks.
This approach is the on-shell amplitude program, see~\cite{Dixon:1996wi,Elvang:2013cua,Arkani-Hamed:2017jhn} for reviews. 
An important part of it is the on-shell spinor-helicity formalism, which seeks to exploit helicity (little group) transformation properties to determine the shape of amplitudes
%governing  the kinematics of particle scattering, thus 
while providing a common framework to formulate scattering amplitudes for all spins and,
 more recently, all masses~\cite{Arkani-Hamed:2017jhn}. Combined with  a recursive method to build higher point amplitudes from lower point ones (BCFW, CSW~\cite{Cachazo:2004kj,Britto:2005fq}), this program aims at a self-contained formulation of quantum field theories. It is not without its own challenges, as determination of off-shell contributions prevents  these methods to extend to arbitrary theories.

If one is to sidestep the Lagrangian formulation, however, care should be taken to ensure that the properties that are naturally implemented in it are satisfied by the formulated amplitudes. Since they are of particular importance in this work, let us briefly review them.
%\begin{align}
%\raisebox{14mm}{\begin{tikzpicture}
%\draw[black,fill=amethyst] (0.,0) circle (15pt);
%%\node(s)[circle, fill=black]{{\tiny }};
%%\draw [thick,blue,->] (-1.45,0.7)node[left]{{\footnotesize{$\mathcal{A}(s_{13},s_{12})$}}}--(-1.45,-0.7);
%\draw [thick,->]  (-0.75,0.75) node [anchor=north east] {$p_1$} --(-0.5,0.5);\draw [thick] (-0.55,0.55)--(-0.37,0.37);
%\draw [thick,->] (-0.75,-0.75) node [anchor= south east] {$p_2$} -- (-.5,-.5);\draw [thick] (-0.55,-0.55)--(-0.37,-0.37);
%\draw [thick,->]  (0.75,0.75) node [anchor=north west] {$p_3$}--(0.5,0.5);\draw [thick] (0.55,0.55)--(0.37,0.37);
%\draw [thick,->] (0.75,-0.75) node [anchor=south west] {$p_4$}--(.5,-.5);\draw [thick] (0.55,-0.55)--(0.37,-0.37);
%%\draw [thick,blue,->] (-0.7,-0.95)node[below right]{{\footnotesize{$\mathcal{A}(s_{12},s_{13})$}}}--(0.7,-0.95);
%\end{tikzpicture} \nonumber
%}
%\includegraphics[width=0.335\textwidth]{Madeltam.pdf} 
%%&~~s_{ij} = (p_i + p_j)^2
%\end{align}
 \textbf{Locality}. Interactions are either point-like or mediated by the exchange of particles propagating between two space-time points.  
This elementary principle dictates the non-analytic 
structure of scattering amplitudes: 
the only singularities occur  when one or more intermediate particles go on-shell. 
%This comprises both tree-level and loop diagrams with singularities given by simple poles and branch cuts, respectively.
\textbf{Unitarity and Causality}. The scattering matrix is a unitary operator as a consequence of probability conservation, and 
causality implies that local observables must commute outside the light-cone in position space~\cite{GellMann:1954db}. Positivity, derived from unitarity, will play a central role in the analysis of this letter.
These principles impose, 
in addition,
stringent constraints on the high-energy behavior of scattering amplitudes~\cite{Adams:2006sv};
of relevance in this work will 
 be the extension of the Froissart bound~\cite{Froissart:1961ux} to a theory with a massless graviton~\cite{Camanho:2014apa} in the forward limit and the Cerulus-Martin bound~\cite{Cerulus:1964cjb,Epstein:2019zdn} in the hard scattering region.
Furthermore, the property of causality results in \textbf{analyticity} of the scattering amplitude thus making possible to represent the latter by means of a dispersion relation via Cauchy's theorem~\cite{Gribov:2009zz}. 

In line with the ``bottom-up'' perspective mentioned above, we shall explore in this work whether it is possible to obtain a UV completion of gravity by adding massive resonances, and, if so, what are the properties required of them. 
To this end, we shall combine the on-shell spinor-helicity formalism with the above-mentioned fundamental properties.
In addition, the discussion will be restricted to {\bf weak coupling} and {\bf tree level}, which is to say in typical EFT language that the new resonances should lay below $\Mpl$.

%Before proceeding, let us make clear that when we talk about UV completion we do not pretend to construct a quantum theory of gravity.
%The goal of this paper is to understand from the ``bottom-up'' what kind of physics is necessary, given the assumptions enumerated before, to tame the bad high-energy behavior of scatterinf amplitudes in %the presence of gravity. 

%It is in summary within the framework of amplitude methods that we will phrase the need to complete gravity while complying with the above stated elementary conditions and studing tree level 4 point amplitudes. In this endeavor, section I presents the formalism and 4-point amplitudes for Hilbert-Einstein and spin J massive mediator, while section II combines the two for a UV completion and section III analyzes the ansatz we propose.

%%%%%%%%%%%%%%%%%%%%%%%%%%%%%%%%%%%%%%%%%%%%%%%%%%
%%%%%%%%%%%%%%%%%%%%%%%%%%%%%%%%%%%%%%%%%%%%%%%%%%
\section{Amplitudes for Gravity}  \label{sec:Elem}
In this section we construct the elementary components of this study, these are three- and four-point amplitudes mediated by gravitational and massive spin $J$ resonance interactions (section~\ref{sec:OrdinaryGravity} and \ref{sec:SpinJExchange}, respectively). It will also serve as introduction to the formalism for the unacquainted reader.

%%%%%%%%%%%%%%%%%%%%%%%%%%%%%%%%%%%%%%%%%%%%%%%%%%
\subsection{Graviton mediated amplitudes}\label{sec:OrdinaryGravity}
On-shell Dirac or Weyl spinors, polarization vectors and polarization tensors are objects that interpolate between the Lorentz group $SO(4)\sim SU(2)_L\times SU(2)_R$ and the little group -- that is $U(1)_{LG}$ for massless or $SU(2)_{LG}$ for massive particles -- and, as such, transform under representations of both. The on-shell spinor-helicity formalism in essence seeks to use group theory {\it in both groups} to determine the shape of amplitudes. The simplest case is that of massless fermions; denote as ${}_\alpha\!\spl{p}$ (${}^{\dot\alpha}\!\spr{p}$) an on-shell momentum $p_\mu$ left-handed (right-handed) spinor with $SU(2)_L$ ($SU(2)_R$) index $\alpha$ ($\dot\alpha$) and anti-symmetric metric $\epsilon^{\alpha\beta}$ ($\epsilon^{\dot{\alpha}\dot{\beta}}$), $\epsilon^{12}=-\epsilon^{21}=1$. The spinor $\spl{p}$ ($\spr{p}$) represents a helicity $h=-1/2$ ($h=1/2$) particle and hence transforms under $U(1)_{LG}$ as $\spl{p}\to \spl{p}e^{-i\phi/2} (\spr{p}e^{i\phi/2})$. Amplitudes are Lorentz invariant quantities but they comprise little group representations, and so a valid amplitude for two left-handed fermions is $\epsilon^{\alpha\beta}{}_{\beta}\!\spl{p_1}{}_\alpha\!\spl{p_2}\equiv\scpl{p_1}{p_2}$. The massive case is obtained upgrading the spinor to the fundamental representation $\spl{p^I}$ with $I$ an $SU(2)_{LG}$  index which again has anti-symmetric metric $\epsilon_{IJ}$.
Here we will typically omit the index $I$ but use boldface $\spl{\bf p}$ to separate it from the massless case following~\cite{Arkani-Hamed:2017jhn}.  The reader accustomed to Dirac spinors will find these variables demystified by the relation $u^I(p)=(\spl{p^I},\spr{p^I})$ in Weyl's basis for $\gamma^\mu$. 

One has then that higher spin is simply built out of the fundamental representations; the familiar polarization vectors read, e.g. $\epsilon_\mu^-(p)=\sprf{\xi}\sigma_\mu\spl{p}/\sqrt{2}\scpr{\xi}{p}$ with $\xi$ an auxiliary spinor and $[\sigma_\mu]_{\alpha\dot\alpha} =(1,\vec\sigma)$ with $\vec \sigma$ the Pauli matrices. 
Objects like polarization vectors or tensors, however, will not appear in the formalism since one rather starts from amplitudes and demands proper little group scaling; 
for instance an amplitude describing a particle with momentum $p_1$ and helicity $-1$, $\mathcal A_{p_1^{-1}}$, scales as $e^{-i\phi}\mathcal A_{p_1^{-1}}$. 
As conventional in amplitude methods we will derive amplitudes with all particles coming in, and we summarize our conventions about 
kinematics in fig.~\ref{fig:schematic}. Given the scattering of particles 1,2 with momenta $p_1,p_2$ and helicities $h_{1,2}$ into particles $3,4$ with momenta $\bar p_{3,4}$ and helicities $\bar h_{3,4}$, denoted here $1^{h_1}2^{h_2}\to \bar{3}^{\bar{h}_3}\bar{4}^{\bar{h}_4}$, the all-incoming amplitude, $1^{h_1}2^{h_2}3^{h_3}4^{h_4}$, is obtained changing the sign of the momenta (and with it the helicities) of the outgoing particles
%Compared to a typical scattering process $1^{h_1}2^{h_2}\to \bar{3}^{\bar{h}_3}\bar{4}^{\bar{h}_4}$ with ingoing initial-state momenta $p_{1,2}$ and helicities $h_{1,2}$ and outgoing final-state momenta $\bar{p}_{3,4}$ and helicities $\bar{h}_{3,4}$ we flip the latter 
 $p_{3,4} = -\bar{p}_{3,4}$, $h_{3,4} = -\bar{h}_{3,4}$. %to obtain the corresponding process with all particles ingoing
Equivalently starting from an all incoming amplitude and taking some of the legs out going, $s_{ij}\equiv(p_i+p_j)^2$ will turn into one of the three Mandelstam variables $s,t,u$ as given in fig.~\ref{fig:schematic} whereas %a negative-argument
 an incoming left-handed spinor $\spl{p}$ will turn into $\spl{-\bar{p}}=e^{i\varphi}\spl{\bar{p}}$ 
 representing an {\it outgoing momentum $\bar p$ helicity} +1/2  particle (our convention for the phase $\varphi$ is in appendix~\ref{app:A}).
Here we find it useful to write amplitudes in terms of $s_{ij}, \spl{i},\spr{j}$ since they make clear the connection between different physical processes related by crossing transformations, symmetries under particle exchange (eg. $1\leftrightarrow2$) and can be viewed as a function with support on 3 different disconnected regions {\color{red}(r)}, {\color{blue}(b)}, {\color{ForestGreen}(g)}  as shown in fig.~\ref{fig:schematic}.
 Finally, for energies close to $M_{\rm Pl}$ the approximation of massless matter (i.e. SM particles), which we shall adopt in the following, is excellent.

 The first step in 
 our quest for the ``bottom-up'' UV completion of gravity with amplitude methods 
 is to build the on-shell three-point amplitudes 
 describing interactions of SM particles with gravity 
  by means of Lorentz-invariant combinations of appropriate powers of spinor variables. 
  As stated in the introduction, we restrict to tree-level amplitudes, and our theory of gravity is General Relativity (GR).
  A graviton of helicity $h_g = 2$ and momentum $q$ coupling to  
a particle of momentum $p_1$, helicity $h$ and a particle $2$ of helicity $\pm h$ reads by little group scaling:
\begin{equation}
 \scpr{1}{2}^{h\pm h-2}\scpr{1}{q}^{2+h\mp h}\scpr{2}{q}^{2-(h\mp h)}\,,
\end{equation}
%\begin{align}
%\raisebox{-9.5mm}{\begin{tikzpicture}
%	\draw [thick]  (-1,1) node [anchor=north east] {1,$\, h$}--(-0.45,0.15);%\draw [thick] (-0.3,0.3)--(-0.12,0.12);
%	\draw [thick] (-1,-1) node [anchor= south east] { 2,$\, -h$} -- (-.45,-.15);%\draw [thick] (-0.5,-0.5)--(-0.12,-0.12);
%	\draw [style={decorate, decoration={snake}},thick] (-1.1,0.034) node [anchor=east] {$q, \pm2$} -- (-0.4,0.034);
%	\draw [style={decorate, decoration={snake}},thick] (-1.1,-0.034) -- (-0.4,-.034);
%	 \draw[black,fill=yellow,thick] (-0.48,0) circle (9pt);
%	\end{tikzpicture}}= c
%\left\{
%\begin{array}{c}
% \scpr{1}{3}^{h_1+h_2-2}\scpr{1}{q}^{h_1-h_2+2}\scpr{2}{q}^{h_2-h_1+2}  \\ 
%	    \\
% \scpl{1}{3}^{-h_1 - h_2 -2}\scpl{1}{q}^{h_2-h_1+2}\scpl{2}{q}^{h_1-h_2+2}
%\end{array}
%\right.\nonumber
%\begin{aligned}
%\end{aligned}
%\end{align}
with the same and opposite $h$ sign yielding, respectively,  mass-dimension of $2h+2$ or $2$, and where the short-hand notation $\spr{p_1}=\spr{1}$ etc. is implied. 
Given that an $n$-point amplitude has mass-dimension $4-n$ and gravity's coupling is $\kappa=\sqrt{8\pi}/\Mpl$ with mass-dimension $-1$, we find that
 only one of the two amplitudes is generated. Explicitly
\begin{align}
\raisebox{-9.5mm}{\begin{tikzpicture}
	\draw [thick]  (-1,1) node [anchor=north east] {1,$\, h$}--(-0.45,0.15);%\draw [thick] (-0.3,0.3)--(-0.12,0.12);
	\draw [thick] (-1,-1) node [anchor= south east] { 2,$-h$} -- (-.45,-.15);%\draw [thick] (-0.5,-0.5)--(-0.12,-0.12);
	\draw [style={decorate, decoration={snake}},thick] (-1.1,0.034) node [anchor=east] {$q, h_g$} -- (-0.4,0.034);
	\draw [style={decorate, decoration={snake}},thick] (-1.1,-0.034) -- (-0.4,-.034);
	 \draw[black,fill=tangerineyellow,thick] (-0.48,0) circle (9pt);
	\end{tikzpicture}}= \frac{\sqrt{8\pi}}{\Mpl}
\left\{
\begin{array}{cc}
 \frac{\scpr{1}{q}^{2h+2}\scpr{2}{q}^{-2h+2}}{\scpr{1}{2}^2}\,, & h_g = 2  \\ 
  &  \\
	   \frac{\scpl{1}{q}^{2-2h} \scpl{2}{q}^{2h+2}}{\scpl{1}{2}^{2}}\,, &  h_g = -2
\end{array}
\right.\nonumber
\begin{aligned}
\end{aligned}
\end{align}
This on-shell amplitude has $p_1+p_2+q=0$ and is gauge invariant, as can be checked by shifting $\spr{q}\to \spr{q+\xi}$ and projecting in terms of the  $\spr{1},\spr{2},\spr{q}$ spinors.
The absence of a $(+h,+h)$ amplitude means that gravity conserves helicity which at this level coincides with any quantum number that the particle might have. As we shall see in the rest of our analysis, this observation plays an important role. 

The four-point amplitude of order $O(\kappa^2)$  is generated by graviton exchange, and hence contains a pole. 
The residue of this pole factorizes into the product of two local three-point on-shell amplitudes, given above. 
One can, therefore, reconstruct the on-shell part of the four-point amplitude 
$1^{h}2^{-h} 3^{-h^{\prime}}4^{h^{\prime}}$
as
\begin{align}
		\raisebox{-18mm}{
	\begin{tikzpicture}
	\draw [thick] (-0.6,1.6) node[anchor= east] {1,$\,\,h$} -- (-.3,1.3);\draw [thick] (-.3,1.3)--(0,1);  
	\draw [thick] (-0.6,-.6+1) node[anchor= east] {$2$,$-h$} -- (-.3,-.3+1);\draw [thick] (-.3,-.3+1)--(0,1);  
	\draw [style={thick,decorate, decoration={snake}}] (0.035,1)  -- (0.035,-1);
	\draw [style={thick,decorate, decoration={snake}}] (-.035,1) -- (-.035,-1);
	\draw [thick] (-0.6,-1.6) node[anchor= east] {$4,\,\,h'$}--(-.3,-1.3);\draw [thick] (-.3,-1.3)--(0,-1); 
	\draw [thick] (-0.6,-.4) node[anchor= east] {$3,-h'$}--(-.3,-0.7);\draw [thick] (-.3,-.7)--(0,-1); 
	\draw[black,fill=tangerineyellow,thick] (0,1) circle (7pt);
	\draw[black,fill=tangerineyellow,thick] (0,-1) circle (7pt);
	\end{tikzpicture}}  %\propto  \scpr{1}{4}^{2h}\scpl{2}{3}^{2h}
&\qquad\begin{aligned}&\frac{\sqrt{8\pi}}{\Mpl}\frac{\scpr{1}{q}^{2h+2}\scpr{2}{q}^{2-2h}}{\scpr{1}{2}^{2}}\\
&\frac{1}{q^2}\\
&\frac{\sqrt{8\pi}}{\Mpl}\frac{\scpl{3}{\bar  q}^{2h'+2}\scpl{4}{\bar q}^{2-2h'}}{\scpl{3}{4}^{2}} \end{aligned}
\end{align}
%\begin{align}
%\mathcal{A}^{J=2\,{\rm graviton}}_{1^h 2^{-h}\to 3^{-h'}4^{h'}}&(s_{12},s_{13}) \nonumber \\ &= %\frac{\scpr{1}{q}^{2h+2}\scpr{2}{q}^{2-2h}}{M_{\rm Pl}\scpr{1}{2}^{2}}\frac{1}{q^2}
%\frac{\scpl{3}{\bar q}^{2-2h'}\scpl{4}{\bar  q}^{2h'+2}}{M_{\rm Pl}\scpl{3}{4}^{2}} \nonumber \\
%&=\frac{ (\splf{2}\hat P_{34}\spr{1})^{2h}(\splf{4}\hat P_{12}\spr{3})^{2h'} }{M_{\rm Pl}^2s_{13}^{2h+2h'-2} s_{12}}\,,\label{eq:SMGrav}
%\end{align}
where $\bar q\equiv -q$ given that the helicity $h_g = +2$ graviton with momentum $q$ enters the second vertex with momentum $-q$ and helicity $h_g = -2$. 
Two currents of possibly different helicity with $h'\geq h$ are considered for arbitrary SM external states. 
Manipulation of the expression above leads, using the on-shell conditions, to (for brevity let us denote $\mathcal A_{1^{h}2^{-h}3^{-h^{\prime}}4^{h^{\prime}}}\equiv\mathcal A_{h,h'}$)
\begin{align}
%\mathcal{A}&^{\rm GR}_{1^{h}2^{-h}3^{-h^{\prime}}4^{h^{\prime}}}=\nonumber\\&
\mathcal{A}_{h,h'}^{\rm GR}=
\frac{s_{13}^{1-h'-r}s_{14}^{1-h'+r}}{(8\pi)^{-1}M_{\rm Pl}^2s_{12}} (\scpr{1}{4}\scpl{2}{3})^{2h}(\splf{3}\!\hat P_{12}\!\spr{4})^{2h'-2h}\,, \label{MstGR}
\end{align}
where $r$ %\in \mathbb{Z}$ or $(2\mathbb{Z}+1)/2$
 is an integer for integer $h'$ or semi-integer is so is $h'$,
and $P_{ij}\equiv (p_i - p_j)/2$, $\hat P_{ij}=P_{ij}^\mu \sigma_\mu$. The appearance of this parameter is related to the extension of the amplitude off-shell; whereas the dependence on spinor variables is fully fixed by the little group, one has that on-shell (i.e. when $s_{12}=0$) $s_{13}=-s_{14}$, and so amplitude methods alone cannot determine $r$.\footnote{Locality bounds $r$ to range in the interval $-1+h<r<1-h$ in order to avoid double poles. This is because we have the 
scaling $\scpr{1}{4}\scpl{2}{3} \sim s_{14}$, $\splf{3}\!\hat P_{12	}\!\spr{4} \sim \sqrt{s_{13}s_{14}}$, and the condition above ensures that no negative powers of $s_{13,14}$ are present.} 
From the scaling with energy of the energy-momentum tensor, we find $-1+h'\leq r\leq1-h'$. 
Furthermore when a scalar current is present ($h=0$) gravity has no handle to distinguish between particles $1$ and $2$, and the amplitude must be $1\leftrightarrow 2$ symmetric. 
The ambiguity in $r$, therefore, reduces to two cases only, $h=h'=0, 1/2$ each characterized by one parameter $a,b$, as displayed explicitly
 in table~\ref{tab:1}.\footnote{Knowledge of the full amplitude in GR can be attained
  through a Feynman rule computation. Here however we keep the contact terms arbitrary. 
  In this sense, note that experimentally we have only tested the pole terms, i.e. the long range interaction} 
In this table we collect all tree-level SM scattering amplitudes mediated by gravity constructed explicitly by means of eq.~(\ref{MstGR}).
We also display gravitational Compton scattering (i.e. scattering among gravitons and SM particles) and graviton-graviton scattering. 
These cases have their helicity structure dictated by the same formula of eq.~(\ref{MstGR}) but now there are poles in all three Mandelstam variables and we find the denominator $s_{12}s_{13}s_{14}$ (i.e. $r=0$).

What is more, one has that the formula in eq.~(\ref{MstGR}), comprises all tree-level four-point amplitudes generated in GR. The amplitudes can be split into matter-matter, matter-graviton and graviton-graviton scattering (here and in the following `matter' generically refers to scalars, fermions and vectors). The fact that gravity does not change the helicity of matter implies that the amplitudes in table~\ref{tab:1} are the only non-vanishing matter-matter cases, one can see this diagrammatically and derive the helicity conservation rule $\sum h_i=0$. 
This is not clear, however, for scattering with gravity where the three point vertex with structure
\begin{align}
\raisebox{-9.5mm}{\begin{tikzpicture}
	\draw [style={decorate, decoration={snake}},thick]  (-0.95,1.05) --(-0.07,0.17);
	\draw [style={decorate, decoration={snake}},thick]  (-1,1) node [anchor=north east] {1,\,$+2$} --(-0.12,0.12);
	\draw [style={decorate, decoration={snake}},thick] (-1,-1) node [anchor= south east] { 2,$\,-2$} -- (-0.12,-0.12);
	\draw [style={decorate, decoration={snake}},thick] (-0.95,-1.05) -- (-0.07,-0.17);
	\draw [style={decorate, decoration={snake}},thick] (-1,0.034) node [anchor=east] {$q,+2$} -- (0.,0.034);
	\draw [style={decorate, decoration={snake}},thick] (-1,-0.034) -- (0.,-.034);
	\draw[black,fill=tangerineyellow,thick] (0,0) circle (9pt);
	\end{tikzpicture}}~
\begin{aligned}= &\quad\frac{\sqrt{8\pi}}{\Mpl}\frac{\scpr{1}{q}^6}{\scpr{1}{2}^2\scpr{2}{q}^2}
\label{eq:ggg}
%\sqrt2 \Lambda \left(\frac{\splf{\bf q}(\hat p_1-\hat p_3)\spr{\bf q}}{2\Lambda M}\right)^h
\end{aligned}
\end{align}
produces a diagram in which the helicity $-2$ is exchanged thus leading to a situation where  $\sum h_i\neq0$. 
The result of summing {\it all} diagrams, however, yields a vanishing amplitude in this case (not only for the pole terms but the full amplitude as can be obtained with Feynman diagrams). The same occurs for graviton scattering which makes table~\ref{tab:1} complete at tree level.
%In conclusion, table~\ref{tab:1} contains the complete set of tree-level  four-point amplitudes in GR.
%%%%%%%%%%%

{\small{
\begin{table*}[t]
\begin{center}
\begin{adjustbox}{max width=1\textwidth}
	\begin{tabular}{!{\vrule width 0.55pt}!{\vrule width 0.55pt}c!{\vrule width 0.55pt}!{\vrule width 0.55pt}c!{\vrule width 0.55pt}!{\vrule width 0.55pt}c!{\vrule width 0.55pt}!{\vrule width 0.55pt}c!{\vrule width 0.55pt}!{\vrule width 0.55pt}c!{\vrule width 0.55pt}!{\vrule width 0.55pt}} \Cline{0.55pt}{1-5} %||c||c||c||c||c||
		\multirow{2}{*}{$\mathcal{A}^{\rm GR}_{1^{h}2^{-h}3^{-h^{\prime}}4^{h^{\prime}}}$}  
	% \left|\mathcal{A}^{\rm GR}_{1^{h}2^{-h}3^{-h^{\prime}}4^{h^{\prime}}}\right|
		& \multirow{2}{*}{\textbf{Scalar}} & \multirow{2}{*}{\textbf{Fermion}} & \multirow{2}{*}{\textbf{Vector}}
		& \multirow{2}{*}{\textbf{Graviton}} \\ 
		&  & & & \\ \Cline{0.55pt}{1-5}
		%%%%%%%%%%%%%%%%%%%%%%%%%%%%%%%% SCALAR ROW
		%%%%%%%%%%%%%%%%%%%%%%%%%%%%%%%%
		\multirow{6}{*}{\textbf{Scalar}}  & 
		\multirow{3}{*}{$\frac{8\pi}{M_{\rm Pl}^2}\left(\frac{s_{13}s_{14}}{s_{12}} - as_{12}\right)$} &  \multirow{6}{*}{$\frac{8\pi\left(\splf{3} \hat P_{12}\spr{4}\right)}{M_{\rm Pl}^2}
		\left(\frac{s_{13}-s_{14}}{2s_{12}}\right) $} & 
		\multirow{6}{*}{$-\frac{8\pi\left(\splf{3} \hat P_{12}\spr{4}\right)^2}{M_{\rm Pl}^2 s_{12}}$} & \multirow{6}{*}{$\frac{8\pi\left(\splf{3} \hat P_{12}\spr{4}\right)^4}{M_{\rm Pl}^2 s_{12}s_{13}s_{14}}$}  \\ 
		 &  & & & \\
		& & & & \\ \cline{2-2}
		& \multirow{3}{*}{$\frac{8\pi}{M_{\rm Pl}^2}\left(\frac{s_{13}s_{14}}{s_{12}}+\frac{s_{12}s_{14}}{s_{13}}+\frac{s_{13}s_{12}}{s_{14}} \right)$}
		 & & & \\
		 &  & & & \\
          	&  & & & \\   \Cline{0.55pt}{1-5}
		%%%%%%%%%%%%%%%%%%%%%%%%%%%%%%%% FERMION ROW
		%%%%%%%%%%%%%%%%%%%%%%%%%%%%%%%%
		\multirow{6}{*}{\textbf{Fermion}}  & 
		%\multirow{4}{*}{$\frac{\lambda_2 P_{34}\tilde{\lambda}_1}{M_{\rm Pl}^2}\left(\frac{s_{13}}{s_{12}} - \frac{s_{14}}{s_{12}}\right)$} 
		& \multirow{3}{*}{$-\frac{8\pi\scpl{2}{3}\scpr{1}{4}}{M_{\rm Pl}^2}\left(\frac{s_{13}}{s_{12}}+ \frac{b}{2}\right)$} & \multirow{6}{*}{$\frac{8\pi\scpl{2}{3}\scpr{1}{4}\left(\splf{3} \hat P_{12}\spr{4}\right)}{M_{\rm Pl}^2 s_{12}}$} & \multirow{6}{*}{$-\frac{8\pi\scpl{2}{3}\scpr{1}{4}\left(\splf{3} \hat P_{12}\spr{4}\right)^3}{M_{\rm Pl}^2 s_{12}s_{13}s_{14}}$} \\ 
		& & & & \\ 
		& & & & \\ \cline{3-3}
		&  & \multirow{3}{*}{$-\frac{8\pi\scpl{2}{3}\scpr{1}{4}}{M_{\rm Pl}^2}\left(\frac{s_{13}}{s_{12}} +\frac{s_{12}}{s_{13}} + b\right)$} & & \\
		& & & & \\ 
		& & & & \\   \Cline{0.55pt}{1-1}\Cline{0.55pt}{3-5}
		%%%%%%%%%%%%%%%%%%%%%%%%%%%%%%%% VECOTOR ROW
		%%%%%%%%%%%%%%%%%%%%%%%%%%%%%%%%
		\multirow{6}{*}{\textbf{Vector}}  & \multicolumn{1}{c}{} 
		%\multirow{4}{*}{$\frac{\left(\lambda_2 P_{34}\tilde{\lambda}_1\right)^2}{M_{\rm Pl}^2 s_{12}}$} 
		&
		%\multirow{4}{*}{$\frac{\scpl{2}{3}\scpr{1}{4}\left(\lambda_2 P_{34}\tilde{\lambda}_1\right)}{M_{\rm Pl}^2 s_{12}}$}
		& 
		\multirow{3}{*}{$-\frac{8\pi\scpl{2}{3}^2\scpr{1}{4}^2}{M_{\rm Pl}^2 s_{12}}$} & 
		\multirow{6}{*}{$\frac{8\pi\scpl{2}{3}^2\scpr{1}{4}^2\left(\splf{3} \hat P_{12}\spr{4}\right)^2}{M_{\rm Pl}^2 s_{12}s_{13}s_{14}}$} \\ 
		& \multicolumn{1}{c}{} & & & \\
		& \multicolumn{1}{c}{} & & & \\ \cline{4-4}
	 	& \multicolumn{1}{c}{}  & & \multirow{3}{*}{$-\frac{8\pi\scpl{2}{3}^2\scpr{1}{4}^2}{M_{\rm Pl}^2}\left(\frac{1}{s_{12}} +\frac{1}{s_{13}}\right)$} &  \\
		& \multicolumn{1}{c}{} & & & \\ 
		& \multicolumn{1}{c}{} & & & \\   \Cline{0.55pt}{1-1}\Cline{0.55pt}{4-5}
		%%%%%%%%%%%%%%%%%%%%%%%%%%%%%%%% GRAVITY ROW
		%%%%%%%%%%%%%%%%%%%%%%%%%%%%%%%%
		\multirow{4}{*}{\textbf{Graviton}}  & \multicolumn{1}{c}{}
		% \multirow{4}{*}{$\frac{\left(\lambda_2 P_{34}\tilde{\lambda}_1\right)^4}{M_{\rm Pl}^2 s_{12}s_{13}s_{14}}$} 
		& \multicolumn{1}{c}{}
		%\multirow{4}{*}{$\frac{\scpl{2}{3}\scpr{1}{4}\left(\lambda_2 P_{34}\tilde{\lambda}_1\right)^3}{M_{\rm Pl}^2 s_{12}s_{13}s_{14}}$} 
		&
		%\multirow{4}{*}{$\frac{\scpl{2}{3}^2\scpr{1}{4}^2\left(\lambda_2 P_{34}\tilde{\lambda}_1\right)^2}{M_{\rm Pl}^2 s_{12}s_{13}s_{14}}$}
		& \multirow{4}{*}{$\frac{8\pi\scpl{2}{3}^4\scpr{1}{4}^4}{M_{\rm Pl}^2 s_{12}s_{13}s_{14}}$} \\
		& \multicolumn{1}{c}{} & \multicolumn{1}{c}{} & & \\
		& \multicolumn{1}{c}{} & \multicolumn{1}{c}{} & & \\
		& \multicolumn{1}{c}{} & \multicolumn{1}{c}{} & & \\  \Cline{0.55pt}{1-1}\Cline{0.55pt}{5-5}
	\end{tabular}
	\end{adjustbox}
	\caption{Complete set of tree-level four-point amplitudes in GR with all particles coming in.
	 To obtain the desired scattering process as a function of the ordinary Mandelstam variables $s,t,u$, one evaluates the amplitude in regions {\color{red}(r)}, {\color{blue}(b)}, {\color{ForestGreen}(g)}  according to the substitution in eqs.~(\ref{eq:Domains1},\,\ref{eq:Domains2},\,\ref{eq:Domains3}) and appendix~\ref{app:A}.
	 Whenever a cell is divided in two, the top (bottom) row refers to scattering of distinguishable (identical) particles. 
	 %The amplitudes in the top row and last column (scalar-scalar, scalar-fermion, scalar-vector and gravitational Compton scattering) are UV-completed by the Virasoro-Shapiro form factor while the %amplitudes in the central triangle (fermion-fermion, fermion-vector and vector-vector) are UV-completed by the Veneziano form factor (see section~\ref{sec:UVcompl} and eq.~(\ref{eq:Dressing})).
	 }
	\label{tab:1}
		\end{center}
\end{table*}
}}

%%%%%%%%%%%%%%%%%%%%%%%%%%%%%%%%%%%%%%%%%%%%%%
\subsection{Massive spin J mediated amplitudes}\label{sec:SpinJExchange}
Consider now the exchange of a massive spin $J$ resonance. Massive spinning particles in the spinor-helicity formalism are represented by symmetric $2J$ tensors on the spinor variables, i.e. for a particle with momentum $q_\mu$, $\spr{{\bf q}_{I_1}}\times \dots \times \spr{{\bf q}_{I_{2J}}}$.
The coupling of this spin $J$ resonance with mass $M$ to (massless) matter with helicities $h_1,h_2$ is given by the following
 three-point amplitude, completely determined by little group scaling
\begin{align}
\raisebox{-4.5mm}{\begin{tikzpicture}
	\draw [thick]  (-1,1) node [anchor=north east] {1,$\, h_1$} --(-0.5,0.5);\draw [thick] (-0.5,0.5)--(-0.12,0.12);
	\draw [thick] (-1,-1) node [anchor= south east] { 2,$\, h_2$} -- (-.5,-.5);\draw [thick] (-0.5,-0.5)--(-0.12,-0.12);
	\draw [style={decorate, decoration={snake}},thick] (-1,0.034) node [anchor=east] {${\bf q}, J, M$} -- (0.,0.034);
	\draw [style={decorate, decoration={snake}},thick] (-1,-0.034) -- (0.,-.034);
	 \draw[black,fill=tangerineyellow,thick] (0,0) circle (9pt);
	\end{tikzpicture}}~
\begin{aligned}= ig_J& \scpl{1}{2}^{J-h_1-h_2} \spr{\bf q}^{2J} \\
&\times \frac{\spr{1}^{J+h_1-h_2} \spr{2}^{J-h_1+h_2}}{M^{3J-h_1 -h_2 -1}}\label{eq:3Ver}
%\sqrt2 \Lambda \left(\frac{\splf{\bf q}(\hat p_1-\hat p_3)\spr{\bf q}}{2\Lambda M}\right)^h
\end{aligned}
\end{align}
where we introduced the coupling constant $g_J$, omitted Lorentz indexes, and we note that, to avoid inverse powers of $\spr{1},\spr{2}$, $J\geq |h_1-h_2|$  since otherwise we would get a vanishing amplitude. 
We now move to construct on-shell four-point amplitudes, and we divide our analysis in three steps.

\subsubsection{Legendre polynomials}\label{sec:Legendre}
%{\bf i) Legendre polynomials}\\
In order to isolate the differences with massive mediators and for exposition purposes we consider first the case with external scalar particles. 
For a conventional diagrammatic derivation of the following results see, e.g.~\cite{Francia:2007qt,Caron-Huot:2016icg}.
 A simple manipulation of eq.~(\ref{eq:3Ver}) in this case brings the spinor structure into the form
\begin{eqnarray}
\scpl{1}{2}^J \spr{\bf q}^{2J} \spr{1}^J\spr{2}^J &=&  (\scpr{{\bf q}}{1}\scpl{1}{2}\scpr{2}{\bf q})^J\nonumber \\
&=& M^J (\splf{\bf q}(\hat p_1-\hat p_2)/2\spr{\bf q})^J\nonumber \\
&=&  M^J(\splf{\bf q}\hat P_{12}\spr{\bf q})^J~,\label{eq:ScalarGeg}
\end{eqnarray}
where  we use matrix notation to omit Lorentz indices. 
We next construct the amplitude contribution from the on-shell exchange of the massive $J$ particle, which decomposes into 
\begin{align}
	\raisebox{-14.5mm}{
	\begin{tikzpicture}
	\draw [thick, dashed] (-1,.5) node[anchor= east] {1} -- (-0,0); 
	\draw [thick, dashed] (-1,-.5) node[anchor= east] {$2$} --(0,0); 
	\draw [style={thick,decorate, decoration={snake}}] (0.025,-0.17)  -- (0.025,-2);
	\draw [style={thick,decorate, decoration={snake}}] (-.045,-0.17) -- (-.045,-2);
	\draw [thick, dashed] (-1,-1.5) node[anchor= east] {$3$} -- (0,-2); 
	\draw [thick,dashed] (-1,-2.5) node[anchor= east] {$4$} -- (0,-2); 
	\draw[black,fill=tangerineyellow,thick] (-.05,0) circle (9pt); 
	\draw[black,fill=tangerineyellow,thick] (-0.05,-2) circle (9pt);
	\end{tikzpicture} }
&\quad \begin{aligned} \nonumber &ig_J M \left(\frac{\splf{\bf q}\hat P_{12}\spr{\bf q}}{M^2}\right)^J \\
 & ~~\frac{i}{s_{12}-M^2} \\
  & ig_J M\left(\frac{\splf{\bf -q}\hat P_{34}\spr{\bf -q}}{ M^2}\right)^J\\ \nonumber
\end{aligned}
%&(-1)^h\sum\prod\left(\frac{\splf{\bf q^{I_{2j}}}P_{13}\spr{\bf q^{I_{2j+1}}}}{\Lambda M}\right)\left(\frac{\splf{\bf q_{J_{2j}}}P_{13}\spr{\bf q_{J_{2j+1}}}}{\Lambda M}\right)
\end{align}
The complication lies in performing the sum over spinor $I$-indexes.
An example of such a configuration can be depicted as (the minus sign in ($\bf -q$) can be pulled out of the spinors to contribute a $(-1)^J$ factor)
%\ra{Checked; $\spr{(-q)^I}=-\spr{q_I}$,
%$\spl{(-q)^I}=\spl{q_I}$}:
\begin{align}\nonumber
&\splf{\bf q_{\color{SeaGreen} I_1}} \hat P_{12} \spr{\bf q_{\color{SeaGreen} I_2}} \,\times\, \splf{\bf q_{\color{SeaGreen} I_3}} \hat P_{12} \spr{\bf q_{\color{SeaGreen} I_4}}\, \times \,\splf{\bf q_{\color{SeaGreen} I_5}}\hat P_{12} \spr{\bf q_{\color{SeaGreen} I_6}}\times...\\ \nonumber
&\quad\begin{tikzpicture}
\draw [thick, dashed,SeaGreen] (0.125,0) -- (1.375,0);
\draw [thick,SeaGreen, rounded corners=4pt] (1.375,0)--(1.5,0) -- (1.5,1) -- (1.375,1);
\draw [thick,SeaGreen, dashed] (0.125,1) -- (1.375,1);
\draw [thick,SeaGreen, rounded corners=4pt] (.125,1)--(0,1) -- (0,0) -- (0.175,0);
%%%%%%
\draw [thick,SeaGreen, rounded corners=4pt] (2.75,1)--(2.625,1) -- (2.625,0) -- (2.75,0);
\draw [thick,SeaGreen, dashed] (2.75,1) -- (4,1);
\draw [thick,SeaGreen, dashed] (2.75,0) -- (4,0);
\draw [thick,SeaGreen, rounded corners=4pt] (4,1)--(4.125,1) -- (5.25,0)-- (5.375,0);
\draw [thick,SeaGreen, rounded corners=4pt] (4,0)--(4.125,0) -- (5.25,1)-- (5.375,1);
\draw [thick,SeaGreen, dashed] (5.375,1) -- (6.625,1);
\draw [thick,SeaGreen, dashed] (5.375,0) -- (6.625,0);
\draw [thick,SeaGreen, rounded corners=4pt] (6.625,0)--(6.75,0) -- (6.75,1) -- (6.625,1);
\end{tikzpicture}
\\ \nonumber
&\splf{\bf q^{\color{SeaGreen} I_1}} \hat P_{34} \spr{\bf q^{\color{SeaGreen} I_2}} \times \splf{\bf q^{\color{SeaGreen} I_3}} \hat  P_{34} \spr{\bf q^{\color{SeaGreen} I_5}} \times \splf{\bf q^{\color{SeaGreen} I_4}}\hat  P_{34} \spr{\bf q^{\color{SeaGreen} I_6}}\times...
\end{align}
Where the solid green lines signal the summed little group indexes while the dotted follows the matrix multiplication in Lorentz indexes.
Using the completion relations $\spl{{\bf q}^I}\sprf{{\bf q}_I}=\hat q$ and $\spr{{\bf q}^I}\sprf{{\bf q}_I}=M$ we can reduce any of the above terms to a product of traces\footnote{Implicit in our notation for  products of $\hat P$'s is the proper contraction of Lorentz indices, $\hat P \hat K$=$\hat P_{\alpha\dot\alpha} \epsilon^{\dot\alpha\dot\beta}\hat K_{\beta\dot\beta}\epsilon^{\gamma\beta}\equiv \hat P\,\bar\sigma^\mu K_\mu$ with $[\bar\sigma_\mu]^{\dot\alpha\alpha} =(1,-\vec\sigma)$} over slashed momenta as $\prod\mbox{tr}[(\hat P_{12}\hat P_{34})^{n_i}]$ with $\sum n_i=2J$ and $n_i$ an ordered array. For instance with $J=3$ one such configuration is $\mbox{tr}[(\hat P_{12}\hat P_{34})^2]\mbox{tr}(\hat P_{12}\hat {P}_{34})$ and its coefficient is given by counting the ways to accommodate a single lasso `o' and double one `$\infty$' in 3 slots; o$\infty, \infty$o, etc; examples for $J=2,3,4$ are given in the appendix, eqs.~(\ref{eq:Lassos},\,\ref{eq:Lassos2},\,\ref{eq:Lassos3}). In a second step we collapse the traces $\mbox{tr}(\hat P_{12}\cdot \hat P_{34})^n$ into polynomials in $P_{12}\cdot P_{34}$, $P_{12(34)}^2$, using the relation in eq.~(\ref{eq:traces}) in the appendix.
 
All in all we get the following expression for the four-point amplitude
\begin{align}\nonumber
%\mathcal A=
%&\frac{g^2_J(2J)!!}{(2J-1)!!}\sum_m\frac{(-1)^{m}(2J-2m)!\left(P_{12}^2P_{34}^2\right)^m\left(P_{12}\!\cdot\! P_{34}\right)^{J-2m}}{M^{2J-2}2^Jm!(J-2m)!(J-m)!\left(s_{12}-M^2\right)}\\
\mathcal{A}_{1234}^{J} &=\frac{g^2_J(2J)!!}{(2J-1)!!}\frac{M^2}{s_{12}-M^2} \times \\
&\sum_m{J \choose m}{{2J-2m}\choose{J}}\frac{\left(-P_{12}^2P_{34}^2\right)^m\left(P_{12}\!\cdot\! P_{34}\right)^{J-2m}}{2^JM^{2J}}\nonumber \\
&  =\frac{g^2_J(2J)!! }{4^J(2J-1)!!}\frac{M^2}{s_{12}-M^2}
 P_J[x(s_{13})]\,, \label{AmpScalarExp}
\end{align}
where $P_J$ are Legendre polynomials and 
\begin{align} \label{eq.x}
x(s_{13})=-\frac{P_{12}\cdot P_{43}}{M^2/4}=\frac{P_{12}\cdot P_{34}}{M^2/4}=1+\frac{2s_{13}}{M^2}\,,
\end{align}
%{\color{red} what is our choice to extend off-shell?  }
where we have used the on-shell condition in  $P_{12}^2=P_{34}^2=-M^2/4$ and $-s_{14}=M^2+s_{13}$.

We consider next the case with equal helicity $h_1 = h_2 \equiv h$ in the three-point amplitude. 
The spinor structure takes the form 
\begin{eqnarray}
\scpl{1}{2}^{J-2h} \spr{\bf q}^{2J} \spr{1}^{J} \spr{2}^{J} &=& \frac{1}{\scpl{1}{2}^{2h}}(\scpr{{\bf q}}{1}\scpl{1}{2}\scpr{2}{\bf q})^J \nonumber\\
&=& \frac{\scpr{1}{2}^{2h}}{M^{4h}}(\scpr{{\bf q}}{1}\scpl{1}{2}\scpr{2}{\bf q})^J\nonumber \\ 
&=&\frac{\scpr{1}{2}^{2h}}{M^{4h - J}}(\splf{\bf q}\hat P_{12}\spr{\bf q})^{J}\,.\label{eq:SameHel}
\end{eqnarray}
This is the same structure we already found in the scalar case, eq.~(\ref{eq:ScalarGeg}). 
This means that in the computation of the four-point amplitude the case with equal helicity $h_1 = h_2 \equiv h$ 
reduces to the previous result in eq.~(\ref{AmpScalarExp}) with Legendre polynomials, with an overall factor 
$\scpr{1}{2}^{2h}\scpr{3}{4}^{2h}/M^{4h}$ that takes into account the helicity of the external particles. 
As an important difference, however, notice that in this type of coupling the interaction with the massive spin $J$ resonance changes any quantum number that matter might have.

\subsubsection{Jacobi polynomials}
%{\bf ii) Jacobi polynomials}\\
 Consider now the case with opposite helicity $h_1 = - h_2 \equiv h$ which corresponds to a helicity and
 quantum number conserving interaction as is the case for the graviton coupling.
The dependence on spinor variables reads
\begin{eqnarray}
\scpl{1}{2}^{J} &\spr{\bf q}^{2J}&  \spr{1}^{J+2h} \spr{2}^{J-2h} \nonumber \\
&=& \scpl{1}{2}^{J-2h}  \scpr{1}{\bf q}^{J-2h} \scpr{2}{\bf q}^{J-2h} \scpr{1}{\bf q}^{4h}\scpl{1}{2}^{2h}\nonumber \\
&=&M^J (\splf{\bf q}\hat P_{12}\spr{\bf q})^{J-2h} (\scpl{{\bf q}}{2}\scpr{1}{\bf q})^{2h}\,.
\end{eqnarray}
The derivation of the four-point amplitude in this case is not a mere rescaling of the scalar case but one can follow the same steps in the computation,
\begin{align}
			\raisebox{-18mm}{
		\begin{tikzpicture}
		\draw [thick] (-0.6,1.6) node[anchor= east] {1,$\,\,h$} -- (-.3,1.3);\draw [thick] (-.3,1.3)--(0,1);  
		\draw [thick] (-0.6,-.6+1) node[anchor= east] {$2$,$-h$} -- (-.3,-.3+1);\draw [thick] (-.3,-.3+1)--(0,1);  
		\draw [style={thick,decorate, decoration={snake}}] (0.035,1)  -- (0.035,-1);
		\draw [style={thick,decorate, decoration={snake}}] (-.035,1) -- (-.035,-1);
		\draw [thick] (-0.6,-1.6) node[anchor= east] {$4,\,\,h'$}--(-.3,-1.3);\draw [thick] (-.3,-1.3)--(0,-1); 
		\draw [thick] (-0.6,-.4) node[anchor= east] {$3,-h'$}--(-.3,-0.7);\draw [thick] (-.3,-.7)--(0,-1); 
		\draw[black,fill=tangerineyellow,thick] (0,1) circle (7pt);
		\draw[black,fill=tangerineyellow,thick] (0,-1) circle (7pt);
		\end{tikzpicture}}
	&\quad \begin{aligned} \nonumber &ig_J  \left(\frac{\splf{\bf q}\hat P_{12}\spr{\bf q}}{M^2}\right)^{J-2h}\left(\frac{\scpl{{\bf q}}{2}\scpr{1}{\bf q}}{M^2}\right)^{2h} \\
	& ~~\frac{i M^2}{s_{12}-M^2} \\
	& ig_J \left(\frac{\splf{\bf \bar{q}}\hat P_{43}\spr{\bf \bar{q}}}{ M^2}\right)^{J-2h'}\left(\frac{\scpl{{\bf \bar{q}}}{3}\scpr{4}{\bf \bar{q}}}{M^2}\right)^{2h'}\\ \nonumber
	\end{aligned}
	\end{align}
	with $\bf \bar{q}=-q$. As in the computation that lead to eq.~(\ref{AmpScalarExp}) one can use the completion relations for the sum in little group indices of $\spl{\bf q}$, $\spr{\bf q}$ to obtain traces over $SU(2)_{L,R}$ indices.
 The computational difference is that traces are not just over chains of $\hat P_{12}$ and $\hat P_{34}$ but factors of $\spr{1}\splf{2}$ and $\spr{4}\splf{3}$ might replace each one of the two factors, e.g. in
tr$(\hat P_{12}\hat P_{34}\spr{1}\splf{2}\hat P_{34})$. The number of possible traces grows much steeply now so instead one can take a faster approach building on the scalar result.
 Take eq.~(\ref{AmpScalarExp}) with $J$ $\hat P_{12}$'s contracted with $J$ $\hat P_{34}$'s as obtained after working out combinatorics and expanding traces in Lorentz scalar products of momenta.
Then one can substitute $\hat P_{12}^J\to \hat P_{12}^{J-2h}(\spr{1}\splf{2})^{2h}$ to get the result for polarized states (note that longitudinal pieces drop
  out, $q\cdot P_{12}=\splf{2}\hat q\spr{1}=q\cdot P_{34}=0$). This means that we need to unfold the sum in eq.~(\ref{AmpScalarExp}) to insert $2h$ factors of $\spr{1}\splf{2}$ and $2h'$ of $\spr{4}\splf{3}$; this is again a combinatorics problem that results in, denoting $\mathcal A^J_{1^h2^{-h}3^{-h'}4^{h'}}\equiv\mathcal A_{h,h'}^J$
\begin{align}	
%\mathcal{A}^J_{1^h2^{-h}3^{-h'}4^{h'}}
\mathcal A_{h,h'}^J & =
\frac{g^2_JM^2(2J)!!}{4^J(2J-1)!!} \times \\ & \left(\frac{\scpr{1}{4}\scpl{3}{2}}{M^2}\right)^{\!\!2h}\left(\frac{\splf{3}\!\hat P_{12}\!\spr{4}}{M^2}\right)^{2h'-2h}\times \nonumber \\ &
{J\choose {2h}}^{\!\!-1}{J\choose {2h'}}^{\!\!-1}\sum_{\ell,m}\frac{c_{\ell,m} (x-1)^{\ell-2h'}x^{J-2m-\ell}}{s_{12}-M^2}\,,\nonumber
 %\sum\frac{2^{2h'-J}(-1)^{m+2h'}(2J-2m)!(x-1)^\ell x^{J-2m-2h'-\ell}}{m!(J-m)!\ell!(2h-\ell)!(2h'-2h+\ell)!(J-2m-2h'-\ell)!}
\label{eq:Combin}
\end{align}
with $x$ as in eq.~(\ref{eq.x}) and $c_{\ell,m}$ as
\begin{align}
&c_{\ell,m}=\\ &\frac{(-1)^{m+2h'}}{2^{J-2h'}} {{J}\choose{m}} {{2J-2m}\choose{J}}{{J-2m}\choose{\ell}} {{\ell}\choose{2h}} {{2h}\choose{\ell-2h'}}\,,\nonumber
\end{align}
%\begin{align}	\label{eq:Combin}
%\mathcal{A}^J_{1^h2^{-h}3^{-h}4^h} &=
%\frac{g^2_JM^2(2J)!!}{(4)^J(2J-1)!!}\times  \\ & {J\choose %{2h}}^{\!\!-2}\!\!\left(\frac{\scpr{1}{4}\scpl{2}{3}}{M^2}\right)^{\!\!2h}\!\sum_{k=0}^{J-2h} 
%\frac{c_{k}^{J,2h} x^{k}}{s_{12}-M^2}\,,
%\nonumber
%\end{align}
%with $x$ as in eq.~(\ref{eq.x}) and $c_k^{i,j}$ as:
%\begin{eqnarray} 
%c_{k}^{i,j}&=&\sum_{b,m}(-1)^{m+k+i}\frac{(i+j+b)!}{i!\,j!\,b!}\\
%&&\times{i\choose m}{j \choose b}{{2i-2m}\choose {i+j+b}}{{b}\choose{k-2m}}\,,
%\end{eqnarray}
where the sum runs over values of $\ell,m$ with %$b=i-j-2m-k,.., i-j-2m$, $2m=J-4h-k,...,J-2h-k$, that is where 
non-negative-entry binomials.
Although written in an unusual form,  we find that the polynomial in $x$
\begin{align}
\sum_{\ell,m} c_{\ell,m}&(x-1)^{\ell-2h'} x^{J-2m-\ell} = \nonumber \\& {{J+2h'}\choose {J} }{J\choose 2h} P_{J-2h'}^{(2h'-2h,2h'+2h)}(x)\,,
\end{align}
is proportional to the Jacobi polynomials $P_n^{(a,b)}(x)$
\begin{align}\nonumber
P^{(a,b)}_n=\sum_k{{n+a}\choose{n-k}}{{n+b}\choose{k}}\left(\frac{x-1}{2}\right)^k\left(\frac{x+1}{2}\right)^{n-k}.
\end{align}
All in all, eq.~(\ref{eq:Combin}) reads
\begin{align}\nonumber
%\mathcal{A}^J&_{1^h2^{-h}3^{-h'}4^{h'}} =
\mathcal A_{h,h'}^J=&
\frac{g^2_J (2J)!!}{4^{J}(2J-1)!!}%\times   \\& 
 \left(\frac{\scpr{1}{4}\scpl{3}{2}}{M^2}\right)^{\!\!2h}\left(\frac{\splf{3}\!\hat P_{12}\!\spr{4}}{M^2}\right)^{2h'-2h}\times  \\ &
\frac{M^2  }{s_{12}-M^2}{{J+2h'}\choose{J}}{{J}\choose{2h'}}^{-1}
P_{J-2h'}^{(2h'-2h,2h'+2h)}(x)\,, \label{eq:Combin2}
\end{align}
and it does contain the scalar and the same helicity  cases as the Jacobi polynomials 
reduce to Legendre polynomials in both limits. Let us also note that the measure for Jacobi polynomials is $\propto(1-x)^a(1+x)^b$, in our case $a=2h'-2h, b=2h'+2h$, and it is related to the helicity scaling of the amplitude as we shall see next.
%The case of the amplitude with different helicities $\mathcal{A}^J_{1^h2^{-h}3^{-h^{\prime}}4^{h^{\prime}}}$  has then a dependence on kinematic
% variables as
%\begin{align}\nonumber
%\left(\frac{\scpr{1}{4}\scpl{2}{3}}{M^2}\right)^{2h}\left(\frac{\splf{3}\!\hat P_{12}\!\spr{4}}{M^2}\right)^{2h'-2h}P_{J-2h'}^{(2h'-2h,2h'+2h)}(x)\,.
%\end{align}
The appearance of Legendre and Jacobi polynomials is indicative of an angular analysis that is, in turn, related to unitarity as we shall make explicit next.

\subsubsection{Wigner $d$-functions and unitarity}
%{\bf iii) Wigner $d$-functions}\\
In order to touch on unitarity and angular analysis, we turn to the red region, {\color{red}(r)}, in fig.~\ref{fig:schematic}, 
that is the kinematic domain where the spin $J$ resonances are kinematically accessible. 
We indicate the corresponding amplitude for the generic process 
$1^{h}2^{-h}\to \bar{3}^{h'}\bar{4}^{h'}$ in red, {\color{red}$\mathcal A_{h,h'}$}, and
 we use the explicit substitutions in fig.~\ref{fig:schematic} which, in terms of the center-of-mass (CM) scattering angle $\theta$, read, cf. appendix~\ref{app:A},
%\begin{align}
%{\color{red} \mathcal A:}\,\quad s_{12}\to s\,,\,\,\, s_{14} \to t\,,\,\,\,s_{13}\to u 
%\end{align}
\begin{gather}\nonumber
s_{13}=-s\,s^2_{\theta/2}, \quad s_{14}=-s\,c^2_{\theta/2},\quad
x=1-\frac{s}{M^2}(1-c_\theta)\\    \label{eq:scpltheta}
\scpl{3}{1}=s_{\theta/2}\sqrt{s}\,,\qquad
\scpr{1}{4}=\scpl{3}{2}=c_{\theta/2}\sqrt{s}\,,
\end{gather} 
where $c_{\theta}\equiv \cos\theta$, $s_{\theta} \equiv \sin\theta$. We see that, on the mass-shell of the resonance, we have the identification  $x(s=M^2)=c_\theta$. 
%It is indeed only on resonance that the have computed the contribution and our for example writting $x=1+2t/M^2$ instead of $x=-2u/M^2-1$ is a `choice' to extend the result off shell, but as such this choice can always be encoded in a contact term. 
To select the resonant contribution while being general we note here that, given our tree-level approximation, the imaginary part of the amplitude comes solely from poles, being explicit
\begin{equation}\label{eq:PV}
\frac{1}{s - M^2 + i\epsilon} = \mathcal{PV}\left[\frac{1}{s - M^2}\right] - i\pi \delta(s - M^2)\,,
\end{equation}
with $\mathcal{PV}$ the Cauchy's principal value, and the delta function explicitly showing that the imaginary part only has support on-shell.
Therefore we can write for the imaginary part%\ra{I have fixed the $\mu$'s to what I think correct}
\begin{align}\nonumber
%&{\rm Im}[ {\color{red}\mathcal A^J_{1^{h}2^{-h}3^{-h^{\prime}}4^{h^{\prime}}}}] 
{\rm Im}[ {\color{red}\mathcal A^J_{h,h'}}]
%\\
 & = %\frac{g^2 (2J)!!}{2^J(2J-1)!!} {{J+2h}\choose{J}}{{J}\choose{2h}}^{-1}
%\pi
 C_J \delta(s-M^2) c_{\theta/2}^{\mu_i+\mu_f}s_{\theta/2}^{\mu_f-\mu_i}%\left(\cos\frac{\theta}{2}\right)^{4h}%\left(\sin\frac{\theta}{2}\right)^{2h'-2h} 
P_{J-\mu_f}^{(\mu_f-
\mu_i,\mu_f+\mu_i)}(c_\theta) \nonumber \\
 & =%\pi
  C_J \delta(s-M^2) d^J_{\mu_i,\mu_f}(\theta)\,,\label{eq:ResWig}
\end{align}
where $C_J$ is a constant, $\mu_i = 2h$, $\mu_f = 2h'$ and $d^J_{\mu_i,\mu_f}$ are the Wigner $d$-functions, in full generality
given by~\cite{Wigner:102713}
\begin{eqnarray}
&&d_{\mu_i,\mu_f}^{J}(\theta) = [(J+\mu_i)!(J-\mu_i)!(J+\mu_f)!(J-\mu_f)!]^{1/2}\times \nonumber\\
&&\sum_{a}\frac{(-1)^{\mu_i - \mu_f +a}(\cos\theta/2)^{2S+\mu_f-\mu_i-2a}
	(\sin\theta/2)^{\mu_i-\mu_f+2a}}{(J+\mu_f-a)!a!(\mu_i-\mu_f+a)!(J-\mu_i-a)!}\,,\nonumber
\end{eqnarray} 
 with the sum taken over values such that the factorials are non-negative. 
 
 Wigner $d$-functions offer a simple generalization for the amplitude generated by the exchange of a massive particle with  spin $J$ 
 in the scattering process $1^{h_1}2^{h_2}\to \bar{3}^{\bar{h}_3}\bar{4}^{\bar{h}_4}$ with $\mu_i=h_1-h_2$, $\mu_f=\bar{h}_3-\bar{h}_4$ where we recall that helicities of out-going particles are minus those of incoming $\bar{h}_{3,4}=-h_{3,4}$. 
 Useful and physically meaningful relations for Wigner $d$-functions are $d_{-a,-b}^J=d_{b,a}^J$ (related to in-out exchange) and $d_{a,-b}^J(\theta)=(-1)^{J+a}d_{a,b}^J(\pi-\theta)$  (related to  $3\leftrightarrow 4$ ($u,t$) exchange).
%\begin{equation}
%{\rm R\overline{es}}(\mathcal A_{1^{h_1}, 2^{h_2}3^{h_3},4^{h_4}})=\bar g^2 d^J_{h_1-h_2,h_4-h_3}(\theta)
%\end{equation}
Given the relations of eq.~(\ref{eq:scpltheta}), one can then work backwards to get the expression in terms of spinor variables and Jacoby polynomials $P^{(\mu_f-\mu_i,\mu_f+\mu_i)}_{{\rm Min}(J-|\mu_{i,f}|)}$ for the general case. Indeed, in generality, a Wigner d-function can be expressed as a Jacobi polynomial times and overall factor of powers of $c_{\theta/2}$,$s_{\theta/2}$, in the present case given by the helicity scaling factors in eq.~(\ref{eq:Combin2}), through substitutions in~(\ref{eq:scpltheta}).% In turn one has that Wigner d-functions have a constant measure as opposed to Jacobin polynomials.

It is indeed in terms of the very same Wigner $d$-functions that the general partial wave expansion is defined as~\cite{Jacob:1959at} 
\begin{gather}\label{eq:PtWvGen}
\begin{tikzpicture}
	\draw [thick]  (-1,1) node [anchor=north east] {1,$\, h_1$} --(-0.5,0.5);\draw [thick] (-0.5,0.5)--(-0.12,0.12);
	\draw [thick] (-1,-1) node [anchor= south east] { 2,$\, h_2$} -- (-.5,-.5);\draw [thick] (-0.5,-0.5)--(-0.12,-0.12);
	\draw [thick]  (1,1) node [anchor=north west] {$\bar{3},\, \bar{h}_3$} --(0.12,0.12);
	\draw [thick] (1,-1) node [anchor= south west] {$ \bar{4},\, \bar{h}_4$} -- (-.12,.12);
	\draw[black,fill=Dandelion,thick] (0,0) circle (15pt);
	\end{tikzpicture}\\  \nonumber
{\color{red}\mathcal A_{1^{h_1}2^{h_2}\to\bar{3}^{\bar{h}_3}\bar{4}^{\bar{h}_4}}} =  16\pi%\sqrt{\frac{s}{s-4m^2}}
\sum_{J\geqslant {\rm Max}(\mu_i,\mu_f)}^{\infty}(2J+1)a_{if}^{J}(s)d^J_{\mu_i,\mu_f}(\theta) 
\end{gather}
with partial-wave amplitudes $a_{if}^{J}(s)$.
The crucial difference compared to eq.~(\ref{eq:ResWig}) is that this decomposition is general, and it applies to  both real and imaginary parts.
In practice, this means that the exchange of a  {\it massive spin $J$ resonance} contributes to the {\it imaginary part of the $J^{\rm \it th}$ partial wave only},  but to the {\it real part of partial waves with} $\leq J$.

Finally, as anticipated, let us connect with unitarity, for completeness and reference. By means of the optical theorem, unitarity of the scattering matrix implies
\begin{equation}\label{eq:UNIT}
2{\rm Im}[\mathcal{A}_{i\rightarrow f}] = \sum_n\int d\Pi_n \mathcal{A}_{f\rightarrow n}^*\mathcal{A}_{i\rightarrow n}\,,
\end{equation}
where on the right-hand side we take the sum over all possible intermediate states $n$ with phase-space measure $d\Pi_n$. In the case of an elastic process $i=f$ the right-hand side is a sum of moduli squared, and we have the positivity constraint $2{\rm Im}[\mathcal{A}_{i\rightarrow i}] \geqslant 0$. 
As a simple special case it follows that in elastic scattering ${\rm Im}[a^{J}_{ii}(s)]\geqslant 0$, as one can infer by taking the forward limit in which $d^J_{a,b}(0) = 1$. 
Given that we have built our amplitude out of three-point amplitudes consequences of unitarity like positivity are implemented, but the unitarity relation does provide a way to rewrite our amplitude in terms of more straightforwardly observables quantities. If the initial state $12$ can convert into 
a single-particle state denoted $J$, the r.h.s. of eq.~(\ref{eq:UNIT}) reads
 \begin{align}
& {\rm Im}[\mathcal A_{12\to 34}] =\pi \sum_J \mathcal A_{34\to J}^* \mathcal A_{12 \to J}\delta(s_{12}-M^2)\,,
 \end{align}
 where we neglected quantum corrections.
Furthermore when  particles $1,2$ and $3,4$ are the same species we have a positive imaginary part that is related to the mediator partial width
\begin{align}
\nonumber
&{\rm Im}[\mathcal A(s_{12},\,\theta=0)] = 16\pi (2J+1)M\Gamma_{J\to 12}\delta(s_{12}-M^2)\,,
\end{align}
which in particular implies in our amplitude~(\ref{eq:Combin2}) for the exchange a spin $J$
\begin{align}
&16\pi(2J+1) \frac{ \Gamma_{J \to 1^h2^{-h}}}{M} =\frac{g_J^2 (2J)!!}{4^{J}(2J-1)!!} 
{{J+2h}\choose{J}}{{J}\choose{2h}}^{-1}\,.
\end{align}
This is to say that the factorial factors and powers of $2$ that relate the three- and four-point amplitudes disappear when rewriting in terms of the mediator width. Let us then define $\alpha^J=\Gamma_J/M$ and write eq.~(\ref{eq:Combin2}) for $h^{\prime}\neq h$ as
\begin{align}\label{eq:MstSpnJ}
%&\mathcal{A}^J_{1^h2^{-h}3^{-h^{\prime}}4^{h^{\prime}}} 
\mathcal A^J_{h,h'}
=&16\pi(2J+1) C_{J,h,h^{\prime}}
\frac{\alpha^J_{h',h} M^2  }{s_{12}-M^2}\left(\frac{\scpr{1}{4}\scpl{2}{3}}{M^2}\right)^{2h}\times
\nonumber \\ & 
\left(\frac{\splf{3}\!\hat P_{12}\!\spr{4}}{M^2}\right)^{2h'-2h}P_{J-2h'}^{(2h'-2h,2h'+2h)}(x)\,,
\end{align}
where we introduced a constant $C_J,h,h^{\prime}$ for normalization
\begin{equation}
C_{J,h,h^{\prime}} \equiv \sqrt{\frac{(J-2h')!(J+2h')!}{(J-2h)!(J+2h)!}}\,,
\end{equation}
and a slight abuse of notation was committed since only for $h'=h$ and $1^h 2^{-h} = 3^{-h} 4^{h}$ we have that $\alpha^J_{h,h}$ is a positive quantity with the interpretation of a decay rate. Aside from the obvious simplifications,  this notation encodes explicitly the strength of the interaction in $\alpha$ and, therefore, its range of validity. 

%%%%%%%%%%%%%%%%%%%%%%%%%%%%%%%%%%%%%%%%%%%%%%%%%%
%%%%%%%%%%%%%%%%%%%%%%%%%%%%%%%%%%%%%%%%%%%%%%%%%%
\section{Completing Gravity in the UV}\label{sec:UVcompl}

In the previous section we computed all four-point amplitudes generated by graviton exchange and
the on-shell contribution to the amplitude due to the exchange of a massive spin $J$ particle.
This section aims at putting these two results together for a consistent theory of gravity in the UV.
The goal is to tame the growth with CM energy by introducing resonances while  satisfying the general properties outlined in the introduction. In this road, we start with scalar particles, in particular
consider the 2-to-2 scattering of distinct scalar particles.\footnote{Similar arguments in the context of the four-graviton scattering were presented in~\cite{HuangTalk}.}

\subsection{Towards a  ``bottom-up'' UV completion}\label{sec:UVconstruction}

%\begin{figure}[!t!]
%\centering
%\includegraphics[width=0.45\textwidth]{LegendrePlot.pdf}\vspace{-0.275cm}  
%\caption{\label{fig:Leg} 
%}
%\end{figure}
The scattering amplitude $\mathcal{A}_{\phi\phi\varphi\varphi}$ for distinct scalar particles $\phi$ and $\varphi$ in GR is
\begin{equation}\label{eq:MasterScalar}
\mathcal{A}_{\phi\phi\varphi\varphi} = \frac{8\pi}{M_{\rm Pl}^2}\left(\frac{s_{13}s_{14}}{s_{12}} - as_{12}\right)\,,
%+ \sum_J \frac{g_{\phi}g_{\varphi}P_J[x(s_{13})]}{s_{12}/M_J^2 - 1}\,,
\end{equation}
and describes two different scattering processes that are related by crossing: 
$\phi(p_1)\phi(p_2) \to  \varphi(\bar{p}_3)\varphi(\bar{p}_4)$ ({\color{red}(r)} in fig.~\ref{fig:schematic}),
 $\phi(p_1)\varphi(p_3) \to  \phi(\bar{p}_2)\varphi(\bar{p}_4)$ {\color{blue}(b)}, (we have {\color{ForestGreen}(g)}={\color{blue}(b)} from crossing symmetry).
%$\phi(p_1)\varphi(p_4) \to \phi(\bar{p}_2)\varphi(\bar{p}_3)$ }.
In the r.h.s. of eq.~(\ref{eq:MasterScalar}) the first term represents the pole contribution, the second term accounts for a possible contact term.
Furthermore, given that we are scattering different particles, we assume that there is only one channel open  in each one of the three scattering processes (as it happens in GR) so that all poles lie in the positive real $s_{12}$ axis (otherwise resonances with $\phi-\varphi$ number would be present in the spectrum).
Finally, notice that the amplitude is symmetric under $1\leftrightarrow 2$ exchange (equivalently, $t\leftrightarrow u$ exchange in {\color{red}(r)}). 

In the following, we would like to emphasize the difficulties that one faces when trying to unitarize 
the amplitude $\mathcal{A}_{\phi\phi\varphi\varphi}$ in all the above scattering processes, and the consequences that 
solving these issues imply. 
A word of warning is pertinent before proceeding;
the solutions that we shall construct and present in this section are not unique, and a number of assumptions are required to obtain them. 
To be pristine, we have enumerated them as {\color{Bittersweet} {\it (i)-(iii)}}.
%\begin{figure}[!t!]
%\centering
%\includegraphics[width=0.45\textwidth]{LegendrePlot.pdf}\vspace{-0.275cm}  
%\caption{\label{fig:Leg} 
%}
%\end{figure}

Take the annihilation process $\phi(p_1)\phi(p_2) \to  \varphi(\bar{p}_3)\varphi(\bar{p}_4)$ 
in the red region {\color{red}(r)} of fig.~\ref{fig:schematic} where, in terms of the familiar Mandelstam variables, we have $s_{12} = s$, $s_{13} = t$ 
and $\mathcal{A}_{\phi\phi\varphi\varphi}^{\rm GR} = 8\pi(tu/s - as)/M_{\rm Pl}^2$.
The GR contribution is problematic in the hard-scattering region 
where it grows with energy as $s/M_{\rm Pl}^2$, eventually violating perturbative unitarity. 
To formalize this aspect, one can compute the partial-wave amplitudes, cf. eq.~(\ref{eq:PtWvGen})
\begin{equation}
a_{\phi\varphi}^J(s) = \frac{1}{32\pi}\int_{-1}^{+1}d(\cos\theta)P_J(\cos\theta)\mathcal{A}_{\phi\phi\varphi\varphi}(s,\cos\theta)\,,
\end{equation}
and impose that they must lie inside the unitarity circle in the Argand plane.
One finds two non-vanishing partial-wave amplitudes
\begin{equation}\label{eq:PartialWaves}
a^{0}_{\phi\varphi}(s) = \frac{s(1-6a)}{12 M_{\rm Pl}^2}\,,%~~~a^{1}_{\phi\varphi}(s) = 0\,,~~~
\qquad a^{2}_{\phi\varphi}(s) =
-\frac{s}{60 M_{\rm Pl}^2}\,,
\end{equation}
where the last term corresponds to $J= 2$ exchange, while the first corresponds to $J= 0$. 
The $J= 0$ component arises from the coupling of the virtual graviton to the trace of the energy-momentum tensor of the scalar field which is 
indeed non-vanishing for a minimally coupled massless scalar field.\footnote{The $a^{0}_{\phi\varphi}(s)$ partial wave vanishes if $a= 1/6$. This special value of $a$ corresponds to a conformally coupled scalar field described (in four space-time dimensions, with $\mathcal{R}$ the Ricci scalar) by the action
\begin{equation}\label{eq:OffShell}
S=\int d^4x \sqrt{-g}\left[
\frac{g^{\mu\nu}}{2}\left(\partial_{\mu}\phi\right)\left(\partial_{\nu}\phi\right) + \frac{\mathcal{R}}{2}a\phi^2
\right]\,,
\end{equation}
which has, for $a=1/6$, vanishing trace $T^{\mu}_{~\mu} = 0$.}
In eq.~(\ref{eq:PartialWaves}) we see that the presence of a contact term could cancel the growth in energy 
of the $J=0$ partial wave amplitude but cannot change the bad high-energy behavior of the highest partial wave $a^{2}_{\phi\varphi}(s)$.
To cure this problem, a first simple guess is the introduction of a spin $J=2$ resonance with mass $M_2$ so that the amplitude results into\footnote{The presence of the 
minus sign in eq.~(\ref{eq:J2Amp}) can be traced back to the minus sign in eq.~(\ref{eq:PV}).}
% (ignoring for the moment the subtraction terms)
\begin{eqnarray}\label{eq:J2Amp}
\mathcal{A}_{\phi\phi\varphi\varphi} &=&\mathcal{A}_{\phi\phi\varphi\varphi}^{\rm GR} -
 \frac{g_{\phi,2}g_{\varphi,2}}{s-M_2^2}
\underbrace{\left[M_2^2P_2\left(1+\frac{2t}{M_2^2}\right)\right.}_{\rm on-shell} \nonumber \\
&+& \underbrace{\left. 
(s-M_2^2)\mathcal{P}_{(1,1)}\left(
\frac{s}{M_2^2},\frac{t}{M_2^2}
\right)
\right]}_{\rm off-shell}\,,
\end{eqnarray}
where the notation $\mathcal{P}_{(i,j)}(x,y)$ indicates a generic polynomial of degree $i$ in the variable $x$
($j$ in the variable $y$) while $g_{\phi,2}$ and $g_{\varphi,2}$ are the couplings of the two scalar fields with the spin $J=2$ resonance.
Compared with eq.~(\ref{eq:MstSpnJ}), the on-shell contribution is accompanied by an off-shell term (just as the GR amplitudes in table~\ref{tab:1} and eq.~(\ref{MstGR})), and the relevant remark is that the off-shell term reduces to a polynomial 
of degree $J-1$ in $s$ and $t$ for a spin $J$ mediator. 
We can now compute the partial-wave amplitudes. 
The contact term does not contribute to the highest partial-wave amplitude $a^{2}_{\phi\varphi}(s)$ while the 
inclusion of the pole term gives 
\begin{equation}
a^{2}_{\phi\varphi}(s) =
-\frac{s}{60 M_{\rm Pl}^2} - \frac{g_{\phi,2}g_{\varphi,2}s}{80\pi M_2^2}\,,
\end{equation}
from which we see that it is possible to compensate gravity if one takes 
$g_{\phi,2}g_{\varphi,2} = -4\pi M_2^2/3M_{\rm Pl}^2$, thus 
implying that the sign of the couplings are opposite. 
However, this possibility -- although not a priori incorrect -- does not extend to more fields. If 
we introduce a third field $\chi$, from the $\phi\chi$ scattering we would deduce that $\chi$ has opposite-sign charge compared to $\phi$, 
but if both $\chi$ and $\varphi$ have opposite charge w.r.t. $\phi$ they must have the same sign w.r.t. each other, and the $\varphi\chi$ scattering is not unitarized.
%Let us now comment about the contact terms $A(t) s+ B(t)$ neglected in the previous argument. 
%We can make explicit use of the partial wave decomposition.
% The graviton contribution to the highest partial wave is $a_{\phi\varphi}^{J=2} = s/60M_{\rm Pl}^2$. 
% For $A(t) s+ B(t)$ to scale with $s$ in the same way 
% one should have $A(t)=A(0)$ or $B(t)=t$, none of which however project onto $a_{\phi\varphi}^{J=2}$. This makes clear that 
% the contact terms cannot cure the bad high-energy behavior of gravity at least as far as its highest partial wave amplitude is concerned.
%A resonance with spin $J=3$ and mass $M_3$ introduces in the partial-wave amplitudes terms that grow with energy as $s^2/M_{3}^4$ 
%that, in turn, would require the introduction of a spin $J=4$ resonance thus kicking off a series in spin that cannot end.
This discussion makes clear that a massive spin $J=2$ resonance alone cannot provide a viable UV completion of gravity.
We turn then to introduce a spin $J=3$ resonance with mass $M_3$.
Similarly to eq.~(\ref{eq:J2Amp}), it will contribute to the scattering amplitude with both an on-shell and an off-shell piece, 
the latter being a polynomial of order $2$ in $s$ and $t$
\begin{eqnarray}\label{eq:J3Amp}
\mathcal{A}_{\phi\phi\varphi\varphi}^{J=3} = -\frac{g_{\phi,3}g_{\varphi,3}M_3^2}{s-M_3^2}&
P_3&\left(1+\frac{2t}{M_3^2}\right)  \\
&+& 
g_{\phi,3}g_{\varphi,3}\mathcal{P}_{(2,2)}\left(
\frac{s}{M_3^2},\frac{t}{M_3^2}
\right)\,. \nonumber
\end{eqnarray}
The contribution to the partial-wave amplitude $a^{2}_{\phi\varphi}(s)$ from the pole term has the form 
$g_{\phi,3}g_{\varphi,3}s^2/16\pi M_3^4$. 
It means that one can compensate the graviton contribution
with same sign couplings $g_{\phi,3}$, $g_{\varphi,3}$ and in particular universal $g_\phi=g_\varphi$, in line with the coupling of the graviton  to matter and as opposed to the $J=2$ case.
However, it is clear that a massive spin $J=3$ resonance cannot suffice since 
its contribution to $a^{2}_{\phi\varphi}(s)$ has a different scaling in $s$ compared to gravity and, more worrisome, 
because it introduces an unbalanced contribution to the partial-wave amplitude $a^{3}_{\phi\varphi}(s)$ that grows with energy as $s^2$.
One then faces the same problematic growth in $a_{\phi\varphi}^3(s)$, and iterates the procedure with a massive spin $J=4$ resonance which, in turn, would require a
massive spin $J=5$ state in a domino effect that yields an infinite tower of increasing massive higher spins. 

This is a sketch of the well-known result that gravity requires an infinite set of resonances, here we would like to underline that the argument was elaborated for the process in region {\color{red}(r)} $\phi\phi\to\varphi\varphi$ but region {\color{blue}(b)} for $\phi\varphi\to\phi\varphi$ presents separate problems.
Indeed the balance of consecutive spins with opposite sign contributions in {\color{red}(r)} does not immediately translate to the region {\color{blue}(b)}  since
one has $s \leftrightarrow t$ and all Legendre polynomials, $P_L(1+2s/M^2)$, and poles, $1/(t-M^2)$, have the same sign in the physical region $ s>0, -s<t<0$. 
One could try to address this issue by adjusting the contact terms but here instead we attempt a summed expression for the amplitude which we write in the form
\begin{equation}
\mathcal{A}_{\phi\phi\varphi\varphi} = \frac{8\pi}{M_{\rm Pl}^2}\left(\frac{tu}{s} - as\right)\frac{N(s,t)}{\prod_{k}^{\infty}(s-M_k^2)}\,,
\end{equation}
where  the numerator  of the highest common denominator  is, at this stage, a polynomial $N(s,t)$ of arbitrary high degree in $s$ and $t$.
Notice that, without loss of generality, we defined it by factoring out the GR amplitude. 
Let us write $N(s,t)$ as the product of its zeros in $t$
\begin{equation}\label{eq:Attempt1}
\mathcal{A}_{\phi\phi\varphi\varphi} =  \frac{8\pi}{M_{\rm Pl}^2}\left(\frac{tu}{s} - as\right)\frac{
\prod_{n}^{\infty}\left[
t - f_n(s)
\right]
}{\prod_{k}^{\infty}(s-M_k^2)}\,.
\end{equation}
We can now use unitarity and locality as a guideline.
Unitarity and locality, as explicitly shown in the previous section, imply that on a pole in $s$ the residue of the amplitude must be a polynomial of finite degree in $t$ corresponding to particle exchange. 
This is not the case if {\it {\color{Bittersweet} (i)}   the functions $f_n(s)$ in eq.~(\ref{eq:Attempt1}) are assumed analytic around} $s=M_\ell^2$, and one is forced to introduce inverse powers of $t$
\begin{equation}
\mathcal{A}_{\phi\phi\varphi\varphi} =  \frac{8\pi}{M_{\rm Pl}^2}\left(\frac{tu}{s} - as\right)\frac{
\prod_{n}^{\infty}\left[
t - f_n(s)
\right]
}{
\prod_{k}^{\infty}(s-M_k^2)\prod_{m}^{\infty}(t-\hat M^2_m)}\,,
\end{equation}
such that the poles in $t$ cancel against the zeros when evaluating at $s=M_\ell^2$, explicitly
\begin{align}\label{eq:tMinf}
\big\{ \hat M_n^2\big\} \subset& \left\{f_{n}(M_\ell^2) \right\}\,,~~~~~ \forall\, \ell
\end{align}
where both sets (poles and zeros) are infinite.
Remarkably, 
even if the starting point of our construction only required $s$-channel resonances, unitarity and our assumption {\color{Bittersweet}{\it (i)}} led us to resonances in the 
dual $t$-channel. 
The same analysis therefore applies to residues when evaluating in $t =\hat M^2_\ell$, which should be polynomials in $s$; in particular  to avoid double poles we have now that
\begin{align}\label{eq:Minfm1}
\left\{ M_n^2\right\} \subset &\, \big \{f_{n}^{-1}(\hat M_\ell^2) \big\}\,,~~~~~ \forall\, \ell
\end{align}
with $f^{-1}_n$ the inverse function and which, as eq.~(\ref{eq:tMinf}), means that the set of zeroes contains the set of poles. The complication in satisfying eq.~(\ref{eq:tMinf}) is that it must hold for all $\ell=0,...,\infty$. 
Here eq.~(\ref{eq:tMinf}) will be satisfied by simply assuming that
  {\it {\color{Bittersweet}(ii)} the set of elements given by $f_n$ evaluated in the $\ell^{th}$ mass $M_\ell^2$, i.e. $\{f_n(M_\ell^2)\}$, contains the set $\{\hat M_n^2\}$ and $\ell$ more elements}  that is
\begin{equation}\label{eq:Anstzii}
	f_{n}(M_\ell^2)\equiv \hat M_{n-\ell}^2\,,
\end{equation}
which, is worth pointing out, means  that also eq.~(\ref{eq:Minfm1}) is automatically satisfied as
\begin{equation}\label{eq:Anstziim1}
f_{n}^{-1}(\hat M_\ell^2)\equiv  M_{n-\ell}^2\,,
\end{equation}
 while the spectrum in $M_n^2,\hat M_n^2$ is still arbitrary.\footnote{The function can be constructed explicitly given $G,\hat G$ such that $G(M_n^2)=n,\hat G(\hat M_n^2)=n,$ as $f_n(x)=\hat G^{-1}(n-G(x))$.}
Complying with eqs.~(\ref{eq:tMinf},\,\ref{eq:Minfm1}) nevertheless only ensures the absence of double poles;  one should also demand finite degree polynomials to respect unitarity and locality. 
Given {\color{Bittersweet} {\it (i)}}, $f_n$ have a Taylor expansion 
%that is truncated at some finite order of the derivative expansion,
$f_n(s)=f_n(s_\ell) + f_n^\prime (s_\ell) (s-s_\ell)+O(f_n^{\prime\prime})$.
If there are second derivatives each function $f_n$ will contribute with one power of $s$, and one has $n=0,..,\infty$. 
One, therefore, is led to impose then that only a finite number of $f_n(s)$ have second derivatives (of course, the same argument applies to higher derivatives). 
The simplest way to address this is to {\color{Bittersweet} {\it (iii)}} {\it assume that all $f_n(s)$ are linear functions}, $f_n(s)=f_n^1s+f_n^0$. This reduces the problem, and in particular eqs.~(\ref{eq:Anstzii},\,\ref{eq:Anstziim1}), 
to a linear system whose solution is strongly over-constrained since we have
\begin{align}\label{eq:LinExp}
f_n^{1}=& - \frac{\hat M^2_{n-\ell}-\hat M^2_\ell}{ M^2_{n-\ell}- M^2_{\ell}}\,, & f_n^0=&\frac{\hat M^2_{n-\ell}M^2_{n-\ell}-\hat M^2_{\ell} M^2_{\ell}}{M^2_{n-\ell}-M^2_{\ell}}\,,
\end{align}
where the r.h.s. should be the same {\bf for  all} $\ell$. This clearly is not true for a general spectrum. 
It is true however for linear spectrum in both $s,t$
 \begin{align}\label{eq:SpectrumSolution}
M_n^2 &= nM^2\,, & 
\hat M_n^2 &= n\hat M^2+\hat M_0^2\,, &
n &=1,2,\dotso \in \mathbb{N}%\mathbb{Z}^+
\,,
\end{align}
where the spectrum in $s$ has no $M_0^2$ due to the pole in $s=0$, as we shall see shortly. 
In this case $f_n^1$ in eq.~(\ref{eq:LinExp}) reduces to the ratio $f_n^1 = -\hat M^2/M^2$ (independent from $n$) while 
$f_n^0 = n\hat M^2 + \hat M_0^2$, and 
the amplitude now  reads 
\begin{align}\label{eq:Attempt2}
\mathcal{A}_{\phi\phi\varphi\varphi} = & \frac{8\pi}{M_{\rm Pl}^2}\left(\frac{tu}{s} -as \right)\mathcal C \times \\ \nonumber&\frac{
\prod_{n}^{\infty}\left[
M^2t +\hat M^2s - M^2(n\hat M^2+\hat M_0^2)
\right]
}{
\prod_{k}^{\infty}(s-kM^2)
\prod_{m}^{\infty}(t-m\hat M^2-\hat M_0^2)
}\,,
\end{align}
where $\mathcal C$ is a normalization constant.
The above solution presents an interplay between resonances in the two channels with the $t$-channel spectrum that 
determines the $s$-channel couplings and vice versa. 
Let us make this explicit and evaluate the second line at $t=\ell \hat M^2+\hat M_0^2$ to obtain the couplings of the $\ell^{th}$ $t$-channel resonance
\begin{equation}\label{eq:UnitarityCancel}
\frac{\prod_n^{\infty}\hat M^2[s- M^2(n-\ell)]}{\prod^{\infty}_k(s-kM^2)\prod^{\infty}_{m} (m-\ell)\hat M^2}\propto \prod_{r=0}^\ell (s+rM^2)\,,
\end{equation}
which depends on the $s$-channel spectrum by means of a finite-degree polynomial in $s$, as dictated by unitarity and locality.
 Note in particular that there is always a power of $s$ to cancel against the pole in $s=0$ from graviton exchange in the first line of eq.~(\ref{eq:Attempt2}), and 
 this is the reason for the absence of the $M^2_0$ term that we anticipated before. 
 If we evaluate eq.~(\ref{eq:Attempt2}) at $s=\ell M^2$ the product above would be on a finite number of terms of the form 
 $(t+r\hat M^2-\hat M^2_0)$, with no zero at $t=0$ since, in this case, there is no GR pole to cancel.
 As for the normalization factor $\mathcal C$ we can address this if we use the Euler definition of the $\Gamma$ function 
\begin{equation}
\Gamma(z) = \frac{1}{z}\prod_{n=1}^{\infty}\frac{(1 + 1/n)^z}{1 + z/n}\,,
\end{equation}
to write 
\begin{align}\nonumber
\frac{\Gamma(1-\tilde s)\Gamma(1-\hat t)}{\Gamma(1-\hat t-\tilde s)}&
= \\ \mathcal C&\frac{\prod_{n}^{\infty}\left[
	M^2t +\hat M^2s - M^2(n\hat M^2+\hat M_0^2)
	\right]
}{
	\prod_{k}^{\infty}(s-kM^2)
	\prod_{m}^{\infty}(t-m\hat M^2-\hat M_0^2)
} \,,\label{eq:VenezianoAtt}
%\\ 
%\tilde s &=\frac{s}{M^2}\,, \qquad\qquad\hat t=\frac{t-\hat M_0^2}{\hat M^2 }\,.\label{eq:VenezianoAtt}
\end{align}
with 
\begin{equation}
\tilde s =\frac{s}{M^2}\,, \qquad\hat t=\frac{t-\hat M_0^2}{\hat M^2 }\,.
\end{equation}
This is the Veneziano amplitude with a linear homogeneous transformation in $s$ and a linear inhomogeneous transformation in $t$. 
It is good to pause at this point, and define
\begin{eqnarray}\label{eq:GenVZ}
\mathcal A_{\rm VZ}^{\eta,\gamma_0}(s,t) &\equiv& \frac{\Gamma(1-\tilde s)\Gamma(1+\eta\gamma_0-\eta \tilde t)}{\Gamma(1+\eta\gamma_0-\eta \tilde t-\tilde s)}\,,\\
\eta &\equiv & \frac{M^2}{\hat M^2}\,,\\
\gamma_0&\equiv &\frac{\hat M^2_0}{M^2}\,,
\end{eqnarray}
where $\tilde t=t/M^2$, and we see the expression is not symmetric in $s,t$ but is compatible with locality. 
Let us now validate this amplitude by checking the high-energy behavior in the hard-scattering limit.
Stirling's approximation for large $s$ yields
\begin{align}
\mathcal A_{\rm VZ}^{\eta,\gamma_0} &\sim e^{R\tilde s}, & R&=\log\left[(1-\eta s_{\theta/2}^2)^{1-\eta s_{\theta/2}^2}(\eta s_{\theta/2}^2)^{\eta s_{\theta/2}^2}\right]\,,
\end{align}
so provided $\eta\leq1$ there is an exponential decrease for large $s$ which makes up for the growth with $s$ of GR.
One has, therefore, that the factor in eq.~(\ref{eq:GenVZ}) does yield a valid UV behavior;  what is more, the exponential fall-off is a faster decrease with $s$ than one can obtain with any finite number of resonances in QFT. 

The non-symmetric behavior in $s\leftrightarrow t$ (or $t\leftrightarrow u$) of eq.~(\ref{eq:GenVZ}) makes it suitable for the UV completion of
 distinguishable particle scattering. This is the case, for instance, of the fermion-vector scattering that we write as
\begin{align}%\nonumber
%\mathcal{A}_{1^{1/2}2^{-1/2} \to 3^{1}4^{-1}} = 
\mathcal{A}_{1/2,1}=
\frac{8\pi\scpl{2}{3}\scpr{1}{4}\splf{3} \hat P_{12}\spr{4}}{M_{\rm Pl}^2 s}\mathcal A_{\rm VZ}^{\eta,\gamma_0}(s,t)\,,%\\
%\mathcal{A}_{1^{1/2}2^{-1/2}3^{-1}4^{1}} = \frac{8\pi\scpl{2}{3}\scpr{1}{4}\splf{3} \hat P_{12}\spr{4}}{M_{\rm Pl}^2 s_{12}}\mathcal A_{VZ}^\gamma(s_{12},s_{13})
\end{align}
whereas if we have $s$-$t$ symmetry enforced by the external states as in the same-fermion scattering we simply set $\eta=1,\gamma_0=0$ in order to get a $s,t$ symmetric result
\begin{align}%\nonumber
%\mathcal{A}_{1^{1/2}2^{-1/2}3^{-1/2}4^{1/2}} = &
\mathcal{A}_{1/2,1/2}&=
\frac{8\pi\scpl{2}{3}\scpr{1}{4}}{M_{\rm Pl}^2 }\left(\frac{s}{t}+\frac{t}{s}+b\right)%\times\\&
\mathcal A_{\rm VZ}^{\eta=1,\gamma_0=0}(s,t)\,.
\end{align}
%The same line of reasoning explained for the scalar case  leads to a UV completion equivalent to eq.~(\ref{eq:Attempt2}). In this case, however, we do not need to impose any symmetry under $1\leftrightarrow 2$ exchange and our final result, using again the Euler definition of the $\Gamma$ function, reads 
%\begin{equation}\label{eq:UVComplqV}
%\mathcal{A}_{q^{+1/2}\bar{q}^{-1/2}V^{-1}V^{+1}} = 4\pi \alpha_{\rm Pl}\tilde{u}\sin\theta~\underbrace{
%	\frac{\Gamma(1-\tilde{s})\Gamma(1-\tilde{t})}{\Gamma(1-\tilde{s}-\tilde{t})}}_{\rm Veneziano\,form\,factor}\,.
%\end{equation}
%Contrary to the previous result in eq.~(\ref{eq:Attempt4}), we do not have  the presence of massive resonances in the $u$-channel 
%(that in this case is the channel that couples, in the convention in which all particles are ingoing, same-helicity states). \\
For simplicity of notation, let us denote $\mathcal A_{\rm VZ}^{\eta=1,\gamma_0=0}\equiv \mathcal A_{\rm VZ}$ in what follows.

The scalar amplitude $\mathcal A_{\phi\phi\varphi\varphi}$, on the contrary, is $t\leftrightarrow u$ symmetric (this is a property of the external states and, as such, it must be respected by the full amplitude), in particular this means that there will be $u$-poles and zeroes as well. The previous amplitude, therefore, must be modified to account for this property.
 Let us start again from the Veneziano factor in eq.~(\ref{eq:VenezianoAtt})
\begin{equation}
\prod_n^{\infty} \frac{(\tilde s+\hat t-n)(\tilde s+\hat u-n)}{(\tilde s-n)(\hat t-n)(\hat u-n)}\,,
\end{equation}
na\"{\i}vely augmented by extra factors to guarantee the $t\leftrightarrow u$ symmetry. To check the validity of this expression, we can use again unitarity and locality as a guideline.
When evaluating at $\hat u=\ell$, the zeroes ($\tilde s+\ell-n$) cancel against the poles $(\tilde s-k)$ but those in  ($\tilde s+\hat t-n$)=$(1-\eta^{-1})\hat t-\gamma_0-\ell-n$  do not contain the poles  $(\hat t-m)$
 so we are forced to introduce extra terms in the numerator. 
 Furthermore, given that the factors $\tilde s+\hat t-n=(1-\eta^{-1})\hat t-\gamma_0-\ell-n$ do not cancel against terms in the denominator they will yield an infinite degree polynomial in $t$ when taking all terms in the product unless $\eta=1$ ($M^2=\hat M^2$) when they reduce to a constant. Therefore we set $\eta=1$ in the following. 
 The extra factors in the numerator can be found by noting that poles in $s$ are all simple and hence we can extend this to $t,u$ by symmetrizing
\begin{equation}\label{eq:VSfirstAttempt}
\prod_n^{\infty}\frac{(\tilde s+\hat t-n)(\tilde s+\hat u-n)(\hat t+\hat u-n)}{(\tilde s-n)(\hat t-n)(\hat u-n)}\,.
\end{equation}
Although the condition $M^2=\hat M^2$ was enforced there is still the $\hat M^2_0$ parameter which is unconstrained so far; let us make it explicit by writing  (this will 
also serve as a definition for the constant $\mathcal C$ in front of the amplitude in eq.~(\ref{eq:VSfirstAttempt}))
\begin{align}
\mathcal C \prod_n^{\infty}&\frac{(\tilde s+\hat t-n)(\tilde s+\hat u-n)(\hat t+\hat u-n)}{(\tilde s-n)(\hat t-n)(\hat u-n)}\nonumber \\ \nonumber
&=\frac{\Gamma(1+2\gamma_0)\Gamma(1+\gamma_0-\tilde u)\Gamma(1+\gamma_0-\tilde t)\Gamma(1-\tilde s)}{\Gamma(1+\tilde u+\gamma_0)\Gamma(1+\tilde t+\gamma_0)\Gamma(1+\tilde s+2\gamma_0)}\\
& \equiv
\mathcal A_{\rm VS}^{\gamma_0}(s,t,u)\,,
\end{align}
%The amplitude in eq.~(\ref{eq:Attempt2}) is not yet our final answer in the case of the scalar scattering $\phi(p_1)\phi(p_2) \to  \varphi(\bar{p}_3)\varphi(\bar{p}_4)$ under scrutiny. The reason is that the amplitude for this process is symmetric under the $t\leftrightarrow u$ exchange (as already clear from the GR result).  This means that in eq.~(\ref{eq:Attempt2}) poles in $t$ will be unavoidably accompanied by poles in $u$. In turn,  absence of higher order poles due to locality modifies the structure of the numerator into
which is a solution symmetric in $t\leftrightarrow u$ but asymmetric in $s\leftrightarrow t$ for $\gamma_0\neq0$.
Consequently, it is well-suited for the UV completion  of the scalar-fermion  scattering amplitude where we have a general $\mathcal A_{\rm VS}^{\gamma_0}$ factor while for scattering of identical scalars we expect
\begin{equation}
\mathcal A_{\phi\phi\phi\phi}=\frac{8\pi}{M_{\rm Pl}^2}\left(\frac{tu}{s}+\frac{su}{t}+\frac{ts}{u}\right) \mathcal A_{\rm VS}^{\gamma_0=0}(s,t,u)\,.
\end{equation}
Let us denote $\mathcal A_{\rm VS}^{\gamma_0=0}\equiv \mathcal A_{\rm VS}$ in what follows. 
If one is to reconstruct the scattering of indistinguishable scalars by symmetrizing the amplitude for distinct scalars as
\begin{equation}
\mathcal A_{\phi\phi\phi\phi}=\mathcal A_{\phi\phi\varphi\varphi}+\mathcal A_{\phi\phi\varphi\varphi}(s\leftrightarrow t)+\mathcal A_{\phi\phi\varphi\varphi}(s\leftrightarrow u)\,,
\end{equation}
then the same factor $\mathcal A_{\rm VS}^{\gamma_0=0}$ should also appear in the case of distinct scalar, and we get to our final result for the $\phi\varphi$
amplitude
 \begin{eqnarray}\label{eq:Attempt3}
\mathcal{A}_{\phi\phi\varphi\varphi} &=&  \frac{8\pi}{M_{\rm Pl}^2}\left(\frac{tu}{s} - as\right) \mathcal A_{\rm VS}(s,t,u)\,.
%\nonumber \\&& \prod_{k=1}^{\infty}\frac{(s + M_k^2)(t + M_k^2)(u + M_k^2)}{(s-M_k^2)(t-M_k^2)(u-M_k^2)}\,,
\end{eqnarray}
The same factor is needed for gravitational Compton scattering this time given that there are poles in all $s,t,u=0$ which -- as shown after eq.~(\ref{eq:UnitarityCancel}) -- requires $\gamma_0=0$. %This case however does not have a $s,t,u$ exchange symmetry as if one where to include the mass of the external matter states the $t,u$ poles would be in $t-m^2$.
%in which indeed all poles are simple and unitarity is guaranteed  by a cancellation analogue to that of eq.~(\ref{eq:UnitarityCancel}).  The Euler definition of the $\Gamma$ function 
%    \begin{equation}
%    \Gamma(z) = \frac{1}{z}\prod_{n=1}^{\infty}\frac{(1 + 1/n)^z}{1 + z/n}\,
%    \end{equation}
%    together with our ansatz for the mass spectrum, gives
% \begin{equation}\label{eq:Attempt4}
%\mathcal{A}_{\phi\phi\varphi\varphi} =  \frac{8\pi M^2}{M_{\rm Pl}^2}\left(\frac{\tilde{t}\tilde{u}}{\tilde{s}} - a\tilde{s}\right)
%\underbrace{\frac{\Gamma(1 - \tilde{s})\Gamma(1 - \tilde{t})\Gamma(1 - \tilde{u})}{
%    \Gamma(1+\tilde{s})\Gamma(1+\tilde{t})\Gamma(1+\tilde{u})
%    }}_{\equiv~{\rm Virasoro-Shapiro\,form\,factor}}\,.
%\end{equation}
%where we introduced the dimensionless variables $\tilde{x} \equiv x/M^2$ for $x=s,t,u$.
%The ratio $\alpha_{\rm Pl} \equiv M^/M^2_{\rm Pl}$ has dimension of a coupling squared, and we expect $\alpha_{\rm Pl}\ll 1$ as a consequence of the assumption of weak coupling.
%The amplitude in eq.~(\ref{eq:Attempt4}) represents our final result as far as the scattering of distinct scalar particles is concerned. 
%In the case of identical scalar particles, the completion of the corresponding GR amplitude, discussed in table~\ref{tab:1}, is analogue to eq.~(\ref{eq:Attempt4}).
%As a second example, we consider the graviton/graviton scattering $g^{+2}g^{-2}\to g^{+2}g^{-2}$ with amplitude in GR given by
Graviton-graviton scattering does also have infrared poles in all three channels and it reads then
\begin{equation}
\mathcal{A}_{g^{+2}g^{-2}g^{-2}g^{+2}} = 
 \frac{8\pi( \scpl{2}{3} \scpr{1}{4})^4}{M_{\rm Pl}^2\,stu}
\mathcal A_{\rm VS}(s,t,u)
%= \frac{8\pi u^4}{M_{\rm Pl}^2\,stu}
\,.
\end{equation}
Notice  that once the helicity structure of the scattering amplitude $( \scpl{2}{3} \scpr{1}{4})^4$ is factorized, the rest of the four-graviton 
amplitude -- stripped of any knowledge about the helicity of the external particles -- is completely symmetric in the three Mandelstam variables $s,t,u$.
%since it basically describes scattering of identical scalar particles. This observation immediately implies that we expect the same UV completion in eq.~(\ref{eq:Attempt4}).

The construction carried out in these examples can be repeated for all the SM scattering amplitudes studied in 
section~\ref{sec:Elem}. 
Before analyzing in detail the properties of our UV completion and its actual validity, therefore, let us summarize our results.

\subsection{Generalization and emergent properties}\label{sec:UVSummary}
In the previous section we have derived what are in practice deformations of Veneziano and Virasoro-Shapiro amplitudes (which reduce to these in a certain limit $\eta=1,\gamma_0=0$) and used them to unitarize GR amplitudes by multiplication.
We also provided a number of examples to illustrate which type of factor is adequate for a given scattering process. In the $s_{ij}$ notation 
our ``bottom-up'' UV completion of gravity  can then be summarized as
\begin{equation}\label{eq:Dressing}
\mathcal{A}_{1^{h_1}2^{h_2}3^{h_3}4^{h_4}} = \mathcal{A}_{1^{h_1}2^{h_2}3^{h_3}4^{h_4}}^{\rm GR}\times
\left\{
\begin{array}{c}
  %\underbrace{
    \mathcal A_{\rm VZ}^{\eta,\gamma_0}(s_{12},s_{13})
%  \frac{\Gamma(1-\tilde{s}_{12})\Gamma(1-\tilde{s}_{13})}{\Gamma(1 + \tilde{s}_{14})}
  %}_ {\equiv \mathcal{A}_{\rm VZ}}   
  \\
  \\
  %\underbrace{
  \mathcal A_{\rm VS}^{\gamma_0}(s_{12},s_{13},s_{14})
  %\frac{\Gamma(1 - \tilde{s}_{12})\Gamma(1 - \tilde{s}_{13})\Gamma(1 - \tilde{s}_{14})}{  \Gamma(1+\tilde{s}_{12})\Gamma(1+\tilde{s}_{13})\Gamma(1+\tilde{s}_{14})}
  %}_{\equiv \mathcal{A}_{\rm VS}}
\end{array}
\right.
\end{equation}
%\begin{equation}\label{eq:Dressing}
%\mathcal{A}_{1^{h_1}2^{h_2}3^{h_3}4^{h_4}} = \mathcal{A}_{1^{h_1}2^{h_2}3^{h_3}4^{h_4}}^{\rm GR}\times 
%\left\{
%\begin{array}{c}
%  \mathcal{A}_{\rm VZ}   \\
%    \\
% \mathcal{A}_{\rm VS}
%\end{array}
%\right.
%\end{equation}
%where the GR amplitude  is completed by a form factor that closely resembles either the 
%Veneziano amplitude ($\mathcal{A}_{\rm VZ}$) or the Virasoro-Shapiro amplitude ($\mathcal{A}_{\rm VS}$) in string theory
%\begin{eqnarray}
%\mathcal{A}_{\rm VZ} &\equiv&  \frac{\Gamma(1-\tilde{s}_{12})\Gamma(1-\tilde{s}_{13})}{\Gamma(1 + \tilde{s}_{14})}\,,\\
%\mathcal{A}_{\rm VS} &\equiv&  \frac{\Gamma(1 - \tilde{s}_{12})\Gamma(1 - \tilde{s}_{13})\Gamma(1 - \tilde{s}_{14})}{
%    \Gamma(1+\tilde{s}_{12})\Gamma(1+\tilde{s}_{13})\Gamma(1+\tilde{s}_{14})}\,,
%\end{eqnarray}
%where $\tilde{s}_{ij}\equiv s_{ij}/M^2$ as before.
%The gravitational amplitudes that are UV-completed by the Veneziano form factor are those
%describing fermion-fermion, fermion-vector and vector-vector scatterings (i.e. all scattering processes in the central triangle in table~\ref{tab:1}) while all amplitudes involving scalars or gravitons (i.e. all scattering processes in the top row and last column in table~\ref{tab:1}) are UV-completed by the Virasoro-Shapiro form factor. 
As for the assignment or $\mathcal A_{\rm VZ}$ or $\mathcal A_{\rm VS}$ we have noted that invariance under particle exchange (i.e crossing symmetry) for certain processes determines the factor, 
and so for identical fermion-fermion and vector-vector  scattering we have $\mathcal A_{\rm VZ}^{\eta=1,\gamma_0=0}$ whereas for identical scalar-scalar and graviton-graviton $\mathcal A_{\rm VS}^{\gamma_0=0}$. 
Gravitational Compton scattering has IR poles in all $s,t,u$ channels, and this led us to choose $\mathcal A_{\rm VS}^{\gamma_0=0}$. 
Scalar-fermion and scalar-vector, on the contrary, only have an IR pole in $s_{12}$ and must be symmetric in $1\leftrightarrow 2$ ($s_{13}\leftrightarrow s_{14}$), and $\mathcal A_{\rm VS}^{\gamma_0}(s_{12},s_{13},s_{14})$ provides a valid UV completion. Equivalently, fermion-vector has a factor $\mathcal A_{\rm VZ}^{\eta,\gamma_0}(s_{12},s_{13})$. 
Finally same-spin matter scattering of distinct particles need not be symmetric under $2\leftrightarrow3$; however, if we expect to obtain undistinguishable same-particle scattering by symmetrization of distinguishable, we also need $\mathcal {A}_{\rm VZ}^{\eta=1,\gamma_0=0}$ for distinguishable fermion-fermion and vector-vector and $\mathcal {A}_{\rm VS}^{\gamma_0=0}$ for distinguishable scalar-scalar. All this is summarized in table~\ref{tab:2}.
\begin{table}[h]
	\begin{tabular}{!{\vrule width 0.55pt}!{\vrule width 0.55pt}c!{\vrule width 0.55pt}!{\vrule width 0.55pt}c!{\vrule width 0.55pt}!{\vrule width 0.55pt}c!{\vrule width 0.55pt}!{\vrule width 0.55pt}c!{\vrule width 0.55pt}!{\vrule width 0.55pt}c!{\vrule width 0.55pt}!{\vrule width 0.55pt}}\hline
	\multirow{2}{*}{$\mathcal A_{\rm UV}$} 	& \multirow{2}{*}{{\bf Scalar}} & \multirow{2}{*}{{\bf Fermion}} & \multirow{2}{*}{{\bf Vector}}  & \multirow{2}{*}{{\bf Graviton}} \\
	  & & & &  \\ \Cline{0.55pt}{1-5}
		\multirow{2}{*}{{\bf Scalar}} & \multirow{2}{*}{$\mathcal A_{\rm VS}$} & \multirow{2}{*}{$\mathcal A_{\rm VS}^{\gamma_0}$} & \multirow{2}{*}{$\mathcal A_{\rm VS}^{\gamma_0}$} & \multirow{2}{*}{$\mathcal A_{\rm VS}$} \\
	  & & & & \\ \hline
		\multirow{2}{*}{{\bf Fermion}} & & \multirow{2}{*}{$\mathcal A_{\rm VZ}$} & \multirow{2}{*}{$\mathcal A_{\rm VZ}^{\eta,\gamma_0}$} & \multirow{2}{*}{$\mathcal A_{\rm VS}$} \\
	  & & & & \\   \Cline{0.55pt}{1-1}\Cline{0.55pt}{3-5}
		{\multirow{2}{*}{\bf Vector}} & \multicolumn{1}{c}{} & & \multirow{2}{*}{$\mathcal A_{\rm VZ}$} & \multirow{2}{*}{$\mathcal A_{\rm VS}$} \\
	  & \multicolumn{1}{c}{} & & & \\   \Cline{0.55pt}{1-1}\Cline{0.55pt}{4-5}
		\multirow{2}{*}{{\bf Graviton}} & \multicolumn{1}{c}{} & \multicolumn{1}{c}{} & & \multirow{2}{*}{$\mathcal A_{\rm VS}$} \\
	  & \multicolumn{1}{c}{} & \multicolumn{1}{c}{} & &  \\  \Cline{0.55pt}{1-1}\Cline{0.55pt}{5-5}
	\end{tabular}
	\caption{UV completion of the GR amplitudes $\mathcal A^{\rm GR}$ collected in table~\ref{tab:1}.
The UV-completed amplitude is given schematically by $\mathcal A = \mathcal A^{\rm GR}\times \mathcal A_{\rm UV}$, with the $\mathcal A_{\rm UV}$ factor 
listed in this table.\label{tab:2}
	}
\end{table}	

The main properties that emerge from our ``bottom-up'' construction are the following.
\begin{itemize}

\item [$\circ$] {\it Infinite resonances with quantized mass.}  
In the ``bottom-up'' approach, these properties follow directly from unitarity and locality. 
We find that the spectrum in eq.~(\ref{eq:SpectrumSolution}) provides a consistent solution.
From a ``top-down'' perspective, the presence of an infinite tower of states with increasing quantized mass is typically the outcome of %clear indicator
 compactification.

\item [$\circ$]  {\it Veneziano or Virasoro-Shapiro form factor.} The UV completion of gravity is realized by dressing the GR amplitude with either the 
Veneziano or the Virasoro-Shapiro form factor, as discussed in eq.~(\ref{eq:Dressing}). These factors change the UV growth with $s$ of GR into an exponential fall-off.

\item [$\circ$]  {\it Resonance duality.} 
As a consequence of unitarity, massive higher-spin resonances are always present at least in two of the three scattering regions in fig.~\ref{fig:schematic}, and they are therefore related by the corresponding crossing transformation. This remains true even for the processes in which the GR scattering  proceeds via the exchange of gravitons in one single channel,
 and hence dual channel resonances will carry SM charge. One also has that an expansion in terms of resonances in a given process, e.g. {\color{red}(r)}, is not a good expansion in the dual crossing related process which has its own expansion in pole terms, ${\color{blue}(b)}$.

\end{itemize}

Some of these properties may be familiar for readers expert in string theory. 
For instance, it is well known that in type-II superstring theory the scattering of four massless bosons is described by an amplitude with 
 a $\Gamma$-structure that  matches the factor $\mathcal{A}_{\rm VS}$ in eq.~(\ref{eq:Dressing})~\cite{Schwarz:1982jn}.
 %\footnote{As an important  remark, notice that type-II superstring theory is only consistent in ten space-time dimensions but we derived eq.\,(\ref{eq:Dressing}) in four %dimensions. 
% There is, however, no contradiction here because it is known that compactification leaves the expression for the string amplitudes unchanged in the tree %approximation~\cite{Schwarz:1982jn}. Furthermore, it is also known that -- following the ``top-down'' approach -- it is possible to match the GR expressions for the four-%graviton scattering starting from the type-II string amplitude~\cite{Sannan:1986tz}.}
Nonetheless, let us stress that our results here are a mere consequence of the ``bottom-up'' approach that only obeys to 
the fundamental properties of unitarity, locality and causality, and is not committed to any particular model.
It is, therefore, important to understand what can be learned from this complementary point of view.
To this end, after outlining in this section the general structure of our result, we shall now move to analyze it in more detail.

%%%%%%%%%%%%%%%%%%%%%%%%%%%%%%%%%%%%%%%%%%%%%%%%%%%%%%%%%%
%%%%%%%%%%%%%%%%%%%%%%%%%%%%%%%%%%%%%%%%%%%%%%%%%%%%%%%%%%
%%%%%%%%%%%%%%%%%%%%%%%%%%%%%%%%%%%%%%%%%%%%%%%%%%%%%%%%%%
%%%%%%%%%%%%%%%%%%%%%%%%%%%%%%%%%%%%%%%%%%%%%%%%%%%%%%%%%%
%%%%%%%%%%%%%%%%%%%%%%%%%%%%%%%%%%%%%%%%%%%%%%%%%%%%%%%%%%
%%%%%%%%%%%%%%%%%%%%%%%%%%%%%%%%%%%%%%%%%%%%%%%%%%%%%%%%%%

%%%%%%%%%%%%%%%%%%%%%%%%%%%%%%%%%%%%%%%%%%%%%%%%%%%%%%%%%%
%%%%%%%%%%%%%%%%%%%%%%%%%%%%%%%%%%%%%%%%%%%%%%%%%%%%%%%%%%
%%%%%%%%%%%%%%%%%%%%%%%%%%%%%%%%%%%%%%%%%%%%%%%%%%%%%%%%%%
%%%%%%%%%%%%%%%%%%%%%%%%%%%%%%%%%%%%%%%%%%%%%%%%%%%%%%%%%%
%%%%%%%%%%%%%%%%%%%%%%%%%%%%%%%%%%%%%%%%%%%%%%%%%%%%%%%%%%
%%%%%%%%%%%%%%%%%%%%%%%%%%%%%%%%%%%%%%%%%%%%%%%%%%%%%%%%%%

\section{
 Analysis and results}\label{sec:Analysis}

We start our analysis by investigating the high-energy behavior of the amplitudes in eq.~(\ref{eq:Dressing}) that are UV-completed by  the Veneziano and 
Virasoro-Shapiro form factors $\mathcal{A}_{\rm VZ}$ and $\mathcal{A}_{\rm VS}$. The asymmetric factors $\mathcal{A}_{\rm VZ}^{\eta,\gamma_0}$ and $\mathcal{A}_{\rm VS}^{\gamma_0}$ are discussed in section~\ref{sec:aVZaVS}. These amplitudes have one or at most two parameters whereas the set of constraints from unitarity are infinitely many so it is non-trivial step to satisfy them.
%Our constructive way of unitarizing gravitational amplitudes has a single parameter yet a wealth of resonances and couplings; see shall next inspect these while checking that all properties of a well behaved amplitude are satisfied.

We consider two bounds in different kinematic regimes.
For a generic  elastic amplitude $\mathcal{A}(s,t)$, at large $s$ and fixed $t$ the Froissart bound~\cite{Froissart:1961ux} does not apply in gravitational theories with a massless graviton since there is no mass gap. However, in this limit causality still implies polynomial boundedness with 
the amplitude that can grow with $s$ but more slowly than $s^{2}$~\cite{Camanho:2014apa}. 
At large $s$ and fixed scattering angle, causality implies that the amplitude can not fall faster than $e^{-\sqrt{s}\ln s}$ (the Cerulus-Martin bound~\cite{Cerulus:1964cjb}) which recenlty has been extended to the more general case~\cite{Epstein:2019zdn}.
We shall start with causality examining the Regge limit $s\to \infty$ with $t$ fixed  (and hence forward scattering $\theta\to 0$)  we find the high-energy behavior 
\begin{align}
 \mathcal{A}_{1^{h}2^{-h}3^{-h^{\prime}}4^{h^{\prime}}}^{\rm GR}\times
\left\{
\begin{array}{cc}
  {\mathcal A}_{\rm VZ}&\sim \frac{\tilde{s}^{2+\tilde{t}}}{\tilde{t}}\,     \\
  &     \\
  {\mathcal A}_{\rm VS} &\sim \frac{\tilde{s}^{2+2\tilde{t}}}{\tilde{t}}\,  
\end{array}
\right.  ~~~~~~\forall\,\, h,h'\,.
 %{\mathcal A}_{\rm VZ}&\sim \frac{s^{2+t}}{t}\,,\\
%  \mathcal{A}_{1^{h}2^{-h}3^{-h^{\prime}}4^{h^{\prime}}}^{\rm GR}{\mathcal A}_{\rm VS} &\sim \frac{s^{2+2t}}{t}\,,
\end{align}
In the physical region $s>0$, $t<0$ one has a power-law milder than $s^2$ and hence compatible with causality as phrased in~\cite{Camanho:2014apa}.
In the hard-scattering limit $s\to \infty$ with $t/s$ fixed (and hence $\theta$ fixed) we find 
\begin{align}
 \mathcal{A}_{1^{h}2^{-h}3^{-h^{\prime}}4^{h^{\prime}}}^{\rm GR}
 \times
\left\{
\begin{array}{cc}
  {\mathcal A}_{\rm VZ} & \sim s_{\theta/2}^{2s_{\theta/2}^2\,\tilde{s}}c_{\theta/2}^{2c_{\theta/2}^2\,\tilde{s}}\,     \\
  &     \\
  {\mathcal A}_{\rm VS} & \sim  s_{\theta/2}^{4s_{\theta/2}^2\,\tilde{s}}c_{\theta/2}^{4c_{\theta/2}^2\,\tilde{s}}  \,  
\end{array}
\right.  ~~\forall\,\, h,h'\,,
%{\mathcal A}_{\rm VZ}&\sim s_{\theta/2}^{2s_{\theta/2}^2s}c	_{\theta/2}^{2c_{\theta/2}^2s} &{\mathcal A}_{\rm VS}&\sim  s_{\theta/2}^{4s_{\theta/2}^2s}c	_{\theta/2}^{4c_{\theta/2}^2s} &&\forall\,%\, h,h'
\end{align}
with an exponential fall-off controlled by $ s^2_{\theta/2}\ln(s^2_{\theta/2})+c^2_{\theta/2}\ln (c^2_{\theta/2})  < 0$~\cite{Polchinski:1998rq}.  
This is a fall-off that is much faster than in any QFT, and the amplitude indeed violates the Cerulus-Martin bound. 
This is due to the presence of an infinite number of resonances, and indeed the lower bound applicable in this case has been extended recently~\cite{Epstein:2019zdn} in agreement with the scaling above. 

Next, we turn to inspecting the properties of the resonances. This is a process that, given the results and conventions of section~\ref{sec:Elem}, can be made automatic.
Let us sketch the generic procedure.
 There can be resonances in all three channels $s_{12},s_{13},s_{14}$; in order  to identify them, 
 we first turn to the domain of the amplitude where they are kinematically accessible {\color{red}(r)}, {\color{blue}(b)}, {\color{ForestGreen}(g)} via the respective substitutions in 
 eqs.~(\ref{eq:Domains1},\,\ref{eq:Domains2},\,\ref{eq:Domains3}),
  and make explicit the CM scattering angle dependence in $t=-ss^2_{\theta/2}$ and the spinor brackets. 
 One then identifies the poles and evaluates the residue as a function of $\theta$ only.
  Finally, the different resonances can be extracted via a projection in the orthogonal set of Wigner $d$-functions. 
  This is to say, when condensed in equations
 % \footnote{Remember that Wigner $d$-functions enjoy  the orthogonality condition
%\begin{equation}
%\int_{-1}^{+1} d^{J}_{m^{\prime},m}(\theta)d^{J^{\prime}}_{m^{\prime},m}(\theta)dc_{\theta}  =
%\frac{2\delta^{JJ^{\prime}}}{2J + 1}\,.
%\end{equation}}
\begin{align}
\alpha^{n,J}_{2h,2h'}&=\frac{1}{32\pi }\int_{-1}^{+1}\underset{s=M^2_n}{\rm R\overline{es}}\left[{\color{red}\mathcal A_{h,h'}}\right]d^{J}_{2h,2h'}(\theta)dc_\theta\,,\nonumber \\\nonumber
\alpha^{n,J}_{h+h',h+h'}&=\frac{1}{32\pi}\int_{-1}^{+1}\underset{ s=M^2_n}{\rm R\overline{es}}\left[{\color{blue}\mathcal A_{h,h'}}\right]d^{J}_{h+h',h+h'}(\theta)dc_\theta\,,\\ \nonumber
\alpha^{n,J}_{h-h',h-h'}&=\frac{1}{32\pi }\int_{-1}^{+1}\underset{ s=M^2_n}{\rm R\overline{es}}\left[{\color{ForestGreen}\mathcal A_{h,h'}}\right]d^{J}_{h-h',h-h'}(\theta)dc_\theta\,,
\end{align}
where the residue R$\rm \overline{es}$ rescaled as
\begin{equation}
\left.\
\begin{array}{c}
\underset{s=M^2_n}{\rm R\overline{es}}\left[{\color{red}\mathcal A_{h,h'}}\right]     \\
 \underset{ s=M^2_n}{\rm R\overline{es}}\left[{\color{blue}\mathcal A_{h,h'}}\right]   \\
   \underset{ s=M^2_n}{\rm R\overline{es}}\left[{\color{ForestGreen}\mathcal A_{h,h'}}\right]  
\end{array}
\right\} \equiv  \lim_{s\to M_n^2}\frac{M_n^2 - s}{M_n^2}
\left\{
\begin{array}{c}
{\color{red}{\mathcal{A}_{1^h2^{-h}3^{-h'}4^{h'}}}} \\ \\
{\color{blue}{\mathcal{A}_{1^h3^{-h'}2^{-h}4^{h'}}}}  \\ \\
{\color{ForestGreen}{\mathcal{A}_{1^h4^{h'}2^{-h}3^{-h'}}}}   
\end{array}
\right.
\end{equation}
 is a function of $(s,\theta)$, and with $\alpha$ given in eq.~(\ref{eq:MstSpnJ}) and generalized to the $h'\pm h$ case. Let us note that regions {\color{blue}(b)}, {\color{ForestGreen}(g)} correspond to elastic scattering and as such the resonances in $s_{13},s_{14}$ variables have couplings subject to positivity; explicitly one has
\begin{align}
\alpha^{J}_{2h,2h}\,,\, \alpha^{J}_{h+h',h+h'}\,,\,\alpha^{J}_{h-h',h-h'}\geq 0\,.
\end{align}
Wigner $d$-functions $d^{J}_{a,b}$ are defined only for $J\geq$Max$(a,b)$ which implies in particular that the lowest spin resonance will have 
$J_{\rm min}=h'-h$ if resonances are present in the $s_{14}$ channel. Therefore, gravitational Compton scattering, scattering with scalars and graviton scattering have resonances of spin $h'-h$; scalar resonances are to be found only for scalar scattering and graviton scattering whereas fermion resonances only for scalar-fermion scattering. The rest of the amplitudes have Venenziano-like factors and a minimum spin $h'+h=1,3/2,2$ for fermion-fermion, fermion-vector and vector-vector.

 Let us be explicit about the angle dependence of the minimum spin Wigner $d$-function
\begin{gather}
d^{J_{\min}}_{h_1-h_2,h_3-h_4}(\theta) \label{dWscaling} \\\nonumber
\begin{tikzpicture}
\draw [very thick, blue, ->] (0,0)--(0,-1);
\draw [very thick, red, ->] (-0.5,0)--(-2.5,-1);
\draw [very thick, ForestGreen, ->] (0.5,0)--(2.5,-1);
\end{tikzpicture}\\
s_{\theta/2}^{2h'-2h}c_{\theta/2}^{2h'+2h}\qquad  c_{\theta/2}^{2h'+2h} \qquad\qquad c_{\theta/2}^{2h'-2h} \nonumber
\end{gather}
This angular dependence has to be compared with that of the amplitudes in eq.~(\ref{eq:Dressing}) which we separate in two factors, namely the 
GR part containing the overall $s_{ij}$ and the helicity factor (cf. eq.~(\ref{MstGR}) and appendix~\ref{app:A})
\begin{gather} \label{thetascaling}
\frac{s_{13}^{1-h'-r}s_{14}^{1-h'+r}}{s_{12}}\left(\scpl{2}{3}\scpr{1}{4}\right)^{2h}\left(\splf{3}\!\hat P_{12}\!\spr{4}\right)^{2h'-2h}\\\nonumber
\begin{tikzpicture}
\draw [very thick, blue, ->] (0,0)--(0,-1);
\draw [very thick, red, ->] (-0.5,0)--(-3,-1);
\draw [very thick, ForestGreen, ->] (0.5,0)--(3,-1);
\end{tikzpicture}\\ \nonumber
ss_{\theta/2}^{2(1-h+r)}c_{\theta/2}^{2(1+h-r)}\quad  ss_{\theta/2}^{-2}c_{\theta/2}^{2(1+h-r)} \quad ss_{\theta/2}^{-2}c_{\theta/2}^{2(1-h+r)}
\end{gather}
and the residue of either the Veneziano form factor $\mathcal{A}_{\rm VZ}$ or the Virasoro-Shapiro form factor $\mathcal{A}_{\rm VS}$. 
For both $\mathcal{A}_{\rm VZ}$ and $\mathcal{A}_{\rm VS}$ the physically accessible resonances are located at the poles of the factor $\Gamma(1-\tilde{s})$, that are 
$\tilde{s}_n = 1+n$, i.e. $M_n^2 = (1+n)M^2$ with $n=0,1,\dotso \in \mathbb{N}$.
%We define 
%\begin{equation}
%\underset{1+n}{\rm R\overline{es}}[\mathcal{A}_{\rm VZ/VS}] \equiv \lim_{\tilde{s} \to 1+n}(1-\tilde{s} + n)\mathcal{A}_{\rm VZ/VS}\,,
%\end{equation}
For brevity we denote $s=(1+n)M^2$ as $1+n$ in the subscript of R$\rm\overline{es}$, and we get for the Veneziano form factor
\begin{align}
\underset{1+n}{\rm R\overline{es}}\left[\mathcal{A}_{\rm VZ}\right] = &\frac{-1}{(1+n)!}\prod_{k=0}^n(k+\tilde{t})\\ =&\frac{-1}{(1+n)!}\prod_{k=0}^n\left[k-(1+n)s_{\theta/2}^2\right]\,,
\end{align}
while in the Virasoro-Shapiro case
\begin{align}
\underset{1+n}{\rm R\overline{es}}\left[\mathcal{A}_{\rm VS}\right]&=\frac{(-1)^n}{(1+n)!^2}\prod_{k=0}^n(k+\tilde{t})(k+\tilde{u})\\ \nonumber
%&=\frac{(-1)^n}{n!}\prod_{k=0}^n\left(k-(1+n)s_{\theta/2}^2\right)\left(k-(1+n)c_{\theta/2}^2\right)\\
%&=\frac{-1}{n!(n+1)!}\prod_{k=0}^n\left[k-(1+n)s_{\theta/2}^2\right]^2\,.
&= \frac{s_{\theta/2}^2c_{\theta/2}^2}{(n!)^2}\prod_{k=1}^n\left[k-(1+n)s_{\theta/2}^2\right]\,.
\end{align}
This means that, on top of the angular powers coming from the helicity factor in eq.~(\ref{thetascaling}), we expect 
at least another factor of $s_{\theta/2}^2c^2_{\theta/2}$ or $s_{\theta/2}^2$ depending on whether we have the Virasoro-Shapiro or Veneziano. Take e.g. the former and restrict to {\color{blue}(b)} and $n=0$ where we have $\mathcal {\color{blue}A_{\rm VS}}\sim c_{\theta/2}^{4+2h}$; for this amplitude to admit a decomposition in terms of Wigner $d$-functions $d^{J}_{h'+h}$ which themselves scale with at least $2h+2h'$ powers of $c_{\theta/2}$ (cf. eq.~(\ref{dWscaling})) one has that $h'\leq 2$.
In the same fashion for Veneziano, comparing eq.~(\ref{dWscaling}) and eq.~(\ref{thetascaling}), we obtain  $h'\leq1$.
Consequently,  we have that this procedure does not apply to higher spin as external states. 

Finally and before moving on, it is pertinent to discuss the quantum numbers of the resonances. 
This can be done by simply combining the representation of the two external states  that annihilate into a resonance. 
It seems one would have all possible combinations of two particles but it is not always the case that we have all $s,t,u$ channels open (this is the case for all the GR amplitudes UV-completed by means of the Veneziano form factor). 
Is there a rule to tell when is a channel open? 
 If we assign {\it helicity charge} $\mathfrak{h} = +1$ to all positive helicity matter particles and $\mathfrak{h} = -1$ to all negative ones, then 
 we find that an inherited consequence of helicity conservation in our scenario is that there are no resonances with helicity charge  $\mathfrak{h} = \pm 2$. 
 This selection rule, for instance, is responsible for the absence of massive higher spin resonances in the $u$-channel for the process 
  $q^{+1/2}\bar{q}^{-1/2}\to V^{+1}V^{-1}$.
%We find that an inherited consequence of MHV in this scenario means that matter {\it helicity sign} is conserved.
%REally what I mean is that assigning $\pm$ charge to helicity sign, there is no odd charge resonances. Except Compton. Maybe is helicity sign parity. No, what it is that there is no charge 2 resonances.

%%%%%%%%%%%%%%%%%%%%%%%%%%%%%
\subsection{Resonance analysis}
After having laid out the tools to extract a resonance spin and couplings the following subsection explores the spectrum within amplitudes in all three classes of matter-matter, matter-graviton and graviton-graviton scattering. Within each category, rather than covering all possibilities we focus on representative cases the emphasis being on positivity.
\subsubsection{Graviton-graviton scattering}
%{\bf i) Graviton-graviton \\}
After the generalities let us turn to a few cases to illustrate the analysis. Consider pure gravity and the process in region {\color{red}(r)} $g^{+2}g^{-2}\to g^{+2}g^{-2}$, which we note is the same process as in {\color{blue}(b)} due to crossing symmetry; one has
\begin{align}
\underset{1+n}{\rm R\overline{es}}\left[{\color{red}\mathcal A_{2,2}}\right]=&
\frac{8\pi M^2(1+n)c_{\theta/2}^8}{M_{\rm Pl}^2 (n!)^2}\prod_{k=1}^n\left[k-(1+n)s_{\theta/2}^2\right]^2\,.
\end{align}
We note that resonances start at $J_{\rm min}=4$. Our definition of $\alpha$ reads now
\begin{align}\label{eq:Gamma0}
\alpha_{4,4}^{n,J}\equiv&\frac{M^2}{M_{\rm Pl}^2}{\rm N}_{4,4}^{n,J}\equiv\alpha_{\rm Pl}{\rm N}_{4,4}^{n,J}\,,
\end{align}
where 
\begin{align}\label{eq:GpGm}
{\rm N}_{4,4}^{n,J}=&\frac{(1+n)}{4n!}\int_{-1}^{+1}c_{\theta/2}^8\,d^J_{4,4}(\theta)\prod_{k=1}^n\left[k-(1+n)s_{\theta/2}^2\right]^2 dc_\theta\,.
\end{align}
With these rewriting of the pseudo-widths $\alpha$ we have separated a universal coupling $\alpha_{\rm Pl}\equiv M^2/M_{\rm Pl}^2$ related to the first new resonance mass and a rational number N, which is tabulated in table~\ref{tab:Dec}.
\begin{table}[htp]
	\begin{center}%\vspace{-0.25cm}
	\begin{adjustbox}{max width=1\textwidth}
		\begin{tabular}{||c||c|c|c|c|c|c|c||c||}\hline
			\multirow{2}{*}{${\rm N}_{4,4}^{n,J}$} & \multirow{2}{*}{$J\leqslant 3$} & \multirow{2}{*}{4} & \multirow{2}{*}{5} & \multirow{2}{*}{6} & \multirow{2}{*}{7} & \multirow{2}{*}{8} & \multirow{2}{*}{$\dots$} \\
			& & & & & & & \\ \hline\hline
			\multirow{2}{*}{$n=0$} & \multirow{2}{*}{\ding{55}} & \multirow{2}{*}{$\frac{1}{18}$} &   &  & & &  \\
			& & & & & & & \\ \hline
			\multirow{2}{*}{$1$} & \multirow{2}{*}{\ding{55}} & \multirow{2}{*}{$\frac{37}{495}$} & \multirow{2}{*}{$\frac{4}{165}$} & \multirow{2}{*}{$\frac{2}{429}$} & & &  \\
			& & & & & & &   \\ \hline
			\multirow{2}{*}{$2$} & \multirow{2}{*}{\ding{55}} & \multirow{2}{*}{$\frac{2723}{34320}$} & \multirow{2}{*}{$\frac{207}{5005}$} & \multirow{2}{*}{$\frac{711}{40040}$} & 
			\multirow{2}{*}{$\frac{81}{14560}$} & \multirow{2}{*}{$\frac{243}{247520}$} & \\
			& & & & & & & \\ \hline
			$\dots$ & \ding{55} & & & & & & \\  \hline
		\end{tabular}
		\end{adjustbox}
	\end{center}\vspace{-0.5cm}\caption{Coefficients of the decomposition in eq.\,(\ref{eq:GpGm}).}\label{tab:Dec}
\end{table}

Consider next the other channel; $g^{+2}g^{+2}\to g^{+2}g^{+2}$, with resonances now starting at $J_{\rm min}=0$ and therefore our basis for the projection being Legendre polynomials. The very same steps lead now to
\begin{align}
\underset{1+n}{\rm R\overline{es}}\left[{\color{ForestGreen}\mathcal A_{2,2}}\right]=&
\frac{8\pi M^2(1+n)}{M_{\rm Pl}^2(n!)^2}
\prod_{k=1}^n\left[k-(1+n)s_{\theta/2}^2\right]^2\,,
\end{align}
and couplings
\begin{equation}\label{eq:Gamma1}
\alpha_{0,0}^{n,J}\equiv \alpha_{\rm Pl}{\rm N}_{0,0}^{n,J}\,,
\end{equation}
where 
\begin{align}\label{eq:GpGp}
{\rm N}_{0,0}^{n,J}=&\frac{(1+n)}{4(n!)^2}\int_{-1}^{+1}P_J(c_\theta)\prod_{k=1}^n\left[k-(1+n)s_{\theta/2}^2\right]^2 dc_\theta\,,
\end{align}
with the corresponding numerical values tabulated in table~\ref{tab:Dec2}.
\begin{figure*}[t]
$$
\includegraphics[width=0.48\textwidth]{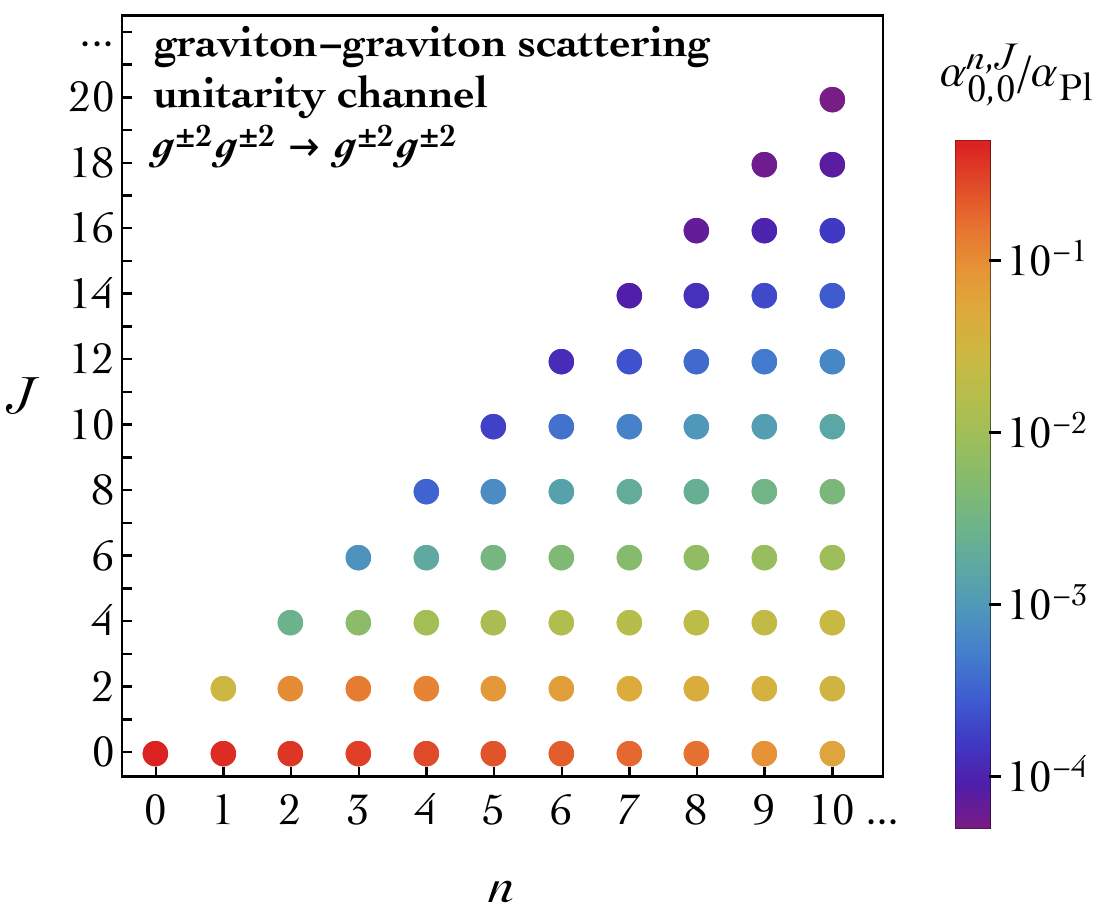}\qquad
\includegraphics[width=0.48\textwidth]{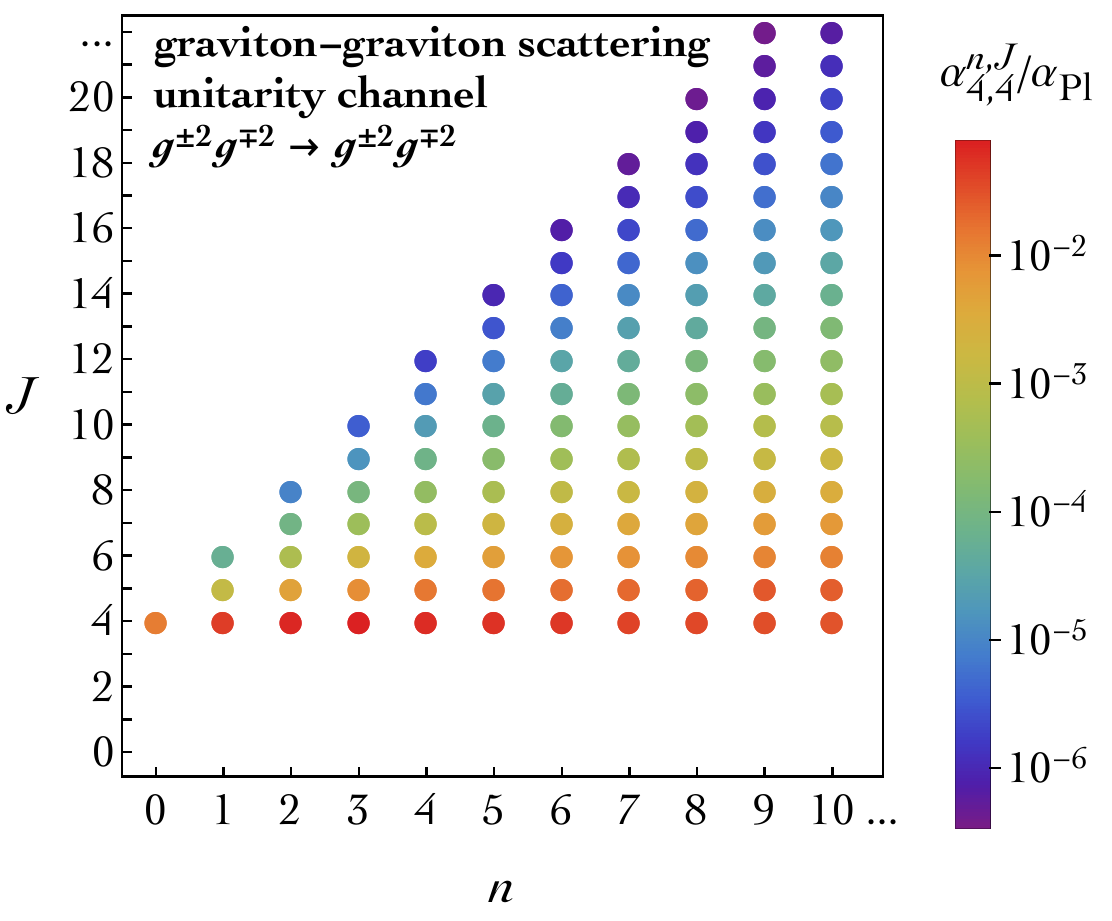}$$
\caption{
Chew-Frautschi spin/mass plot of the spectrum of exchanged resonances in the four-graviton scattering  $g^{\pm 2}g^{\pm 2} \to g^{\pm 2}g^{\pm 2} $ (left panel) and 
$g^{\pm 2}g^{\mp 2} \to g^{\pm 2}g^{\mp 2}$ (right panel). The mass is quantized according to the relation $M_n^2 = (1+n)M^2$ with 
$n=0,1,\dotso \in \mathbb{N}$ on the $x$-axes. The rainbow colors in the legend mark, for each resonance, the value of $\Gamma/M$ in units of $\alpha_{\rm Pl}\equiv M^2/M_{\rm Pl}^2$ (see eq.~(\ref{eq:Gamma0}) and eq.~(\ref{eq:Gamma1})).
\label{fig:Res}}
\end{figure*}
\begin{table}[htp]
\begin{center}%\vspace{-0.25cm}
\begin{tabular}{||c||c|c|c|c|c|c|c|c||c|}\hline
\multirow{2}{*}{N$_{0,0}^{n,J}$} & \multirow{2}{*}{$J = 0$} & \multirow{2}{*}{1} & \multirow{2}{*}{2} & \multirow{2}{*}{3} & \multirow{2}{*}{4} & \multirow{2}{*}{5} & \multirow{2}{*}{6} & \multirow{2}{*}{$\dots$} \\
 & & & & & & & &    \\ \hline\hline
\multirow{2}{*}{$n=0$} & \multirow{2}{*}{$\frac12$} & & & & & & &  \\
 & & & & & & & &  \\ \hline
\multirow{2}{*}{$1$} & \multirow{2}{*}{$\frac{1}{3}$} & \multirow{2}{*}{\ding{55}} & \multirow{2}{*}{$\frac{2}{15}$} & & & & &  \\
 & & & & & & & & \\ \hline
\multirow{2}{*}{$2$} & \multirow{2}{*}{$\frac{21}{80}$} & \multirow{2}{*}{\ding{55}} & \multirow{2}{*}{$\frac{9}{56}$} & \multirow{2}{*}{\ding{55}} & \multirow{2}{*}{$\frac{27}{560}$} & & &  \\
 & & & & & & & & \\ \hline
%$3$ & $\frac{428}{105}$ & \ding{55} & $\frac{104}{7}$ & \ding{55} & $\frac{4864}{385}$ & \ding{55} & $\frac{1024}{231}$ & \\ \hline
$\dots$ & & \ding{55} & & \ding{55} & & \ding{55} & & \ding{55} \\  \hline
\end{tabular}
\end{center}\vspace{-0.5cm}\caption{Coefficients of the decomposition in eq.\,(\ref{eq:GpGp}).}\label{tab:Dec2}
\end{table}
The lack of odd spins can be traced to the amplitude being even under $\theta\to\pi-\theta$.
The positivity of all entries in this tables is in accordance with unitarity and indeed the $\alpha's$ correspond to tree-level decay widths. 

For illustrative purposes, we show in fig.~\ref{fig:Res} the Chew-Frautschi  plot 
yielding the spin of the resonances (on the $y$-axes) versus the square of their quantized mass (on the $x$-axes, labeled with the integer $n$) for the two channels $g^{+ 2}g^{+ 2} \to g^{+ 2}g^{+ 2} $ (left panel) and 
$g^{+ 2}g^{- 2} \to g^{+ 2}g^{- 2}$ (right panel).  Each resonance is colored according to the corresponding value of tree-level decay width ($\alpha_{0,0}^{n,J}$ and $\alpha_{4,4}^{n,J}$, respectively) as indicated in the legend. Resonances are more and more weakly coupled for increasing values of spin.

%Before proceeding, let us try to summarize the conceptual importance of our findings. We discussed a possible UV completion of the four-graviton scattering in the allowed helicity configurations. What is remarkable is that this was possible--by means of the on-shell spinor helicity formalism--without the introduction of any Lagrangian density but just as a consequence of fundamental principles of QFT. The spinor-helicity formalism gives access to the on-shell part of the amplitude. Nonetheless, despite the fact that it is not possible to reconstruct the full amplitude, we have shown that important pieces of information about the physics that is responsible for the unitarization can be derived.Apart from the four-graviton scattering, it is interesting to look at other four-point amplitudes with SM external states in thepresence of gravity. This is what we shall do next.

%%%%%%%%%%%%%%%%%%%%%%%%%%%%%%%%%%%%%%%%%%%%%%%%%%
%%%%%%%%%%%%%%%%%%%%%%%%%%%%%%%%%%%%%%%%%%%%%%%%%%
\subsubsection{Matter-matter scattering}
%{\bf ii) Matter-matter\\}
The analysis carried out in the previous section presents no difference when applied to matter as external states. 
There is, however,  a difference in the amplitudes themselves since some of them present an independent parameter other than $\alpha_{\rm Pl}$. 
These are fermion-fermion and distinguishable scalar scattering (see table~\ref{tab:1}). 
In the following, we shall focus on these cases to show what can the analysis reveal about these parameters.
%, and what can the analysis reveal about these parameters.
 
 %%%%%%%%%%%%%%%%%%%%
$\circ$~ Consider first distinct scalar in the elastic channel so that
\begin{align}
\underset{1+n}{\rm R\overline{es}}\left[{\color{blue}\mathcal A_{0,0}}\right]=&\frac{8\pi M^2(1+n)}{M_{\rm Pl}^2(n!)^2}
\left(\frac{c_{\theta/2}^2}{s_{\theta/2}^2}+as_{\theta/2}^2\right)s_{\theta/2}^2c_{\theta/2}^2\times\nonumber\\
&\prod_{k=1}^n\left[k-(1+n)s_{\theta/2}^2\right]^2\,,
\end{align}
 and again 
 \begin{align}\label{eq:S1S2}
 {\rm N}_{0,0}^{n,J}=&\frac{(1+n)}{4(n!)^2}\int_{-1}^{+1}\left( \frac{c_{\theta/2}^2}{s_{\theta/2}^2}+as_{\theta/2}^2  \right)
 s_{\theta/2}^2c_{\theta/2}^2P_J(c_\theta)\times \nonumber \\ &\prod_{k=1}^n\left[k-(1+n)s_{\theta/2}^2\right] dc_\theta\,.
 \end{align}
 Positivity demands $\alpha_{0,0}^{n,J},{\rm N}_{0,0}^{n,J}\geq 0$, and as table~\ref{tab:NScalar} shows this is not the case for all values of $a$ but only those in the interval
\begin{align}\label{eq:abound}
0\leq a\leq 2\,,
\end{align}
which is compatible with 0 (that is with the case of a scalar field minimally coupled to Einstein-Hilbert gravity) and $1/6$ for the conformal case. It is interesting to remark that the parameter $a$ does not appear in same-scalar scattering since we symmetrize to find $a(s_{12}+s_{13}+s_{14})=0$.
  \begin{table}[htp]
 	\begin{center}%\vspace{-0.25cm}
	\begin{adjustbox}{max width=.48\textwidth}
 		\begin{tabular}{||c||c|c|c|c|c|c|c||c||}\hline
 			\multirow{2}{*}{N$_{0,0}^{n,J}$} & \multirow{2}{*}{$J=0$} & \multirow{2}{*}{1} & \multirow{2}{*}{2} & \multirow{2}{*}{3} & \multirow{2}{*}{4} & \multirow{2}{*}{5} & \multirow{2}{*}{$\dots$} \\
			& & & & && &    \\ \hline\hline
 			\multirow{2}{*}{$n=0$} &  \multirow{2}{*}{$\frac{4+a}{24}$} & \multirow{2}{*}{$\frac{10-a}{120}$} & \multirow{2}{*}{$\frac{2-a}{120}$} & \multirow{2}{*}{$\frac{a}{280}$} & & & \\
			& & & & && & \\ \hline
 			\multirow{2}{*}{$1$} &  \multirow{2}{*}{$\frac{8+a}{60}$} & \multirow{2}{*}{$\frac{14-a}{140}$} & \multirow{2}{*}{$\frac{26+a}{420}$} & \multirow{2}{*}{$\frac{36+a}{1260}$} & \multirow{2}{*}{$\frac{2-a}{315}$} & \multirow{2}{*}{$\frac{a}{693}$}  & \\
			& & & & && & \\ \hline
 			\multirow{2}{*}{$2$} &  \multirow{2}{*}{$\frac{256+19a}{2240}$} & \multirow{2}{*}{$\frac{218-11a}{2240}$} & \multirow{2}{*}{$\frac{166+7a}{2240}$} & \multirow{2}{*}{$\frac{1188-21a}{24640}$} & \multirow{2}{*}{$\frac{636-21a}{24640}$} & \multirow{2}{*}{$\frac{702+69a}{64064}$}   &
 			%$\frac{162-81}{64064}$ &\frac{27a}{45760}
 			\dots \\
			& & & & && & \\ \hline
 			$\dots$ &  & & & & & & \\  \hline
 		\end{tabular}
		\end{adjustbox}
 	\end{center}\vspace{-0.5cm}\caption{Coefficients of the decomposition in eq.~(\ref{eq:S1S2}).}\label{tab:NScalar}
 \end{table}

$\circ$~We now move to consider the 2-to-2 fermion scattering.
Consider first the distinguishable case, e.g. a fermion and a lepton $q^{1/2}\ell^{-1/2}\to q^{1/2}\ell^{-1/2}$. Residues in the s-channel have quantum number under the SM gauge group
 $(3,2,1/6)$, $(3,1,5/3)$, $(\bar 3,2,5/6)$, etc. or consider $q_R^{1/2}q_L^{-1/2}\to q_R^{1/2}q_L^{-1/2}$ with more elaborate color $(3\oplus6,2,5/6)$, $(3\oplus6,2,-1/6)$ and baryon number 2/3. Each of these will have a residue at $M^2(1+n)$ as
\begin{align}
\underset{1+n}{\rm R\overline{es}}\left[{\color{blue}\tilde{\mathcal{A}}_{1/2,1/2}}\right]=&\frac{8\pi  M^2}{M_{\rm Pl}^2n!}
\left(-\frac{c_{\theta/2}^2}{s_{\theta/2}^2}+\frac{bc^2_{\theta/2}}{2}\right)\times \nonumber \\ &
\prod_{k=0}^n\left[k-(1+n)s_{\theta/2}^2\right]\,,
\end{align}
with minimum spin $J_{\rm min}=1$ and coefficients
 \begin{align}\label{eq:F1F2}
\tilde {\rm N}_{1,1}^{n,J}=&\frac{1}{4n!}\int_{-1}^{+1}
\left(-\frac{c_{\theta/2}^2}{s_{\theta/2}^2} + \frac{bc^2_{\theta/2}}{2} \right)d^J_{1,1}(\theta)\times \nonumber \\ & \prod_{k=0}^n\left[k-(1+n)s_{\theta/2}^2\right] dc_\theta\,,
\end{align}
that are listed in table~\ref{tab:DecQ}.
 \begin{table}[htp]
	\begin{center}%\vspace{-0.25cm}
		\begin{tabular}{||c||c|c|c|c|c|c|c||c||}\hline
			\multirow{2}{*}{$\tilde {\rm N}_{1,1}^{n,J}$} & \multirow{2}{*}{$J=0$} & \multirow{2}{*}{1} & \multirow{2}{*}{2} & \multirow{2}{*}{3} & \multirow{2}{*}{4} & \multirow{2}{*}{5} & \multirow{2}{*}{$\dots$} \\
			& & & & & & &  \\ \hline\hline
			\multirow{2}{*}{$n=0$} & \multirow{2}{*}{\ding{55}} & \multirow{2}{*}{$\frac{8-b}{48}$} & \multirow{2}{*}{$\frac{b}{80}$} & & & & \\
			& & & & & & & \\ \hline
			\multirow{2}{*}{$1$} & \multirow{2}{*}{\ding{55}} & \multirow{2}{*}{$\frac{1}{6}-\frac{b}{120}$} & \multirow{2}{*}{$\frac{12-b}{120}$} & \multirow{2}{*}{$\frac{b}{105}$} &  & &  \\
			& & & & & & & \\ \hline
			\multirow{2}{*}{$2$} &  \multirow{2}{*}{\ding{55}} & \multirow{2}{*}{$\frac{26-b}{160}$} &  \multirow{2}{*}{$\frac{3(42-b)}{1120}$} & \multirow{2}{*}{$\frac{9(16-b)}{2240}$} & \multirow{2}{*}{$\frac{3b}{448}$} & & \\
			& & & & & & &  \\ \hline
			$\dots$ & \ding{55} & & & & & & \\  \hline
		\end{tabular}
	\end{center}\vspace{-0.5cm}\caption{Coefficients of the decomposition in eq.\,(\ref{eq:F1F2}).}\label{tab:DecQ}
\end{table}

The residue for same fermion species scattering is instead
\begin{align}\nonumber
\underset{1+n}{\rm R\overline{es}}\left[{\color{blue}\mathcal A_{1/2,1/2}}\right]=
&\frac{8\pi M^2}{M_{\rm Pl}^2n!}c_{\theta/2}^2\left(-\frac{1}{s_{\theta/2}^2}-s_{\theta/2}^2+b\right) \times \\
&\prod_{k=0}^n\left[k-(1+n)s_{\theta/2}^2\right]\,,
\end{align}
with minimum spin $J_{\rm min}=1$ and coefficients
\begin{align}\nonumber
{\rm N}_{1,1}^{n,J}=&\frac{1}{4n!}\int_{-1}^{+1}c_{\theta/2}^2\left(-\frac{1}{s_{\theta/2}^2}-s_{\theta/2}^2+b\right)d^J_{1,1}(\theta)\times 
\\&\prod_{k=0}^n\left[k-(1+n)s_{\theta/2}^2\right] dc_\theta\,. \label{eq:FF}
\end{align}
We remark that the coefficient $b$ affects the residues of the massive resonances whereas it does not alter the graviton pole where it enters as a free parameter. 
{\it All} the coefficients ${\rm N}_{1,1}^{n,J}$  (as well as the coefficients $\tilde {\rm N}_{1,1}^{n,J}$) are subject to positivity.
 \begin{table}[htp]
 \begin{center}%\vspace{-0.25cm}
\begin{tabular}{||c||c|c|c|c|c|c|c||c||}\hline
\multirow{2}{*}{N$_{1,1}^{n,J}$} & \multirow{2}{*}{$J=0$} & \multirow{2}{*}{1} & \multirow{2}{*}{2} & \multirow{2}{*}{3} & \multirow{2}{*}{4} & \multirow{2}{*}{5} & \multirow{2}{*}{$\dots$}  \\
 & & & & & & &  \\ \hline\hline
\multirow{2}{*}{$n=0$} & \multirow{2}{*}{\ding{55}} & \multirow{2}{*}{$\frac{22-5b}{120}$} & \multirow{2}{*}{$\frac{3b-2}{120}$} & \multirow{2}{*}{$\frac{1}{210}$} & & &  \\
 & & & & & & & \\ \hline
\multirow{2}{*}{$1$} & \multirow{2}{*}{\ding{55}} & \multirow{2}{*}{$\frac16-\frac{b}{60}$} & \multirow{2}{*}{$\frac{46-7b}{420}$} & \multirow{2}{*}{$\frac{8b-5}{420}$} & \multirow{2}{*}{$\frac{1}{252}$} & &  \\
 & & & & & & &  \\ \hline
\multirow{2}{*}{$2$} & \multirow{2}{*}{\ding{55}} & \multirow{2}{*}{$\frac{92-7b}{560}$} & \multirow{2}{*}{$\frac{62-3b}{560}$}  & \multirow{2}{*}{$\frac{79-9b}{1120}$}  & \multirow{2}{*}{$\frac{15b-9}{1120}$} & \multirow{2}{*}{$\frac{9}{3080}$} &  \\
 & & & & & & & \\ \hline
$\dots$ & \ding{55} & & & & & & \\  \hline
\end{tabular}
\end{center}\vspace{-0.5cm}\caption{Coefficients of the decomposition in eq.\,(\ref{eq:FF}).}\label{tab:DecQ}
\end{table}
We show the first few values of the decomposition in table~\ref{tab:DecQ}.
Clearly, unitarity provides a non-trivial condition since not all coefficients are positive for any value of $b$.
\begin{figure}[!t!]
\centering
\includegraphics[width=0.45\textwidth]{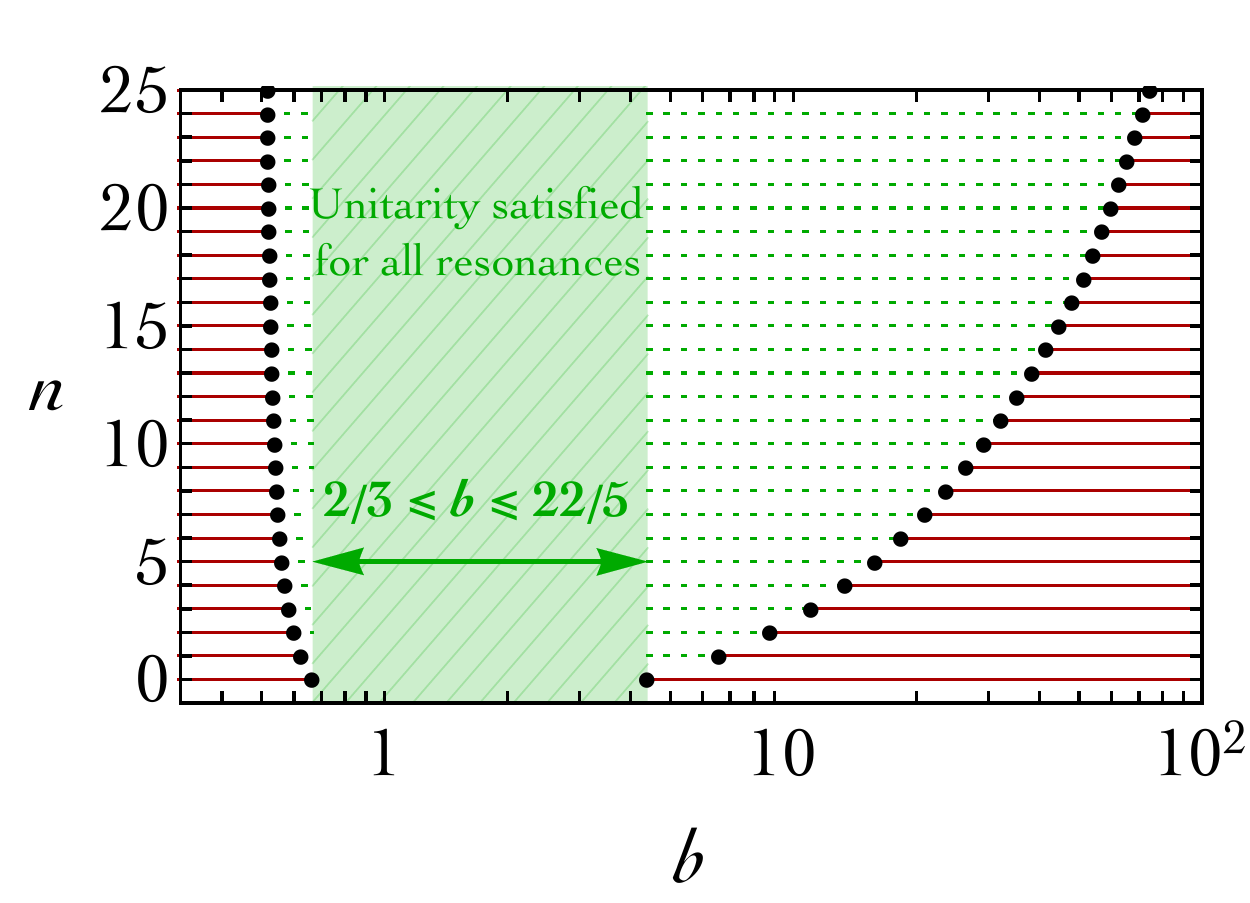}\vspace{-0.275cm}  
\caption{\label{fig:Unitarityb} 
Unitarity bound (allowed region shaded in green) on the parameter $b$ in table~\ref{tab:1} obtained 
from the positivity of the coefficients in the 
expansion in eq.~(\ref{eq:FF}).
The left-side of the bound is described by the simple analytical formula 
$b \geqslant (4+n)/2(3+n)$, which asymptotes to $1/2$ for $n\to \infty$. 
}
\end{figure}
Numerically, we find the situation illustrated in fig.~\ref{fig:Unitarityb}.
For each value of $n$ on the y-axis we find an allowed range of values for $b$ (dotted green lines, while solid red lines indicate intervals excluded by unitarity).
What is non-trivial is that all these constraints must be satisfied simultaneously, all the way up to $n \to \infty$. 
This is true in the region shaded in green, that is set by the first resonance of the spectrum with $n=0$.
We note, in particular, that the value $b=0$ is not compatible with unitarity.
Hence unitarity bounds the a priori free parameter to lie in the range
\begin{equation}\label{eq:Boundb}
\frac23\leq b \leq\frac{22}{5}\,.
\end{equation}
The bound that we get from the positivity of the coefficients $\tilde {\rm N}_{1,1}^{n,J}$ is milder, $0 \leqslant b\leqslant 8$.
This very simple result, despite being a trivial consequence of unitarity, has profound implications for the structure of the theory. 
We shall discuss this important issue in section~\ref{sec:GRLimit}.

\subsubsection{Matter-graviton scattering}
%{\bf iii) Matter-graviton}
For completeness let us treat here the mixed gravity-matter case. 
This case is different in the sense that only in gravitational Compton scattering the intermediate state can be classified as `matter' in the following sense. 
In all previous processes resonances had helicity charge $\mathfrak{h} = 0$. Furthermore, as already discussed,
 channels with charge $\mathfrak{h} = \pm 2$  are closed since no resonance has such charge. The case that is left is the one in which the exchanged resonances have charge $\mathfrak{h} = \pm 1$.
 This situation is realized in gravitational Compton scattering, where resonances have the charge of the matter we scatter.

Let us select fermion-graviton scattering to discuss semi-integer resonances and focus on region {\color{ForestGreen}(g)} to obtain the lowest spin resonance at $J_{\rm min}=3/2$. We have
\begin{table}[htp]
	\begin{center}%\vspace{-0.25cm}
		\begin{tabular}{||c||c|c|c|c|c|c|c||c||}\hline
			\multirow{2}{*}{${\rm N}_{3/2,3/2}^{n,J}$} & \multirow{2}{*}{$J=1/2$} & \multirow{2}{*}{3/2} & \multirow{2}{*}{5/2} & \multirow{2}{*}{7/2} & \multirow{2}{*}{9/2} & \multirow{2}{*}{11/2} & \multirow{2}{*}{$\dots$}  \\
			& & & & & & & \\ \hline\hline
			\multirow{2}{*}{$n=0$} & \multirow{2}{*}{\ding{55}} & \multirow{2}{*}{$\frac{1}{8}$} &   &  & & & \\
			& & & & & & & \\ \hline
			\multirow{2}{*}{$1$} & \multirow{2}{*}{\ding{55}} & \multirow{2}{*}{$\frac{7}{60}$} & \multirow{2}{*}{$\frac{2}{35}$} & \multirow{2}{*}{$\frac{1}{42}$} & & & \\
			& & & & & & &  \\ \hline
			\multirow{2}{*}{$2$} & \multirow{2}{*}{\ding{55}} & \multirow{2}{*}{$\frac{237}{2240}$} & \multirow{2}{*}{$\frac{81}{1120}$} & \multirow{2}{*}{$\frac{99}{2240}$} & \multirow{2}{*}{$\frac{243}{12320}$} & \multirow{2}{*}{$\frac{27}{3520}$} & \\
			& & & & & & & \\ \hline
			%$3$ & \ding{55} & $\frac{8000}{63}$ & $\frac{5504}{45}$ & $\frac{33344}{385}$ & $\frac{1792}{45}$ & $\frac{4096}{495}$ &  \\ \hline
			$\dots$ & \ding{55} & & & & & & \\  \hline
		\end{tabular}
	\end{center}\vspace{-0.5cm}\caption{Coefficients of the decomposition in eq.\,(\ref{eq:FG}).}\label{tab:DecV}
\end{table}
\begin{align}
\underset{1+n}{\rm R\overline{es}}\left[{\color{ForestGreen}\mathcal A_{1/2,2}}\right]=&\frac{8\pi M^2 (1+n)}{M_{\rm Pl}^2(n!)^2}\,c_{\theta/2}^3
\prod_{k=1}^n\left[k-(1+n)s_{\theta/2}^2\right]^2\,.
\end{align}
%This decomposition has purely imaginary coefficients for semi-integer resonances 
The couplings are given by
\begin{align}\label{eq:FG}
{\rm N}&_{-3/2,-3/2}^{n,J}=\frac{(1+n)}{4(n!)^2}\times 
\\&\int_{-1}^{+1} c_{\theta/2}^3d^J_{-3/2,-3/2}(\theta) \prod_{k=1}^n\left[k-(1+n)s_{\theta/2}^2\right]^2 dc_\theta\nonumber\,,
\end{align}
and the first few of them are listed in table~\ref{tab:DecV}. We find all positive coefficients, as demanded by unitarity, and we have a tower of infinite resonances with increasing half-integer spin.

%where the factor of $i$ corresponds to the scalar spinor contractions $\scpl{1}{3}=\sqrt{s_{13}<0}$ and is absorbed when turning to the amplitude in terms of $s_{ij}$ and all incoming particles.\\
\subsection{Asymmetric Veneziano and Virasoro-Shapiro form factor}\label{sec:aVZaVS}
We now move to investigate the special cases of the scattering amplitudes that are completed by the form factors $\mathcal{A}_{\rm VZ}^{\eta,\gamma_0}$ and $\mathcal{A}_{\rm VS}^{\gamma_0}$. 
The goal is to check whether the positivity constraint coming from unitarity  provides  non-trivial input on the free parameters $\eta$ and $\gamma_0$.
To this end, let us consider first fermion-vector scattering in the elastic channel, $q^{1/2}V^{-1}\to q^{1/2}V^{-1}$ in  region {\color{blue} (b)}, that is UV-completed by the form factor (notice the exchange $s\leftrightarrow t$ w.r.t. eq.~(\ref{eq:GenVZ}))
\begin{equation}
{\color{blue}{\mathcal A_{\rm VZ}^{\eta,\gamma_0}}} =\frac{\Gamma(1-\tilde t)\Gamma(1+\eta\gamma_0-\eta \tilde s)}{\Gamma(1+\eta\gamma_0-\eta \tilde s-\tilde t)}\,.
\end{equation}
The values in $s$ for which the $\Gamma$ hits the pole $\Gamma(-n)$ read
\begin{equation}\label{eq:snAVZ}
 \tilde s_n=\eta^{-1}(n+1+\eta \gamma_0)=(n+1)\eta^{-1}+\gamma_0\,.
\end{equation}
We can  compute the residues 
\begin{align}
\underset{\tilde{s}_n}{\rm R\overline{es}}\left[{\color{blue}\mathcal A^{\eta,\gamma_0}_{1/2,1}}\right] = & \frac{8\pi M^2 \tilde{s}_n}{M_{\rm Pl}^2 n!\eta}c_{\theta/2}^3
\prod_{k=1}^n\left(
k-\tilde{s}_n s_{\theta/2}^2
\right)\,,
\end{align}
and couplings 
\begin{equation}\label{eq:AsyVZ}
\left.{\rm N}_{3/2,3/2}^{n,J}\right|_{\eta,\gamma_0}= \frac{\tilde{s}_n}{4n!\eta}\int_{-1}^{+1}c_{\theta/2}^3
\prod_{k=1}^n\left(
k-\tilde{s}_n s_{\theta/2}^2
\right)dc_{\theta}\,,
\end{equation}
with minimal spin given in this case by $J_{\rm min} = 3/2$. The first few values are listed in table~\ref{tab:DecVI}. 
As clear from eqs.~(\ref{eq:snAVZ},\,\ref{eq:AsyVZ}),  the parameters $\eta, \gamma_0$ only enter the decomposition thorough the combination $\eta^{-1}(n+1)+\gamma_0$ which we define to be $\tilde \gamma_0$.
\begin{table}[htp]
	\begin{center}%\vspace{-0.25cm}
		\begin{adjustbox}{max width=.48\textwidth}
		\begin{tabular}{||c||c|c|c|c|c||}\hline
			\multirow{2}{*}{$\left.{\rm N}_{3/2,3/2}^{n,J}\right|_{\eta,\gamma_0}$} & \multirow{2}{*}{$J=1/2$} & \multirow{2}{*}{3/2}& \multirow{2}{*}{5/2} & \multirow{2}{*}{$7/2$} & \multirow{2}{*}{$\dots$}   \\
			& & & & &   \\ \hline\hline
			\multirow{2}{*}{$n=0$} & \multirow{2}{*}{\ding{55}} & \multirow{2}{*}{$\frac{1}{8}\tilde{\gamma}_0$} & & &   \\
			& & & & & \\ \hline
			\multirow{2}{*}{$1$} & \multirow{2}{*}{\ding{55}} & \multirow{2}{*}{$\frac{1}{40}(5\tilde{\gamma}_0-\tilde{\gamma}_0^2)$} & 
			\multirow{2}{*}{$\frac{1}{60}\tilde{\gamma}_0^2$} & &  \\
			& & & & &   \\ \hline
			\multirow{2}{*}{$2$} & \multirow{2}{*}{\ding{55}} & \multirow{2}{*}{$\frac{\tilde\gamma_0}{240}(30-9\tilde\gamma_0+\tilde\gamma_0^2)$} & \multirow{2}{*}{$\frac{\tilde\gamma_0^2}{840}(21-4\tilde\gamma_0)$} &  \multirow{2}{*}{$\frac{\tilde\gamma_0^3}{672}$} &  \\ 
									& & & & &   \\  \hline
			$\dots$ & \ding{55} & & & & \\  \hline
		\end{tabular}
		\end{adjustbox}
	\end{center}\vspace{-0.5cm}\caption{Coefficients of the decomposition in eq.\,(\ref{eq:AsyVZ}). 
	We use the short-hand notation $\tilde{\gamma}_0 \equiv \gamma_0+\eta^{-1}(n+1)$.}\label{tab:DecVI}
\end{table}
%For generic $n >1$, we find that the region allowed by the positivity constraint  is given by 
%\begin{equation}\label{eq:Etan}
%\eta > \eta_n \equiv 1 - \frac{3(1+n)}{2n\gamma_0}\,,~~~n>1\,,
%\end{equation}
%which asymptotes to $\eta_{\infty} = 1-3/2\gamma_0$.
%\begin{figure}[!h!]
%\centering
%\includegraphics[width=0.45\textwidth]{UniBoundVZ}\vspace{-0.275cm}  
%\caption{\label{fig:Unitarityb2} 
%Positivity constraint on the parameters $\eta$ and $\gamma_0$ in the Veneziano form factor $\mathcal{A}_{\rm VZ}^{\eta,\gamma_0}$
%in eq.~(\ref{eq:GenVZ}). The red region ($\eta < \eta_{\infty} = 1-3/2\gamma_0$) is excluded. For illustrative purposes we also mark with  
%dashed lines different values of $\eta_{n=1,\dots,5}$ (from top to bottom) which asymptote to $\eta_{\infty}$.
%}
%\end{figure}
%The situation is illustrated in fig.~\ref{fig:Unitarityb2} where we plot different values of $\eta_{n=1,\dots,5}$ in eq.~(\ref{eq:Etan}) together with its asymptotic value $\eta_{\infty}$.
It is easy to see that the first Regge trajectory automatically satisfies positivity since it has the form $\tilde\gamma_0^n$. The sub-leading trajectory however requires for positive entries that $\eta,\gamma_0$ satisfy
\begin{align}
%&(n+1)\eta^{-1}+ \gamma_0\leq\frac{(3+2n)(n+1)}{2n} \\
&(\eta^{-1}-1)+ \frac{\gamma_0}{n+1}\leq\frac{3}{2n}\,.
\end{align}
Where one can see that whereas the bound on $\gamma_0$ converges onto $3/2$, the value of $\eta$ has to be ever closer to 1. Hence one has that positivity not only restricts deformations from Veneziano in the 
 `$\eta$ -direction', it forces $\eta$ onto the value of Veneziano amplitude, i.e.
\begin{equation}
{\rm Unitarity\,\,\, enforces}~~~ \eta=1\,, \,\,\, {\gamma_0<3/2}\,.
\end{equation}
The analysis can be now repeated for $\mathcal A_{\rm VS}^{\gamma_0}$ again in region ${\color{blue} (b)}$ for the scalar-fermion and scalar-vector scattering. We have the Veneziano-Shapiro form factor
\begin{equation}
{\color{blue} \mathcal  A_{\rm VS}^{\gamma_0}}=\frac{\Gamma(1+2\gamma_0)\Gamma(1-\tilde t)\Gamma(1+\gamma_0-\tilde s)\Gamma(1+\gamma_0-\tilde u)}{\Gamma(1+\tilde t+2\gamma_0)\Gamma(1+\gamma_0+\tilde s)\Gamma(1+\gamma_0+\tilde u)}\,.
\end{equation}
The residues at $\tilde s_n=\gamma_0+1+n$ read
\begin{align}\nonumber
\underset{\tilde{s}_n}{\rm R\overline{es}}\left[{\color{blue}\mathcal  A_{\rm VS}^{\gamma_0}}\right]&=\frac{\Gamma(1+2\gamma_0)s_{\theta/2}^2(\tilde s_nc^2_{\theta/2}+\gamma_0)}{\Gamma(2+n+2\gamma_0)n!} \times \\ \label{eq:aVSResb}
& \prod_{k=1}^n(k-\tilde s_ns^2_{\theta/2})(k+2\gamma_0-\tilde s_ns^2_{\theta/2})\,,
\end{align}
to be supplemented by the GR amplitudes for $(h,h') = (0,1/2), (0,1)$ evaluated at $\tilde{s}_n$
\begin{align} \label{eq:aVS-GR}
{\color{blue} \mathcal  A^{\rm GR}_{0,1/2}} &=\frac{8\pi M^2 \tilde s_n}{M_{\rm Pl}^2}\left(\frac{c_{\theta/2}+c^3_{\theta/2}}{2s_{\theta/2}^2}\right)\,,\\
{\color{blue} \mathcal  A^{\rm GR}_{0,1}} &=\frac{8\pi M^2 \tilde s_n c^2_{\theta/2}}{M_{\rm Pl}^2 s_{\theta/2}^2}\,.
\end{align}
The product of eq.~(\ref{eq:aVSResb}) and~(\ref{eq:aVS-GR}) is projected onto Wigner $d$-functions $d^J_{1/2,1/2}$ (with $J_{\rm min} = 1/2$) and $d_{1,1}^J$ (with $J_{\rm min} = 1$) respectively for the coefficients N$_{h',h'}^{n,J}$. One again has that the sub-leading Regge trajectory is not guaranteed positive and it sets upper limits on $\gamma_0$ the most stringent of which comes from the first term and yields, respectively
\begin{align}
&{\rm Scalar-Fermion}\,,\,{\rm N}^{1,1/2}_{1/2,1/2}&&\gamma_0<2.5215\,,\label{eq:VSbound1} \\
&{\rm Scalar-Vector}\,,\,{\rm N}^{1,1}_{1,1}&&\gamma_0<3.1169\,,\label{eq:VSbound2}
\end{align}
where the values are the positive solution to a third degree polynomial approximated here, but with exact values as given in the appendix (cf. eqs.~(\ref{eq:VSBound1},\,\ref{eq:VSBound2})) and they imply a squared mass for the first ($n=0$) state lighter than $3.5215\,M^2$ and $4.1169\,M^2$ respectively.
\subsection{Low energy limit and matching to GR}\label{sec:GRLimit}
%{\bf Low energy limit and matching to GR}
In the low energy limit, $s_{ij}\ll M^2$, the expansion in energy yields an EFT that should na\"{\i}vely reproduce the results of  GR at the leading order.
%with what one would expect to be Hilbert Einstein Gravity as the leading term. 
This limit for the amplitudes in eq.~(\ref{eq:Dressing}) is straightforward to take, and one finds that the corrections  enter at order $M^{-2}=\alpha_{\rm Pl}^{-1}M_{\rm Pl}^{-2}$
\begin{align}
&\mathcal A_{\rm VZ}^{\gamma_0}=1+\left[\frac{\Gamma^\prime(1+\gamma_0)}{\Gamma^\prime(1+\gamma_0)}-\frac{\Gamma^\prime(1)}{\Gamma(1)}\right]\frac{s_{12}}{\alpha_{\rm Pl}M_{\rm Pl}^2}\,,\\ \nonumber
&\mathcal A_{\rm VS}^{\gamma_0}=1+\left[\frac{2\Gamma^\prime(1+\gamma_0)}{\Gamma(1+\gamma_0)}-\frac{\Gamma^\prime(1)}{\Gamma(1)}-\frac{\Gamma^\prime(1+2\gamma_0)}{\Gamma(1+2\gamma_0)}\right]\frac{s_{12}}{\alpha_{\rm Pl}M_{\rm Pl}^2}\,.
%&\frac{\Gamma(1-\tilde{s}_{12})\Gamma(1-\tilde{s}_{13})}{\Gamma(1+\tilde{s}_{14})}= 1-\frac{\pi^2}{6}\frac{s_{12} s_{13}}{M^4}+ O(M^{-6})\,,\\
%&\frac{\Gamma(1-\tilde{s}_{12})\Gamma(1-\tilde{s}_{13})\Gamma(1-\tilde{s}_{14})}{\Gamma(1+\tilde{s}_{12})\Gamma(1+\tilde{s}_{13})\Gamma(1+\tilde{s}_{14})}= 1+ O(M^{-6})\,.
\end{align}
This in particular means that the entries in table~\ref{tab:1} 
should match the tree-level computation of Einstein-Hilbert gravity coupled to matter to order $s/M^2$. This is the case for most of the amplitudes but crucially not all. 
All the amplitudes do reproduce the on-shell contribution of the graviton pole (which is after all the only way we have tested gravity) 
but we recall %the interesting part comes with 
that there are unspecified
off-shell pieces in the scalar-scalar and fermion-fermion scattering, cf. table~\ref{tab:1}. Let us compare these amplitudes  with the result in GR minimally coupled to matter obtained by means of Feynman rules. For the scattering of distinguishable scalars and identical fermions we find
\begin{align}
&\frac{8\pi}{M_{\rm Pl}^2}\left(\frac{s_{13}s_{14}}{s_{12}} \right)\,,\\
 &\frac{8\pi\scpl{2}{3}\scpr{1}{4}}{M_{\rm Pl}^2}\left(\frac{s_{13}}{s_{12}} +\frac{s_{12}}{s_{13}} + \frac12\right)\,,\label{eq:EHFermion}
\end{align}
and so $a=0,$ $b=1/2$. The rather unexpected finding is that whereas $a$ is compatible with the positivity bounds, (\ref{eq:abound},\ref{eq:Boundb}), $b$ in the fermion case is not. This means that the current UV completion differs from Einstein-Hilbert GR at the tree level in the infrared $s/M^2\to0$. 

The difference is however in $b$, a contact term, which does not modify gravity at long distance; it does nevertheless require modifying Hilbert-Einstein gravity as our IR theory. Let us put forward a possible modification.

%It is legitimate to ask what deformation of Einstein-Hilbert gravity could generate  a correction to eq.~(\ref{eq:EHFermion}) of the same order  $O(1/M_{\rm Pl}^2)$ such that the bound on $b$ in eq.~(\ref{eq:Boundb}) is satisfied. In the following, we shall put forward and discuss one interesting possibility.
The value $b=1/2$ in eq.~(\ref{eq:EHFermion}) arises from the Lagrangian density  of a right-handed fermion minimally coupled to gravity of the form 
\begin{equation}\label{eq:MinimalFermion}
\mathcal{L} = \frac{i\sqrt{-g}}{2}
\left[
\bar{\psi}\sigma^{a}e^{\mu}_a(\nabla_{\mu}\psi) - 
(\overline{\nabla_{\mu}\psi})\sigma^a e^{\mu}_a\psi
\right]\,,
\end{equation}
where %$\gamma^a$ are the Dirac matrices and 
the vierbein $e^{\mu}_a$ links global coordinates with those in a locally flat space, 
$g^{\mu\nu} = e^{\mu}_ae^{\nu}_b\eta^{ab}$. 
The torsion-free  spin connection $\omega_{\mu}^{ab}$ enters via the covariant derivative 
$\nabla_{\mu}\psi = \partial_{\mu}\psi + \omega_{\mu}^{ab}\sigma_{ab}\psi/4$, with $\sigma_{ab} \equiv (\bar\sigma_a\sigma_b - \bar\sigma_b\sigma_a)/2$, and it can be derived in terms of vierbein $e^{\mu}_a$ and the Christoffel symbols as 
$\omega_{\mu}^{ab} = e_{\nu}^a\Gamma^{\nu}_{\sigma\mu}g^{\sigma\rho}e_{\rho}^b - g^{\nu\rho}e_{\rho}^b(\partial_{\mu}e_{\nu}^a)$. 

Similarly to the scalar case (in which a minimally coupled scalar field has $a=0$ but a non-zero value can be obtained by adding a non-minimal coupling), modifying the value $b=1/2$ requires going beyond this minimal picture. 
A non-minimal coupling can be obtained by generalizing eq.~(\ref{eq:MinimalFermion}) to 
\begin{equation}\label{eq:NonMinimalFermion}
\mathcal{L} = \frac{i\sqrt{-g}}{2}
\left[(1-i\alpha)
\bar{\psi}\sigma^{a}e^{\mu}_a(\tilde{\nabla}_{\mu}\psi) -{\rm h.c.}
% (1+i\alpha)(\overline{\tilde{\nabla}_{\mu}\psi})\gamma^a e^{\mu}_a\psi
\right]\,,
\end{equation}
where $\alpha$ is a real parameter (note that choosing a left-handed fermion makes $\alpha\to-\alpha$) and the covariant derivative 
$\tilde{\nabla}_{\mu}\psi = \partial_{\mu}\psi + \tilde{\omega}_{\mu}^{ab}\sigma_{ab}\psi/4$ is now written in terms of a 
modified spin connection  that takes into account the presence of a non-zero torsion 
by means of the definition $\tilde{\omega}_{\mu}^{ab} \equiv \omega_{\mu}^{ab} + \mathcal{K}_{\mu}^{ab}$  
where $\omega_{\mu}^{ab}$ is the torsion-free spin connection defined above and $\mathcal{K}_{\mu}^{ab}$ is the so called contorsion tensor~\cite{Hehl:1976kj}.
The effect of the real constant $\alpha$ introduced in eq.~(\ref{eq:NonMinimalFermion}) reduces to a total derivative if the theory is torsion-free but it becomes 
 non-trivial in the presence of non-zero torsion. Torsion is encoded in the term
 \begin{equation}\label{eq:Immz}
S_{\gamma} = \frac{M_{\rm Pl}^2}{16\pi\gamma}\int ({\rm d}\tilde\omega ^{ab}  +\tilde \omega^{ac}\wedge \tilde \omega_{c}^{\,\,\,b})\wedge e_a \wedge e_b\,,
 \end{equation}
where d and $\wedge$ are, respectively,  the external derivative (d$x^\mu\partial_\mu\wedge$) and product, $\gamma$ is the Immirzi parameter (that is, in full generality, a complex number) and $\tilde \omega^{ab}=\tilde \omega^{ab}_\mu dx^\mu$, $e^a=e^a_\mu dx^\mu$.
%\footnote{The presence of the Immirzi parameter is related to the fact that the Holst term in the gravitational action~\cite{Holst:1995pc} -- proportional to the inverse of the Immirzi parameter -- does %contribute to the equation of motion of the connection in the presence of fermions (or, more in general, matter fields that couple to the connection)~\cite{Jacobson:1988qta}.
%}. 
The relevant property about eq.~(\ref{eq:Immz}) is that in the limit $\gamma\to 0$ the torsion vanishes. 
To obtain the contribution of torsion to fermion scattering we
integrate out the contorsion\footnote{It is indeed only in the case of fermions coupled to gravity that the effects of torsions become manifest.
	As all other scattering processes with fermions (fermion-scalar, fermion-vector and gravitational Compton scattering with fermions) do not leave any room for contact terms, as explained in section~\ref{sec:Elem}.} tensor (by means of the equation of motion of the connection $\tilde{\omega}_{\mu}^{ab}$).
%\footnote{Notice that the gravitational action now includes 
%not only the familiar Einstein-Hilbert term  but also the so-called Holst term~\cite{Holst:1995pc}. In tetrad formalism, we have 
%\begin{equation}
%\mathcal{S}= -\frac{1}{16\pi G_N}\int d^4 x \sqrt{-g}e^{\mu}_a e^{\nu}_b
%\left(
%R_{\mu\nu}^{ab} - \frac{1}{\gamma}\tilde{R}_{\mu\nu}^{ab}
%\right)\,,
%\end{equation}
%where $R_{\mu\nu}^{ab}$ is the curvature of the spin-connection, $\tilde{R}_{\mu\nu}^{ab}$ its dual,  and $\gamma$ is the so-called Immirzi parameter.
%}
This procedure generates the four-fermion effective operators~\cite{Hehl:1976kj,Perez:2005pm,Freidel:2005sn,Magueijo:2012ug}
\begin{equation}\label{eq:ImmOp}
\mathcal{L}= 
\frac{3\pi G_N}{8}\left(
\frac{\gamma^2}{\gamma^2 + 1}
\right)\left(1-\alpha^2+2\frac{\alpha}{\gamma}
\right)(\psi^\dagger \sigma_\mu\psi)^2\,,
\end{equation}
%\begin{equation}
%\mathcal{L}= 
%\frac{3\pi G_N}{2}\left(
%\frac{\gamma^2}{\gamma^2 + 1}
%\right)\left(
%A_{\mu}A^{\mu} + \frac{2\alpha}{\gamma}A_{\mu}V^{\mu} - \alpha^2V_{\mu}V^{\mu}
%\right)\,,
%\end{equation}
%where 
%$A^{\mu} \equiv \psi\gamma^5 \gamma^{\mu} \psi$, 
%$V^{\mu} \equiv \psi \gamma^{\mu} \psi$
and the parameter $b$  becomes
\begin{equation}\label{eq:Immb}
b =  \frac12
+\frac{3\gamma}{8(1+\gamma^2)}\left[
 \gamma - 
\alpha(2+\alpha\gamma)
\right]\,,
\end{equation}
and the bound in eq.~(\ref{eq:Boundb}) can be now satisfied. For instance, if one takes $\alpha = 0$ and restricts the analysis to real and positive values of the Immirzi parameter, the bound on $b$ is satisfied if $\gamma \geqslant 2/\sqrt{5}$. 
%In conclusion, the incompatibility of the minimal coupling in eq.~(\ref{eq:MinimalFermion}) with the unitarity bound in eq.~(\ref{eq:Boundb}) 
%could indicate that space-time presents both curvature and torsion. 
%or, in equivalent words, that space-time in the presence of gravity is a Riemann-Cartan geometry  (as opposed to the Riemannian geometry of GR in which torsion, by construction, always vanishes). 

An equivalent way to obtain a modification of $b$ slightly more familiar to the working class particle theorist is through the introduction of a Kalb-Ramond field, a three form $H\equiv 3 $d$ B$ with $B=B_{\mu\nu}$d$x^\mu\wedge $d$x^\nu$. This field occurs in string theory and its presence is associated with gravitational torsion~\cite{Candelas:1985en,Shapiro:2001rz,Ortin:2004ms}. Let it couple to a fermion as
\begin{equation}\label{eq:KR}
S_{\rm int}=-\frac{g}{  M}\int H \wedge (\psi^\dagger \sigma_\mu\psi dx^\mu)\,,%=-\frac{3gi}{M} \int d B\wedge (\psi^\dagger \sigma_\mu\psi dx^\mu)
\end{equation}
%In the low-energy limit of string theory, it is conceivable to obtain a gravitational theory that differs from GR since it is expected to contain, in addition to the graviton, 
%also a scalar field, known as the dilaton, and an antisymmetric tensor field strength $H_{\mu\nu\rho}$, which is a three-form, known as the Kalb-Ramond field.
%Gravitational torsion is associated to the presence of the latter~\cite{Candelas:1985en,Shapiro:2001rz}. 
%Let us discuss a very simple example.
%Suppose to have some low-energy realization of string theory that generates the coupling 
%\begin{equation}\label{eq:KR}
%\mathcal{L}_{\psi H} = -\frac{g}{M}\bar{\psi}\sigma_{\mu\nu\lambda}H^{\mu\nu\lambda}\psi\,,~~{\rm with}~\sigma_{\mu\nu\lambda} \equiv i\epsilon_{\mu\nu\lambda\alpha}\gamma^5\gamma^{\alpha}\,,
%\end{equation}
%between the Kalb-Ramond three-form field and a Dirac fermion $\psi$.
%This coupling plays a role analogue to that of the contorsion tensor in eq.~(\ref{eq:NonMinimalFermion})\cite{Pilling:2002ij,Pilling:2002dz}.
 %In eq.~(\ref{eq:KR})
 where we can identify $M$ as the mass scale in our UV completion and the coupling $g$ is real and dimensionless.
 % and if both are assumed real the interaction is anti-hermitian. 
 %Alternatively the particle phenomenologist could use instead an axion field $a$ with $\partial_\mu a =\epsilon_{\mu\nu\rho\kappa} H^{\nu\rho\kappa}/6$, 
 %and purely imaginary decay constant $f_a=iM/6g$.
 One expects in string theory $g\sim M/M_{\rm Pl}$ but the precise value of $g$ depends on the expectation values of the moduli fields and, therefore, on the details of the compactification procedure.
Consequently, it is  important to remark that the coupling $g$ in the low-energy limit of string theory is in principle computable but, on balance, a free parameter.

%In the spirit of this section, we can now compute the  amplitude for the tree level exchange of an antisymmetric Kalb-Ramond tensor field.
%As customary, the latter  can  be  related to a potential $B_{\mu\nu}$ that encodes the propagating degree of freedom as
% $H_{\mu\nu\rho} = \partial_{\mu}B_{\nu\rho} +  \partial_{\nu}B_{\rho\mu} + \partial_{\rho}B_{\mu\nu}$.\footnote{The propagator can be obtained from the kinetic term of the free Kalb-Ramond field $\mathcal{L}_{\rm KR} = -H_{\mu\nu\rho}H^{\mu\nu\rho}/12$ after adding an appropriate gauge-fixing term.}
% The tree-level exchange of the Kalb-Ramond field only generates a contact interaction because the pole of the propagator cancels against the combination of momenta in the numerator. 
Integrating out the two-form $B$  one obtains a contact interaction (the equation of motion for $H$ is algebraic or, alternatively,  the propagator pole of $B$ cancels against the derivative coupling) which reads
 \begin{equation}
 b_{\rm KR} = \frac{18 g^2 M_{\rm Pl}^2}{\pi M^2}\,,%=\frac{M_{\rm Pl}^2}{2\pi f_a^2}\,,
 \end{equation}
and translates through eq.~(\ref{eq:Immb}) the Immirzi parameter into a ratio of scales. In order to satisfy the unitarity bound in eq.~(\ref{eq:Boundb}) we obtain an upper and lower bound on the 
 mass ratio $gM_ {\rm Pl}/M$
\begin{equation}
\frac{1}{108}\leq  \frac{g^2 M_{\rm Pl}^2}{\pi M^2} \leq \frac{13}{60}\,.
\end{equation}
% that would correct the GR value $b=1/2$ in the direction predicted by our unitarity bound. 
%Hence translating the effect of torsion into interactions with an auxiliary field which nonetheless has imaginary coupling.
 
 This simple discussion teaches us an important prerogative of the ``bottom-up'' approach to the UV completion of gravity pursued in this paper. 
 From a typical ``top-down'' perspective like that of string theory, 
 it is difficult to make firm statements about what we expect in our phenomenological four-dimensional world. 
 In this sense, the possible presence of space-time torsion is a prototypical example, with deviations from GR well motivated theoretically but difficult to compute without introducing free parameters 
 encompassing  the complicated step of compactification.
 Our  ``bottom-up'' perspective, on the contrary, {\it predicts} a deviation from GR, and gives a precise indication about what we should expect for it.
 The above analysis exposes in plain sight the true complementarity of the two approaches.

%%%%%%%%%%%%%%%%%%%%%%%%%%%%%%%%%%%%%%%%%%%%%%%%%%
%%%%%%%%%%%%%%%%%%%%%%%%%%%%%%%%%%%%%%%%%%%%%%%%%%

\section{Grand Finale}
The on-shell amplitude program has paved the road for the formulation and
 analysis of amplitudes featuring  resonances of arbitrary spin and mass,  clearing the computational obstacles of the conventional Lagrangian approach.
 %The on-shell spinor-helicity formalism is tailor-made for the analysis of scattering amplitude in the presence of massive higher-spin force carriers since the conventional  approach with Feynman diagrams becomes utterly  complicated and eventually unmanageable for increasing spin. 
 
 In this letter, this formalism was applied to study the UV completion of gravity following a ``bottom-up'' approach. 
The derivation of a viable UV completion started from the need to tame the growth with energy of the scattering amplitudes mediated by gravity, which at energies comparable to the Planck scale  grow above the unitarity bound. 
With typical particle physicist demeanor, we approached the problem by introducing massive resonances and the solution obtained was validated against unitarity, locality and causality. 
In this regard, we do not put forth in this work a full quantum theory of gravity 
but rather a UV-complete formulation of tree-level amplitudes. These amplitudes nonetheless yield a great deal of information on the full theory, 
and have non-trivial implications in the infrared.

 In more depth, the main results of our analysis are:
 
{\it (i)} After enumerating our working assumptions -- locality, unitarity and causality, weak coupling limit -- in section~\ref{sec:Intro}, in section~\ref{sec:OrdinaryGravity} we constructed the most general tree-level
 scattering amplitude mediated by gravitational interactions  in the context of GR with  SM particles (including gravitons) on the external states, eq.~(\ref{MstGR}).
 We considered not only the contributions to the scattering amplitudes coming from the poles but also, in full generality, the possible presence of contact (polynomial) terms compatible with our working assumptions.

{\it (ii)} In section~\ref{sec:SpinJExchange} we computed the contribution due to the exchange of a massive resonance with arbitrary spin $J$, eq.~(\ref{eq:MstSpnJ}). 
The result of this computation was already quoted in~\cite{Arkani-Hamed:2017jhn} in terms of `spinning Gegenbauer polynomials'.
Here we derived an equivalent expression in terms of Jacobi polynomials that makes more transparent the helicity structure of the amplitude.
Furthermore, by combining the Jacobi polynomials with the helicity factor, we related the scattering amplitudes to Wigner $d$-functions. 
%in a manner that can be related to the Jacob-Wick partial-wave expansion.
Finally, the study puts in  the foreground the role of unitarity that, in the case of elastic processes, relates the residue of the amplitude to the decay width of resonances.  

{\it (iii)} In section~\ref{sec:UVcompl} we combined the results of {\it (i)} and {\it (ii)} to obtain a valid UV completion of GR amplitudes. 
To this end, we only used as guidelines unitarity, locality and causality.
A number of properties -- i.e. quantization of the mass spectrum and  duality relations among resonances exchanged in channels related by crossing transformations -- elegantly arises  from the mathematical consistency dictated by these three fundamental principles. The obtained solutions follow from three ansatz, clearly displayed in section~\ref{sec:UVconstruction}, put forward to solve 
the mathematical consistency conditions that follow from locality and unitarity.
The UV completion dresses  the GR amplitudes with form factors that are product of Euler $\Gamma$ functions, thus closely resembling the Veneziano and Virasoro-Shapiro amplitudes as summarized in section~\ref{sec:UVSummary}. 
%It is important to remark that we do not have a uniqueness theorem about our results. 
%On the contrary, they were obtained as a consequence of three specific ansatz, clearly enumerated in section~\ref{sec:UVconstruction}, put forward to solve 
%the mathematical consistency conditions that follow from locality and unitarity. 
%Obtaining alternative solutions and testing their validity is an interesting direction of research that is worth exploring 

%In particular, we find that the mass of the resonances is quantized.
% and that the resonances are exchanged  in different channels related by crossing transformations. 
 
{\it (iv)} Section~\ref{sec:Analysis} is devoted to the analysis of the physical properties of the resonances. By performing an angular decomposition in terms of Wigner $d$-functions, 
we extracted couplings and spin of the resonances. Unitarity enforces positivity constraints on the  couplings of resonances kinematically accessible in elastic scattering processes.
The  ``bottom-up'' approach showcases the power of this basic principle. 
As a consequence of unitarity, for instance, we find that the proposed UV completion {\it predicts} leading order deviations from GR minimally coupled to fermions. 
These deviations needed to restore unitarity are present if space-time has torsion in addition to curvature. 
 Unitarity also imposes non-trivial constraints on the contact terms introduced in point {\it (ii)} 
 as well as on possible deformations of the Veneziano and Virasoro-Shapiro amplitudes found at point {\it (iii)}.

The ambitious and overarching question lurking in the background is uniqueness: {\it are fundamental principles like unitarity and locality so demanding as to select one theory of gravity only?}
%There is potential for an answer in the ``bottom-up'' approach and the exploration of possible solutions, see~\cite{HuangTalk} for a related discussion. 
We believe that the ``bottom-up'' approach may help in shedding light on this question,  see~\cite{HuangTalk} for a related discussion. 
In this regard, our findings have glaring similarities with string theory,
%to the  UV completion of gravity proposed in this paper 
% led us to scattering amplitudes that
% share many similarities  with what one expects, following a reversed  ``top-down'' logic, in string theory.
tantalizingly yet inconclusively in line with the belief that string theory is the only consistent quantum theory of gravity.
A more detailed  exploration of this connection is left for future work since it would have diluted the main aim of this letter.
The main aim being the model-independent ``bottom-up'' and unbiased approach to the UV completion of gravity with fundamental principles like unitarity and locality as sole reference.
%In this respect, what we think is important to stress as a final remark 
%is that our construction does not rely on any string theory notion, and provides complementary physical insights. 
This road has been shown here to lead to relevant results in complementarity to the ``top-down'' prevailing trend, an outstanding example
are the constraints obtained from unitarity in  {\it (iv)} 
% In this optic, the analysis carried out at point {\it (iv)}, and in particular the constraints obtained from unitarity, 
given that studying the same subject in the  ``top-down'' perspective of string theory requires overcoming the obstacle of compactification to make contact with our observable
 four-dimensional Universe, losing predictivity  along the way.\\

%%%%%%%%%%%%%%%%%%%%%%%%%%%%%%%%%%%%%%%%%%%%%%%%%%
%%%%%%%%%%%%%%%%%%%%%%%%%%%%%%%%%%%%%%%%%%%%%%%%%%
\paragraph*{Acknowledgments:} 
We thank Marco Fabbrichesi and Simeon Hellerman for discussions and Yu-tin Huang for discussions and for bringing to our attention the work in~\cite{HuangTalk}.
A.U. thanks the Kavli IPMU, 
where  this  project  was  completed, and the city of Tokyo for  the  warm  and kind hospitality. R.A. acknowledges Kendrick Lamar for reminding him with his lyrics of Compton scattering. A.U. acknowledges financial support from the H2020-MSCA-RISE project ``InvisiblesPlus'' and from the INFN grant SESAMO. 
This work was supported by World Premier International Research Center Initiative (WPI Initiative), MEXT, Japan.

%%%%%%%%%%%%%%%%%%%%%%%%%%%%%%%%%%%%%%%%%%%%%%%%%%
%%%%%%%%%%%%%%%%%%%%%%%%%%%%%%%%%%%%%%%%%%%%%%%%%%

\bibliography{AmpResUVGr-al}

\begin{thebibliography}{31}%
\makeatletter
\providecommand \@ifxundefined [1]{%
 \@ifx{#1\undefined}
}%
\providecommand \@ifnum [1]{%
 \ifnum #1\expandafter \@firstoftwo
 \else \expandafter \@secondoftwo
 \fi
}%
\providecommand \@ifx [1]{%
 \ifx #1\expandafter \@firstoftwo
 \else \expandafter \@secondoftwo
 \fi
}%
\providecommand \natexlab [1]{#1}%
\providecommand \enquote  [1]{``#1''}%
\providecommand \bibnamefont  [1]{#1}%
\providecommand \bibfnamefont [1]{#1}%
\providecommand \citenamefont [1]{#1}%
\providecommand \href@noop [0]{\@secondoftwo}%
\providecommand \href [0]{\begingroup \@sanitize@url \@href}%
\providecommand \@href[1]{\@@startlink{#1}\@@href}%
\providecommand \@@href[1]{\endgroup#1\@@endlink}%
\providecommand \@sanitize@url [0]{\catcode `\\12\catcode `\$12\catcode
  `\&12\catcode `\#12\catcode `\^12\catcode `\_12\catcode `\%12\relax}%
\providecommand \@@startlink[1]{}%
\providecommand \@@endlink[0]{}%
\providecommand \url  [0]{\begingroup\@sanitize@url \@url }%
\providecommand \@url [1]{\endgroup\@href {#1}{\urlprefix }}%
\providecommand \urlprefix  [0]{URL }%
\providecommand \Eprint [0]{\href }%
\providecommand \doibase [0]{http://dx.doi.org/}%
\providecommand \selectlanguage [0]{\@gobble}%
\providecommand \bibinfo  [0]{\@secondoftwo}%
\providecommand \bibfield  [0]{\@secondoftwo}%
\providecommand \translation [1]{[#1]}%
\providecommand \BibitemOpen [0]{}%
\providecommand \bibitemStop [0]{}%
\providecommand \bibitemNoStop [0]{.\EOS\space}%
\providecommand \EOS [0]{\spacefactor3000\relax}%
\providecommand \BibitemShut  [1]{\csname bibitem#1\endcsname}%
\let\auto@bib@innerbib\@empty
%</preamble>
\bibitem [{\citenamefont {Woodard}(2009)}]{Woodard:2009ns}%
  \BibitemOpen
  \bibfield  {author} {\bibinfo {author} {\bibfnamefont {R.~P.}\ \bibnamefont
  {Woodard}},\ }\href {\doibase 10.1088/0034-4885/72/12/126002} {\bibfield
  {journal} {\bibinfo  {journal} {Rept. Prog. Phys.}\ }\textbf {\bibinfo
  {volume} {72}},\ \bibinfo {pages} {126002} (\bibinfo {year} {2009})},\
  \Eprint {http://arxiv.org/abs/0907.4238} {arXiv:0907.4238 [gr-qc]}
  \BibitemShut {NoStop}%
%%CITATION = ARXIV:0907.4238;%%
\bibitem [{\citenamefont {DeWitt}(1967)}]{DeWitt:1967uc}%
  \BibitemOpen
  \bibfield  {author} {\bibinfo {author} {\bibfnamefont {B.~S.}\ \bibnamefont
  {DeWitt}},\ }\href {\doibase 10.1103/PhysRev.162.1239} {\bibfield  {journal}
  {\bibinfo  {journal} {Phys. Rev.}\ }\textbf {\bibinfo {volume} {162}},\
  \bibinfo {pages} {1239} (\bibinfo {year} {1967})}\BibitemShut {NoStop}%
%%CITATION = PHRVA,162,1239;%%
\bibitem [{\citenamefont {Berends}\ and\ \citenamefont
  {Gastmans}(1975)}]{Berends:1974gk}%
  \BibitemOpen
  \bibfield  {author} {\bibinfo {author} {\bibfnamefont {F.~A.}\ \bibnamefont
  {Berends}}\ and\ \bibinfo {author} {\bibfnamefont {R.}~\bibnamefont
  {Gastmans}},\ }\href {\doibase 10.1016/0550-3213(75)90528-3} {\bibfield
  {journal} {\bibinfo  {journal} {Nucl. Phys.}\ }\textbf {\bibinfo {volume}
  {B88}},\ \bibinfo {pages} {99} (\bibinfo {year} {1975})}\BibitemShut
  {NoStop}%
%%CITATION = NUPHA,B88,99;%%
\bibitem [{\citenamefont {Grisaru}\ \emph {et~al.}(1975)\citenamefont
  {Grisaru}, \citenamefont {van Nieuwenhuizen},\ and\ \citenamefont
  {Wu}}]{Grisaru:1975bx}%
  \BibitemOpen
  \bibfield  {author} {\bibinfo {author} {\bibfnamefont {M.~T.}\ \bibnamefont
  {Grisaru}}, \bibinfo {author} {\bibfnamefont {P.}~\bibnamefont {van
  Nieuwenhuizen}}, \ and\ \bibinfo {author} {\bibfnamefont {C.~C.}\
  \bibnamefont {Wu}},\ }\href {\doibase 10.1103/PhysRevD.12.397} {\bibfield
  {journal} {\bibinfo  {journal} {Phys. Rev.}\ }\textbf {\bibinfo {volume}
  {D12}},\ \bibinfo {pages} {397} (\bibinfo {year} {1975})}\BibitemShut
  {NoStop}%
%%CITATION = PHRVA,D12,397;%%
\bibitem [{\citenamefont {Sannan}(1986)}]{Sannan:1986tz}%
  \BibitemOpen
  \bibfield  {author} {\bibinfo {author} {\bibfnamefont {S.}~\bibnamefont
  {Sannan}},\ }\href {\doibase 10.1103/PhysRevD.34.1749} {\bibfield  {journal}
  {\bibinfo  {journal} {Phys. Rev.}\ }\textbf {\bibinfo {volume} {D34}},\
  \bibinfo {pages} {1749} (\bibinfo {year} {1986})}\BibitemShut {NoStop}%
%%CITATION = PHRVA,D34,1749;%%
\bibitem [{\citenamefont {Dixon}(1996)}]{Dixon:1996wi}%
  \BibitemOpen
  \bibfield  {author} {\bibinfo {author} {\bibfnamefont {L.~J.}\ \bibnamefont
  {Dixon}},\ }in\ \href
  {http://www-public.slac.stanford.edu/sciDoc/docMeta.aspx?slacPubNumber=SLAC-PUB-7106}
  {\emph {\bibinfo {booktitle} {{QCD and beyond. Proceedings, Theoretical
  Advanced Study Institute in Elementary Particle Physics, TASI-95, Boulder,
  USA, June 4-30, 1995}}}}\ (\bibinfo {year} {1996})\ pp.\ \bibinfo {pages}
  {539--584},\ \Eprint {http://arxiv.org/abs/hep-ph/9601359}
  {arXiv:hep-ph/9601359 [hep-ph]} \BibitemShut {NoStop}%
%%CITATION = HEP-PH/9601359;%%
\bibitem [{\citenamefont {Elvang}\ and\ \citenamefont
  {Huang}(2013)}]{Elvang:2013cua}%
  \BibitemOpen
  \bibfield  {author} {\bibinfo {author} {\bibfnamefont {H.}~\bibnamefont
  {Elvang}}\ and\ \bibinfo {author} {\bibfnamefont {Y.-t.}\ \bibnamefont
  {Huang}},\ }\href@noop {} {\  (\bibinfo {year} {2013})},\ \Eprint
  {http://arxiv.org/abs/1308.1697} {arXiv:1308.1697 [hep-th]} \BibitemShut
  {NoStop}%
%%CITATION = ARXIV:1308.1697;%%
\bibitem [{\citenamefont {Arkani-Hamed}\ \emph {et~al.}(2017)\citenamefont
  {Arkani-Hamed}, \citenamefont {Huang},\ and\ \citenamefont
  {Huang}}]{Arkani-Hamed:2017jhn}%
  \BibitemOpen
  \bibfield  {author} {\bibinfo {author} {\bibfnamefont {N.}~\bibnamefont
  {Arkani-Hamed}}, \bibinfo {author} {\bibfnamefont {T.-C.}\ \bibnamefont
  {Huang}}, \ and\ \bibinfo {author} {\bibfnamefont {Y.-t.}\ \bibnamefont
  {Huang}},\ }\href@noop {} {\  (\bibinfo {year} {2017})},\ \Eprint
  {http://arxiv.org/abs/1709.04891} {arXiv:1709.04891 [hep-th]} \BibitemShut
  {NoStop}%
%%CITATION = ARXIV:1709.04891;%%
\bibitem [{\citenamefont {Cachazo}\ \emph {et~al.}(2004)\citenamefont
  {Cachazo}, \citenamefont {Svrcek},\ and\ \citenamefont
  {Witten}}]{Cachazo:2004kj}%
  \BibitemOpen
  \bibfield  {author} {\bibinfo {author} {\bibfnamefont {F.}~\bibnamefont
  {Cachazo}}, \bibinfo {author} {\bibfnamefont {P.}~\bibnamefont {Svrcek}}, \
  and\ \bibinfo {author} {\bibfnamefont {E.}~\bibnamefont {Witten}},\ }\href
  {\doibase 10.1088/1126-6708/2004/09/006} {\bibfield  {journal} {\bibinfo
  {journal} {JHEP}\ }\textbf {\bibinfo {volume} {09}},\ \bibinfo {pages} {006}
  (\bibinfo {year} {2004})},\ \Eprint {http://arxiv.org/abs/hep-th/0403047}
  {arXiv:hep-th/0403047 [hep-th]} \BibitemShut {NoStop}%
%%CITATION = HEP-TH/0403047;%%
\bibitem [{\citenamefont {Britto}\ \emph {et~al.}(2005)\citenamefont {Britto},
  \citenamefont {Cachazo}, \citenamefont {Feng},\ and\ \citenamefont
  {Witten}}]{Britto:2005fq}%
  \BibitemOpen
  \bibfield  {author} {\bibinfo {author} {\bibfnamefont {R.}~\bibnamefont
  {Britto}}, \bibinfo {author} {\bibfnamefont {F.}~\bibnamefont {Cachazo}},
  \bibinfo {author} {\bibfnamefont {B.}~\bibnamefont {Feng}}, \ and\ \bibinfo
  {author} {\bibfnamefont {E.}~\bibnamefont {Witten}},\ }\href {\doibase
  10.1103/PhysRevLett.94.181602} {\bibfield  {journal} {\bibinfo  {journal}
  {Phys. Rev. Lett.}\ }\textbf {\bibinfo {volume} {94}},\ \bibinfo {pages}
  {181602} (\bibinfo {year} {2005})},\ \Eprint
  {http://arxiv.org/abs/hep-th/0501052} {arXiv:hep-th/0501052 [hep-th]}
  \BibitemShut {NoStop}%
%%CITATION = HEP-TH/0501052;%%
\bibitem [{\citenamefont {Gell-Mann}\ \emph {et~al.}(1954)\citenamefont
  {Gell-Mann}, \citenamefont {Goldberger},\ and\ \citenamefont
  {Thirring}}]{GellMann:1954db}%
  \BibitemOpen
  \bibfield  {author} {\bibinfo {author} {\bibfnamefont {M.}~\bibnamefont
  {Gell-Mann}}, \bibinfo {author} {\bibfnamefont {M.~L.}\ \bibnamefont
  {Goldberger}}, \ and\ \bibinfo {author} {\bibfnamefont {W.~E.}\ \bibnamefont
  {Thirring}},\ }\href {\doibase 10.1103/PhysRev.95.1612} {\bibfield  {journal}
  {\bibinfo  {journal} {Phys. Rev.}\ }\textbf {\bibinfo {volume} {95}},\
  \bibinfo {pages} {1612} (\bibinfo {year} {1954})}\BibitemShut {NoStop}%
%%CITATION = PHRVA,95,1612;%%
\bibitem [{\citenamefont {Adams}\ \emph {et~al.}(2006)\citenamefont {Adams},
  \citenamefont {Arkani-Hamed}, \citenamefont {Dubovsky}, \citenamefont
  {Nicolis},\ and\ \citenamefont {Rattazzi}}]{Adams:2006sv}%
  \BibitemOpen
  \bibfield  {author} {\bibinfo {author} {\bibfnamefont {A.}~\bibnamefont
  {Adams}}, \bibinfo {author} {\bibfnamefont {N.}~\bibnamefont {Arkani-Hamed}},
  \bibinfo {author} {\bibfnamefont {S.}~\bibnamefont {Dubovsky}}, \bibinfo
  {author} {\bibfnamefont {A.}~\bibnamefont {Nicolis}}, \ and\ \bibinfo
  {author} {\bibfnamefont {R.}~\bibnamefont {Rattazzi}},\ }\href {\doibase
  10.1088/1126-6708/2006/10/014} {\bibfield  {journal} {\bibinfo  {journal}
  {JHEP}\ }\textbf {\bibinfo {volume} {10}},\ \bibinfo {pages} {014} (\bibinfo
  {year} {2006})},\ \Eprint {http://arxiv.org/abs/hep-th/0602178}
  {arXiv:hep-th/0602178 [hep-th]} \BibitemShut {NoStop}%
%%CITATION = HEP-TH/0602178;%%
\bibitem [{\citenamefont {Froissart}(1961)}]{Froissart:1961ux}%
  \BibitemOpen
  \bibfield  {author} {\bibinfo {author} {\bibfnamefont {M.}~\bibnamefont
  {Froissart}},\ }\href {\doibase 10.1103/PhysRev.123.1053} {\bibfield
  {journal} {\bibinfo  {journal} {Phys. Rev.}\ }\textbf {\bibinfo {volume}
  {123}},\ \bibinfo {pages} {1053} (\bibinfo {year} {1961})}\BibitemShut
  {NoStop}%
%%CITATION = PHRVA,123,1053;%%
\bibitem [{\citenamefont {Camanho}\ \emph {et~al.}(2016)\citenamefont
  {Camanho}, \citenamefont {Edelstein}, \citenamefont {Maldacena},\ and\
  \citenamefont {Zhiboedov}}]{Camanho:2014apa}%
  \BibitemOpen
  \bibfield  {author} {\bibinfo {author} {\bibfnamefont {X.~O.}\ \bibnamefont
  {Camanho}}, \bibinfo {author} {\bibfnamefont {J.~D.}\ \bibnamefont
  {Edelstein}}, \bibinfo {author} {\bibfnamefont {J.}~\bibnamefont
  {Maldacena}}, \ and\ \bibinfo {author} {\bibfnamefont {A.}~\bibnamefont
  {Zhiboedov}},\ }\href {\doibase 10.1007/JHEP02(2016)020} {\bibfield
  {journal} {\bibinfo  {journal} {JHEP}\ }\textbf {\bibinfo {volume} {02}},\
  \bibinfo {pages} {020} (\bibinfo {year} {2016})},\ \Eprint
  {http://arxiv.org/abs/1407.5597} {arXiv:1407.5597 [hep-th]} \BibitemShut
  {NoStop}%
%%CITATION = ARXIV:1407.5597;%%
\bibitem [{\citenamefont {Cerulus}\ and\ \citenamefont
  {Martin}(1964)}]{Cerulus:1964cjb}%
  \BibitemOpen
  \bibfield  {author} {\bibinfo {author} {\bibfnamefont {F.~A.}\ \bibnamefont
  {Cerulus}}\ and\ \bibinfo {author} {\bibfnamefont {A.}~\bibnamefont
  {Martin}},\ }\href {\doibase 10.1016/0031-9163(64)90807-8} {\bibfield
  {journal} {\bibinfo  {journal} {Phys. Lett.}\ }\textbf {\bibinfo {volume}
  {8}},\ \bibinfo {pages} {80} (\bibinfo {year} {1964})}\BibitemShut {NoStop}%
%%CITATION = PHLTA,8,80;%%
\bibitem [{\citenamefont {Epstein}\ and\ \citenamefont
  {Martin}(2019)}]{Epstein:2019zdn}%
  \BibitemOpen
  \bibfield  {author} {\bibinfo {author} {\bibfnamefont {H.}~\bibnamefont
  {Epstein}}\ and\ \bibinfo {author} {\bibfnamefont {A.}~\bibnamefont
  {Martin}},\ }\href@noop {} {\  (\bibinfo {year} {2019})},\ \Eprint
  {http://arxiv.org/abs/1903.00953} {arXiv:1903.00953 [hep-th]} \BibitemShut
  {NoStop}%
%%CITATION = ARXIV:1903.00953;%%
\bibitem [{\citenamefont {Gribov}(2012)}]{Gribov:2009zz}%
  \BibitemOpen
  \bibfield  {author} {\bibinfo {author} {\bibfnamefont {V.~N.}\ \bibnamefont
  {Gribov}},\ }\href
  {http://cambridge.org/catalogue/catalogue.asp?isbn=9780521856096} {\emph
  {\bibinfo {title} {{Strong interactions of hadrons at high emnergies: Gribov
  lectures on Theoretical Physics}}}},\ edited by\ \bibinfo {editor}
  {\bibfnamefont {Y.~L.}\ \bibnamefont {Dokshitzer}}\ and\ \bibinfo {editor}
  {\bibfnamefont {J.}~\bibnamefont {Nyiri}},\ Vol.~\bibinfo {volume} {27}\
  (\bibinfo  {publisher} {Cambridge University Press},\ \bibinfo {year}
  {2012})\BibitemShut {NoStop}%
%%CITATION = CMPCE,27,;%%
\bibitem [{\citenamefont {Francia}\ \emph {et~al.}(2007)\citenamefont
  {Francia}, \citenamefont {Mourad},\ and\ \citenamefont
  {Sagnotti}}]{Francia:2007qt}%
  \BibitemOpen
  \bibfield  {author} {\bibinfo {author} {\bibfnamefont {D.}~\bibnamefont
  {Francia}}, \bibinfo {author} {\bibfnamefont {J.}~\bibnamefont {Mourad}}, \
  and\ \bibinfo {author} {\bibfnamefont {A.}~\bibnamefont {Sagnotti}},\ }\href
  {\doibase 10.1016/j.nuclphysb.2007.03.021} {\bibfield  {journal} {\bibinfo
  {journal} {Nucl. Phys.}\ }\textbf {\bibinfo {volume} {B773}},\ \bibinfo
  {pages} {203} (\bibinfo {year} {2007})},\ \Eprint
  {http://arxiv.org/abs/hep-th/0701163} {arXiv:hep-th/0701163 [hep-th]}
  \BibitemShut {NoStop}%
%%CITATION = HEP-TH/0701163;%%
\bibitem [{\citenamefont {Caron-Huot}\ \emph {et~al.}(2017)\citenamefont
  {Caron-Huot}, \citenamefont {Komargodski}, \citenamefont {Sever},\ and\
  \citenamefont {Zhiboedov}}]{Caron-Huot:2016icg}%
  \BibitemOpen
  \bibfield  {author} {\bibinfo {author} {\bibfnamefont {S.}~\bibnamefont
  {Caron-Huot}}, \bibinfo {author} {\bibfnamefont {Z.}~\bibnamefont
  {Komargodski}}, \bibinfo {author} {\bibfnamefont {A.}~\bibnamefont {Sever}},
  \ and\ \bibinfo {author} {\bibfnamefont {A.}~\bibnamefont {Zhiboedov}},\
  }\href {\doibase 10.1007/JHEP10(2017)026} {\bibfield  {journal} {\bibinfo
  {journal} {JHEP}\ }\textbf {\bibinfo {volume} {10}},\ \bibinfo {pages} {026}
  (\bibinfo {year} {2017})},\ \Eprint {http://arxiv.org/abs/1607.04253}
  {arXiv:1607.04253 [hep-th]} \BibitemShut {NoStop}%
%%CITATION = ARXIV:1607.04253;%%
\bibitem [{\citenamefont {Wigner}(1959)}]{Wigner:102713}%
  \BibitemOpen
  \bibfield  {author} {\bibinfo {author} {\bibfnamefont {E.~P.}\ \bibnamefont
  {Wigner}},\ }\href {http://cds.cern.ch/record/102713} {\emph {\bibinfo
  {title} {{Group theory and its application to the quantum mechanics of atomic
  spectra}}}},\ Pure Appl. Phys.\ (\bibinfo  {publisher} {Academic Press},\
  \bibinfo {address} {New York, NY},\ \bibinfo {year} {1959})\ \bibinfo {note}
  {trans. from the German}\BibitemShut {NoStop}%
\bibitem [{\citenamefont {Jacob}\ and\ \citenamefont
  {Wick}(1959)}]{Jacob:1959at}%
  \BibitemOpen
  \bibfield  {author} {\bibinfo {author} {\bibfnamefont {M.}~\bibnamefont
  {Jacob}}\ and\ \bibinfo {author} {\bibfnamefont {G.~C.}\ \bibnamefont
  {Wick}},\ }\href {\doibase 10.1016/0003-4916(59)90051-X} {\bibfield
  {journal} {\bibinfo  {journal} {Annals Phys.}\ }\textbf {\bibinfo {volume}
  {7}},\ \bibinfo {pages} {404} (\bibinfo {year} {1959})}\BibitemShut {NoStop}%
%%CITATION = APNYA,7,404;%%
\bibitem [{\citenamefont {tin Huang}()}]{HuangTalk}%
  \BibitemOpen
  \bibfield  {author} {\bibinfo {author} {\bibfnamefont {Y.}~\bibnamefont {tin
  Huang}},\ }\href {http://www2.yukawa.kyoto-u.ac.jp/~qft.web/2016/index.html}
  {\enquote {\bibinfo {title} {Lessons from perturbative unitarity in graviton
  scattering amplitudes},}\ }\bibinfo {note} {YITP Workshop Strings and Fields
  2016}\BibitemShut {NoStop}%
\bibitem [{\citenamefont {Schwarz}(1982)}]{Schwarz:1982jn}%
  \BibitemOpen
  \bibfield  {author} {\bibinfo {author} {\bibfnamefont {J.~H.}\ \bibnamefont
  {Schwarz}},\ }\href {\doibase 10.1016/0370-1573(82)90087-4} {\bibfield
  {journal} {\bibinfo  {journal} {Phys. Rept.}\ }\textbf {\bibinfo {volume}
  {89}},\ \bibinfo {pages} {223} (\bibinfo {year} {1982})}\BibitemShut
  {NoStop}%
%%CITATION = PRPLC,89,223;%%
\bibitem [{\citenamefont {Polchinski}(2007)}]{Polchinski:1998rq}%
  \BibitemOpen
  \bibfield  {author} {\bibinfo {author} {\bibfnamefont {J.}~\bibnamefont
  {Polchinski}},\ }\href {\doibase 10.1017/CBO9780511816079} {\emph {\bibinfo
  {title} {{String theory. Vol. 1: An introduction to the bosonic string}}}},\
  Cambridge Monographs on Mathematical Physics\ (\bibinfo  {publisher}
  {Cambridge University Press},\ \bibinfo {year} {2007})\BibitemShut {NoStop}%
%%CITATION = INSPIRE-487240;%%
\bibitem [{\citenamefont {Hehl}\ \emph {et~al.}(1976)\citenamefont {Hehl},
  \citenamefont {Von Der~Heyde}, \citenamefont {Kerlick},\ and\ \citenamefont
  {Nester}}]{Hehl:1976kj}%
  \BibitemOpen
  \bibfield  {author} {\bibinfo {author} {\bibfnamefont {F.~W.}\ \bibnamefont
  {Hehl}}, \bibinfo {author} {\bibfnamefont {P.}~\bibnamefont {Von Der~Heyde}},
  \bibinfo {author} {\bibfnamefont {G.~D.}\ \bibnamefont {Kerlick}}, \ and\
  \bibinfo {author} {\bibfnamefont {J.~M.}\ \bibnamefont {Nester}},\ }\href
  {\doibase 10.1103/RevModPhys.48.393} {\bibfield  {journal} {\bibinfo
  {journal} {Rev. Mod. Phys.}\ }\textbf {\bibinfo {volume} {48}},\ \bibinfo
  {pages} {393} (\bibinfo {year} {1976})}\BibitemShut {NoStop}%
%%CITATION = RMPHA,48,393;%%
\bibitem [{\citenamefont {Perez}\ and\ \citenamefont
  {Rovelli}(2006)}]{Perez:2005pm}%
  \BibitemOpen
  \bibfield  {author} {\bibinfo {author} {\bibfnamefont {A.}~\bibnamefont
  {Perez}}\ and\ \bibinfo {author} {\bibfnamefont {C.}~\bibnamefont
  {Rovelli}},\ }\href {\doibase 10.1103/PhysRevD.73.044013} {\bibfield
  {journal} {\bibinfo  {journal} {Phys. Rev.}\ }\textbf {\bibinfo {volume}
  {D73}},\ \bibinfo {pages} {044013} (\bibinfo {year} {2006})},\ \Eprint
  {http://arxiv.org/abs/gr-qc/0505081} {arXiv:gr-qc/0505081 [gr-qc]}
  \BibitemShut {NoStop}%
%%CITATION = GR-QC/0505081;%%
\bibitem [{\citenamefont {Freidel}\ \emph {et~al.}(2005)\citenamefont
  {Freidel}, \citenamefont {Minic},\ and\ \citenamefont
  {Takeuchi}}]{Freidel:2005sn}%
  \BibitemOpen
  \bibfield  {author} {\bibinfo {author} {\bibfnamefont {L.}~\bibnamefont
  {Freidel}}, \bibinfo {author} {\bibfnamefont {D.}~\bibnamefont {Minic}}, \
  and\ \bibinfo {author} {\bibfnamefont {T.}~\bibnamefont {Takeuchi}},\ }\href
  {\doibase 10.1103/PhysRevD.72.104002} {\bibfield  {journal} {\bibinfo
  {journal} {Phys. Rev.}\ }\textbf {\bibinfo {volume} {D72}},\ \bibinfo {pages}
  {104002} (\bibinfo {year} {2005})},\ \Eprint
  {http://arxiv.org/abs/hep-th/0507253} {arXiv:hep-th/0507253 [hep-th]}
  \BibitemShut {NoStop}%
%%CITATION = HEP-TH/0507253;%%
\bibitem [{\citenamefont {Magueijo}\ \emph {et~al.}(2013)\citenamefont
  {Magueijo}, \citenamefont {Zlosnik},\ and\ \citenamefont
  {Kibble}}]{Magueijo:2012ug}%
  \BibitemOpen
  \bibfield  {author} {\bibinfo {author} {\bibfnamefont {J.}~\bibnamefont
  {Magueijo}}, \bibinfo {author} {\bibfnamefont {T.~G.}\ \bibnamefont
  {Zlosnik}}, \ and\ \bibinfo {author} {\bibfnamefont {T.~W.~B.}\ \bibnamefont
  {Kibble}},\ }\href {\doibase 10.1103/PhysRevD.87.063504} {\bibfield
  {journal} {\bibinfo  {journal} {Phys. Rev.}\ }\textbf {\bibinfo {volume}
  {D87}},\ \bibinfo {pages} {063504} (\bibinfo {year} {2013})},\ \Eprint
  {http://arxiv.org/abs/1212.0585} {arXiv:1212.0585 [astro-ph.CO]} \BibitemShut
  {NoStop}%
%%CITATION = ARXIV:1212.0585;%%
\bibitem [{\citenamefont {Candelas}\ \emph {et~al.}(1985)\citenamefont
  {Candelas}, \citenamefont {Horowitz}, \citenamefont {Strominger},\ and\
  \citenamefont {Witten}}]{Candelas:1985en}%
  \BibitemOpen
  \bibfield  {author} {\bibinfo {author} {\bibfnamefont {P.}~\bibnamefont
  {Candelas}}, \bibinfo {author} {\bibfnamefont {G.~T.}\ \bibnamefont
  {Horowitz}}, \bibinfo {author} {\bibfnamefont {A.}~\bibnamefont
  {Strominger}}, \ and\ \bibinfo {author} {\bibfnamefont {E.}~\bibnamefont
  {Witten}},\ }\href {\doibase 10.1016/0550-3213(85)90602-9} {\bibfield
  {journal} {\bibinfo  {journal} {Nucl. Phys.}\ }\textbf {\bibinfo {volume}
  {B258}},\ \bibinfo {pages} {46} (\bibinfo {year} {1985})}\BibitemShut
  {NoStop}%
%%CITATION = NUPHA,B258,46;%%
\bibitem [{\citenamefont {Shapiro}(2002)}]{Shapiro:2001rz}%
  \BibitemOpen
  \bibfield  {author} {\bibinfo {author} {\bibfnamefont {I.~L.}\ \bibnamefont
  {Shapiro}},\ }\href {\doibase 10.1016/S0370-1573(01)00030-8} {\bibfield
  {journal} {\bibinfo  {journal} {Phys. Rept.}\ }\textbf {\bibinfo {volume}
  {357}},\ \bibinfo {pages} {113} (\bibinfo {year} {2002})},\ \Eprint
  {http://arxiv.org/abs/hep-th/0103093} {arXiv:hep-th/0103093 [hep-th]}
  \BibitemShut {NoStop}%
%%CITATION = HEP-TH/0103093;%%
\bibitem [{\citenamefont {Ortin}(2004)}]{Ortin:2004ms}%
  \BibitemOpen
  \bibfield  {author} {\bibinfo {author} {\bibfnamefont {T.}~\bibnamefont
  {Ortin}},\ }\href {\doibase 10.1017/CBO9780511616563} {\emph {\bibinfo
  {title} {{Gravity and strings}}}},\ Cambridge Monographs on Mathematical
  Physics\ (\bibinfo  {publisher} {Cambridge Univ. Press},\ \bibinfo {year}
  {2004})\BibitemShut {NoStop}%
%%CITATION = INSPIRE-648696;%%
\end{thebibliography}%

%%%%%%%%%%%%%%%%%%%%%%%%%%%%%%%%%%%%%%%%%%%%%%%%%%
%%%%%%%%%%%%%%%%%%%%%%%%%%%%%%%%%%%%%%%%%%%%%%%%%%

\onecolumngrid
\appendix
\section{Basics of massless kinematics and spinor-helicity formalism}
\label{app:A}

We consider the massless 2-to-2 scattering $1(p_1) + 2(p_2) \to \bar{3}(\bar{p}_3) + \bar{4}(\bar{p}_4)$ in the CM frame with four-momenta
\begin{equation}
p_1 = \left(E,0,0,E\right)~,~~~p_2 = \left(E,0,0,-E\right)~,~~~
\bar{p}_3 = (E,E\sin\theta,0,E\cos\theta)~,~~~
\bar{p}_4 = (E,-E\sin\theta,0,-E\cos\theta)\,,\label{eq:Momenta}
\end{equation}
where $\theta$ is the scattering angle. The physical region for this process is the red region {\color{red}{(r)}} in fig.~\ref{fig:schematic}.
The four-momenta in eq.~(\ref{eq:Momenta}) describe ingoing initial-state and outgoing finale-state particles.
The Mandelstam variables are given by 
\begin{eqnarray}
s &\equiv& (p_1 + p_2)^2 = (\bar{p}_3 + \bar{p}_4)^2 = 4E^2\,,\\
t&\equiv & (p_1 - \bar{p}_3)^2 = (p_2 - \bar{p}_4)^2 = -\frac{4E^2}{2}\left(1-\cos\theta\right)= -\frac{s}{2}\left(1-\cos\theta\right) = -s\sin^2\frac{\theta}{2}\,,\label{eq:Mandt}\\
u&\equiv & (p_1 - \bar{p}_4)^2 = (p_2 - \bar{p}_3)^2 = -\frac{4E^2}{2}\left(1+\cos\theta\right)= -\frac{s}{2}\left(1+\cos\theta\right) = -s\cos^2\frac{\theta}{2} = -s-t\,,
\end{eqnarray}
and, consequently, we have for the scattering angle $\cos\theta = 1+2t/s$. For the on-shell production of a resonance with mass $M$ we have $s=4E^2 = M^2$.

For a massless particle of momentum $p^{\mu}\equiv(p^0,p^1,p^2,p^3)$, $p^2=0$, the map into a $(1/2,1/2)$ representation of the Lorentz group of the momentum can be written as the outer product of two two-component spinors (we use the mostly-minus flat Minkowski metric)
\begin{eqnarray}
\hat{p}_{\alpha\dot{\alpha}} \equiv p_{\mu}(\sigma^{\mu})_{\alpha\dot{\alpha}} = 
\left(
\begin{array}{cc}
p^0 - p^3  & -p^1 + ip^2   \\
-p^1 -ip^2  &   p^0 + p^3
\end{array}
\right) ={}_{\alpha}\!\spl{p}\sprf{p}_{\dot{\beta}}
%\lambda_{\alpha}\tilde{\lambda}_{\dot{\alpha}}\,,
\end{eqnarray}
given that det$(\hat p)=p^2 = 0$. For real momenta, $p_{\alpha\dot{\alpha}}$ is Hermitian and we have the reality condition $\sprf{p}_{\dot{\alpha}} = \pm({}_{\alpha}\!\spl{p})^*$. 
Indices are raised and lowered as $\varepsilon^{\alpha\beta}{}_{\beta}\!\spl{p} = \splf{p}\!^{\alpha}$ and $\sprf{p}_{\dot{\alpha}}=\varepsilon_{\dot{\alpha}\dot{\beta}}{}^{\dot{\beta}}\!\spr{p}$, 
with the Levi-Civita symbol in two dimensions $\varepsilon^{\alpha\beta} = \varepsilon^{\dot{\alpha}\dot{\beta}} = ((0,1),(-1,0)) = -\varepsilon_{\alpha\beta} = -\varepsilon_{\dot{\alpha}\dot{\beta}}$.

In the main text, we used explicitly the spinor-helicity formalism following the convention according to which all momenta are ingoing.
This means that, compared to eq.~(\ref{eq:Momenta}), we have $\bar{p}_{3,4} \to p_{3,4} =-\bar{p}_{3,4}$ (notice that these flipped momenta 
enter in fig.~\ref{fig:schematic} in the definition of $s_{ij}$).
In this case, we find that one possible explicit choice of spinors is
\begin{eqnarray}
(\hat{p}_1)_{\alpha\dot{\alpha}} &=& 
\left(
\begin{array}{cc}
 0 &  0  \\
 0 &  2E
\end{array}
\right) = {}_{\alpha}\!\spl{p_1}\sprf{p_1}_{\dot{\beta}}
%\lambda_{1,\alpha}\tilde{\lambda}_{1,\dot{\alpha}}
~~~\Rightarrow~~~
\spl{p_1} =\sprf{p_1}=  \left(
\begin{array}{c}
 0   \\
 \sqrt{2E}
\end{array}
\right)\,,\\
(\hat{p}_2)_{\alpha\dot{\alpha}} &=& 
\left(
\begin{array}{cc}
 2E &  0  \\
 0 &  0
\end{array}
\right) = {}_{\alpha}\!\spl{p_2}\sprf{p_2}_{\dot{\beta}}~~~\Rightarrow~~~
\spl{p_2} =\sprf{p_2}=  \left(
\begin{array}{c}
  \sqrt{2E}   \\
0
\end{array}
\right)\,,
\end{eqnarray}
and
\begin{align}
(\hat{p}_3)_{\alpha\dot{\alpha}} &= \left(
\begin{array}{cc}
 -2E\sin^2\frac{\theta}{2} &  E\sin\theta  \\
 E\sin\theta &  -2E\cos^2\frac{\theta}{2}
\end{array}
\right) = {}_{\alpha}\!\spl{p_3}\sprf{p_3}_{\dot{\beta}}~~~\Rightarrow~~~
\spl{p_3} = - \sprf{p_3} = \sqrt{2E} \left(
\begin{array}{c}
 -\sin\frac{\theta}{2}   \\
\cos\frac{\theta}{2}
\end{array}
\right)\,,\label{eq:p3Flipped}\\
%~~\sprf{p_3} = -(\spl{p_3})^*\,,\label{eq:p3Flipped}\\
(\hat{p}_4)_{\alpha\dot{\alpha}} &= \left(
\begin{array}{cc}
 -2E\cos^2\frac{\theta}{2} &  -E\sin\theta  \\
 -E\sin\theta &  -2E\sin^2\frac{\theta}{2}
\end{array}
\right) ={}_{\alpha}\!\spl{p_4}\sprf{p_4}_{\dot{\beta}}~~~\Rightarrow~~~
\spl{p_4}= - \sprf{p_4} = \sqrt{2E} \left(
\begin{array}{c}
 \cos\frac{\theta}{2}   \\
\sin\frac{\theta}{2}
\end{array}
\right)\,.\label{eq:p4Flipped}
%~~\sprf{p_4} = -(\spl{p_4})^*\,.\label{eq:p4Flipped}
\end{align}
One can now compute explicitly the angle/square brackets that enter in the helicity structure of the scattering amplitudes studied in the main text (see table~\ref{tab:1}) for region {\color{red}(r)} whereas the case for regions {\color{blue}(b)} and {\color{ForestGreen}(g)} is obtained by a reshuffling of the $1$-$2$-$3$-$4$ indexes.
In summary we have, in each of the regions,
\begin{gather}
\scpl{2}{3}\scpr{1}{4} \\\nonumber
\begin{tikzpicture}
\draw [very thick, blue, ->] (0,0)--(0,-1);
\draw [very thick, red, ->] (-0.5,0)--(-2.5,-1);
\draw [very thick, ForestGreen, ->] (0.5,0)--(2.5,-1);
\end{tikzpicture}\\
-s\cos^{2}\frac{\theta}{2}\quad\quad\quad s\cos^{2}\frac{\theta}{2}\quad\quad\quad\quad-s \nonumber
\end{gather}
and
\begin{gather}
\splf{3} \hat P_{12}\spr{4} \\\nonumber
\begin{tikzpicture}
\draw [very thick, blue, ->] (0,0)--(0,-1);
\draw [very thick, red, ->] (-0.5,0)--(-2.5,-1);
\draw [very thick, ForestGreen, ->] (0.5,0)--(2.5,-1);
\end{tikzpicture}\\
\frac{s}{2}\sin\theta\quad\quad\quad -s\cos\frac{\theta}{2}\quad\quad\quad s\cos\frac{\theta}{2} \nonumber
\end{gather}
from which it is possible to write explicitly all amplitudes in terms of conventional Mandelstam variables.\\

Finally, useful formulas used but not quoted in the text are the following.

\begin{itemize}

\item [$\circ$] In section~\ref{sec:Legendre} an example of the contractions that give rise to the spin $J$ mediated four-point amplitude with external scalars 
 is given by (for $J = 2,\,3,\,4$)
\begin{align}\label{eq:Lassos}
%&\frac{1}{(2h)!}\mbox{Cy}(\{n_i\})\prod_i\mbox{tr}((P_{13}P_{24})^{n_i})\\
J=2\,,\quad&\frac{1}{4!}\left[8\,
\raisebox{-1.5mm}{\begin{tikzpicture}
	\draw [thick, rounded corners=2pt] (0.125,0)--(.25,0)--(0.25,.5)--(0,.5)--(0,0)--(0.125,0);
	\end{tikzpicture}
	\begin{tikzpicture}
	\draw [thick, rounded corners=2pt] (0.125,0)--(.25,0)--(0.25,.5)--(0,.5)--(0,0)--(0.125,0);
	\end{tikzpicture}
}+16\,
\raisebox{-1.5mm}{\begin{tikzpicture}
	\draw [thick, rounded corners=2pt] (0.175,0)--(.25,0)--(0.5,.5) -- (0.625,.5)--(.625,0) -- (.5,0) -- (.25,.5)-- (0.125,.5)-- (.125,0) --(0.2,0);
	\end{tikzpicture}
}
\right]\\ \label{eq:Lassos2}
J=3\,,\quad&\frac{3!2^3}{6!}\left[\left(\raisebox{-1.5mm}{\begin{tikzpicture}
	\draw [thick, rounded corners=2pt] (0.125,0)--(.25,0)--(0.25,.5)--(0,.5)--(0,0)--(0.125,0);
	\end{tikzpicture}
}\!\right)^3+ 6\,
\raisebox{-1.5mm}{\begin{tikzpicture}
	\draw [thick, rounded corners=2pt] (0.125,0)--(.25,0)--(0.25,.5)--(0,.5)--(0,0)--(0.125,0);
	\end{tikzpicture}
	\begin{tikzpicture}
	\draw [thick, rounded corners=2pt] (0.175,0)--(.25,0)--(0.5,.5) -- (0.625,.5)--(.625,0) -- (.5,0) -- (.25,.5)-- (0.125,.5)-- (.125,0) --(0.2,0);
	\end{tikzpicture}
}
+8\,
\raisebox{-1.5mm}{
	\begin{tikzpicture}
	\draw [thick, rounded corners=2pt] (0.175,0)--(.25,0)--(0.5,.5) -- (0.625,.5)--(.875,0) -- (1,0) -- (1,.5)-- (.875,.5)--(.625,0)--(.5,0)--(.25,.5)--(.125,.5)--(.125,0)--(.175,0);
	\end{tikzpicture}
}
\right]\\  \label{eq:Lassos3}
J=4\,,\quad&\frac{4!2^4}{8!}\Big[
\left(\raisebox{-1.5mm}{\begin{tikzpicture}
	\draw [thick, rounded corners=2pt] (0.125,0)--(.25,0)--(0.25,.5)--(0,.5)--(0,0)--(0.125,0);
	\end{tikzpicture}
}\!\right)^4+12\left(
\raisebox{-1.5mm}{\begin{tikzpicture}
	\draw [thick, rounded corners=2pt] (0.175,0)--(.25,0)--(0.5,.5) -- (0.625,.5)--(.625,0) -- (.5,0) -- (.25,.5)-- (0.125,.5)-- (.125,0) --(0.2,0);
	\end{tikzpicture}
}\!\right)^2+12\,
\left(\raisebox{-1.5mm}{\begin{tikzpicture}
	\draw [thick, rounded corners=2pt] (0.125,0)--(.25,0)--(0.25,.5)--(0,.5)--(0,0)--(0.125,0);
	\end{tikzpicture}
}\!\right)^2
\raisebox{-1.5mm}{\begin{tikzpicture}
	\draw [thick, rounded corners=2pt] (0.175,0)--(.25,0)--(0.5,.5) -- (0.625,.5)--(.625,0) -- (.5,0) -- (.25,.5)-- (0.125,.5)-- (.125,0) --(0.2,0);
	\end{tikzpicture}
}+32\,
\raisebox{-1.5mm}{\begin{tikzpicture}
	\draw [thick, rounded corners=2pt] (0.125,0)--(.25,0)--(0.25,.5)--(0,.5)--(0,0)--(0.125,0);
	\end{tikzpicture}
}\!\!
\raisebox{-1.5mm}{
	\begin{tikzpicture}
	\draw [thick, rounded corners=2pt] (0.175,0)--(.25,0)--(0.5,.5) -- (0.625,.5)--(.875,0) -- (1,0) -- (1,.5)-- (.875,.5)--(.625,0)--(.5,0)--(.25,.5)--(.125,.5)--(.125,0)--(.175,0);
	\end{tikzpicture}
}+48\,
\raisebox{-1.5mm}{
	\begin{tikzpicture}
	\draw [thick, rounded corners=2pt] (0.175,0)--(.25,0)--(0.5,.5) -- (0.625,.5)--(.875,0) -- (1,0) -- (1.25,.5)-- (1.375,.5) -- (1.375,0)--(1.25,0) -- (1,.5)--(.875,.5)-- (.625,0)--(.5,0)--(.25,.5)--(.125,.5)--(.125,0)--(.15,0);
	\end{tikzpicture}
}
\Big]
\end{align}
Each term schematically represents a product of traces over slashed momenta, which itself one writes in terms of Lorentz scalar products as
\begin{align}\label{eq:traces}
\mbox{tr}(\bar\sigma^\mu p_\mu\sigma^\mu k_\mu)^n=2^n\sum_m
{n\choose {2m}} (p\cdot k)^{n-2m}\left[(p\cdot k)^2-p^2k^2\right]^m\,.
\end{align}

\item [$\circ$] In section~\ref{sec:aVZaVS} we extracted in eqs.~(\ref{eq:VSbound1},\,\ref{eq:VSbound2}) an upper bound on the coefficient $\gamma_0$ of the modified 
Virasoro-Shapiro form factor $\mathcal A_{\rm VS}^{\gamma_0}$ as a consequence of unitarity applied to the scalar-fermion and scalar-vector scattering processes.
In more detail, in the scalar-fermion case the bound follows from the positivity condition imposed on the coefficient ${\rm N}^{1,1/2}_{1/2,1/2}$ while in
 the scalar-vector the same happens for ${\rm N}^{1,1}_{1,1}$. We find
\begin{align}
{\rm Scalar-fermion:}~~~~
&{\rm N}^{1,1/2}_{1/2,1/2} >0~~~&30 + \gamma_0\left(
101  + 46\gamma_0 -36\gamma_0^2
\right) >0~~~&\Longrightarrow~~~\gamma_0 < 2.5215\,, \label{eq:VSBound1} \\
{\rm Scalar-vector:}~~~~&{\rm N}^{1,1}_{1,1}>0~~~
 &14 +\gamma_0\left[
47-\gamma_0\left(
-24 + 13\gamma_0
\right)
\right] > 0~~~&\Longrightarrow~~~\gamma_0 < 3.1169\,.\label{eq:VSBound2}
\end{align}
%The solutions to N=0 for VS
%\begin{align}
%&\frac{23}{54}+\frac{2\sqrt{814}}{27}\cos\left(\frac13{\rm ArcTan}\left[\frac{135\sqrt{9039}}{31247}\right]\right) &&\frac{13}{8}+\frac{2\sqrt{803}}{13\sqrt{3}}\cos\left(\frac13{\rm ArcTan}%\left[\frac{26\sqrt{81710}}{12417\sqrt{3}}\right]\right)
%\end{align}

\end{itemize}

\end{document}